\documentclass[11pt]{article}

\usepackage[a4paper,margin=0.75in]{geometry}
\usepackage[T1]{fontenc}

\usepackage{color,array}
\usepackage{soul, xcolor}
\usepackage{graphicx}
\usepackage{amsmath, amssymb, amsthm}
\usepackage{tabularx}
\usepackage{multirow}
\newcolumntype{Y}{>{\centering\arraybackslash}X}

\definecolor{myred}{HTML}{000000}
\usepackage{booktabs}
\usepackage{subfig}
\usepackage{float}
\usepackage{cite}
\usepackage{hyperref}
\usepackage{cleveref}

\definecolor{revisioncolor}{HTML}{000000} 

\newcommand{\revision}[1]{\textcolor{revisioncolor}{#1}}  

\begin{document}

\thispagestyle{empty}
\noindent
This paper has been accepted for publication in \textit{IEEE Transactions on Aerospace and Electronic Systems}.

\begin{center}
DOI: \href{https://doi.org/10.1109/TAES.2026.3702657}{10.1109/TAES.2026.3702657}

\medskip
IEEE Xplore: \href{https://ieeexplore.ieee.org/document/11561815}{https://ieeexplore.ieee.org/document/11561815}
\end{center}
\vspace{2em}

\noindent
\copyright\ 2026 IEEE. Personal use of this material is permitted. Permission from IEEE must be obtained for all other uses, in any current or future media, including reprinting/republishing this material for advertising or promotional purposes, creating new collective works, for resale or redistribution to servers or lists, or reuse of any copyrighted component of this work in other works.

\clearpage
\setcounter{page}{1}

\title{Deep Uncertainty-aware Tracking for Maneuvering Targets} 

\author{Shuyang Zhang$^{1}$, Chang Gao$^{1}$, Bo Chen$^{1}$, Qingfu Zhang$^{2}$, Shefeng Yan$^{3}$, and Hongwei Liu$^{1}$\\[0.4em]
\small $^{1}$National Key Laboratory of Radar Signal Processing, Xidian University, Xi'an 710071, China\\
\small $^{2}$Department of Computer Science, City University of Hong Kong, Hong Kong, China\\
\small $^{3}$Institute of AI for Industries, Chinese Academy of Sciences, Nanjing 211135, China\\[0.4em]
\small Corresponding authors: Chang Gao (\href{mailto:changgao@xidian.edu.cn}{changgao@xidian.edu.cn}) and Hongwei Liu (\href{mailto:hwliu@xidian.edu.cn}{hwliu@xidian.edu.cn})\\
\small This work was supported in part by the National Natural Science Foundation of China (62192714, 62192711, 62501446).}
\date{}
\maketitle

\begin{abstract}
Traditional target tracking approaches typically presuppose a target motion model. 
Once the target maneuvers, the model may get mismatched and is incapable of depicting the complex motion characteristics of the target.
Most data-driven methods are designed based on a regression-only framework, which increases the difficulty of network training. 
Moreover, these methods are typically designed under the assumption of a fixed measurement noise distribution. 
However, the measurement noise distribution of actual radar is influenced by fluctuations in the target's radar cross section.
Such fluctuations may lead to a degradation in tracking performance. 
In this paper, we propose a representation method for radar measurements, referred to as the measurement uncertainty projection operator (MUPO). 
It projects the measurement onto a target state space to capture the measurement noise distribution.
Additionally, we propose a framework that reformulates maneuvering target tracking as a hybrid task of classification and regression. 
To realize the functionality of this framework, we establish a MUPO-based target tracking network.
The rationality of the proposed network structure is supported by the validation experiment. 
Numerical results demonstrate that the proposed method achieves superior estimation precision compared with existing data-driven methods during target maneuvers. 
Furthermore, the proposed algorithm retains its superior tracking performance when extended to 3D maneuvering \revision{target} tracking and multi-target tracking scenarios.

\end{abstract}


\section{INTRODUCTION}

T{\scshape arget} tracking can typically be described as the process of using sensors to collect information about a target of interest and leveraging this information to estimate the target's precise position, velocity, and other relevant parameters.
Target tracking comprises two fundamental components: data association, which resolves measurement-origin uncertainty by associating measurements with the appropriate targets in the presence of clutter and false alarms, and target state estimation, which manages target motion uncertainty to accurately estimate the target's state \cite{cite3, cite4}.  
These components are equally vital for achieving robust target tracking.
This paper primarily focuses on the filtering aspect, emphasizing accurate state estimation from noisy measurements.
%
The majority of the tracking approaches are designed within the Bayesian framework, such as the Kalman filter (KF) and its derivatives, namely the extended Kalman filter (EKF) and the unscented Kalman filter (UKF) \cite{cite6},\cite{cite7},\cite{cite8}. 
Nevertheless, these methods function well only when the target motion model matches the target dynamic behavior.  
In the real world, the dynamic behavior of the target typically changes over time.  
It is often challenging for the sensors to perceive these changes in advance.  
Particularly for a non-cooperative target, when the target's dynamic behavior undergoes an unknown random change, the target motion model of the filter may become mismatched. 
It gives rise to an adverse effect in the estimation of target states, and ultimately the radar system may lose the target.  
The most commonly used methods for handling the target motion uncertainty can be categorized into two types: target tracking after maneuver detection and adaptive target tracking without maneuver detection \cite{cite9}.

Target tracking after maneuver detection typically necessitates the detection of the target's sudden change, and subsequently adjusts the filter gain or the filter structure to deal with the target's maneuver. 
Among them, the process noise adaptive approach regards the target maneuver as the process noise. 
When the target maneuver is detected, this method adjusts the filter gain by changing the power of the filter's process noise \cite{cite10},\cite{cite11},\cite{cite12}. 
The input estimation method and the variable-dimension filtering method, on the other hand, change the filter structure to accommodate the tracking of the maneuvering target after detecting the target maneuver \cite{cite13, cite14}. 
Such methods typically have a time delay in the detection of the target maneuver. 
When the time delay is similar to the autocorrelation time constant of the target maneuver, these methods do not perform effectively. 

Adaptive target tracking conducts state estimation and modifies the filter gain concurrently without maneuver detection. 
In the adaptive maneuvering target tracking algorithm, the Singer model method presumes that the noise process is not white Gaussian noise and models the target acceleration as a zero-mean exponential autocorrelated random process \cite{cite15}. 
Meanwhile, the current statistical model method estimates the mean of the target acceleration and utilizes the estimated value to rectify the acceleration distribution in the Singer model, eventually transferring the information of target acceleration to the subsequent filter gain in the form of variance \cite{cite16}.
It is unrealistic to anticipate that one model can comprehensively describe the target maneuver. 
The multiple model (MM) algorithm supposes several target motion models and computes the current probability of each model. 
The algorithm selects the model based on the probability and conducts the estimation of the target state through weighted summation \cite{cite17},\cite{cite18},\cite{cite19}. 
On the basis of MM, the interacting multiple model (IMM) algorithm assumes that the transitions between different models obey a known finite-state Markov chain, and it realizes the tracking of maneuvering targets by using multiple models through the interaction of input and output \cite{cite20}, \cite{cite21}. 
When the model set covers the dynamic behavior of the target, this type of method can have better accuracy in estimating the target state. 
In practical applications, in order to make the model set contain more models, the model set needs to be expanded, but this will bring about a computational burden, and the tracking accuracy may decrease when more models are used \cite{cite18}. 

However, the aforementioned model-based tracking methods remain unsuitable for complex real-world scenarios due to model mismatch.
The data-driven methods are capable of directly learning the mapping from measurements to target states via a large amount of training data. 
The research conducted in the field of target tracking has primarily focused on recurrent neural networks (RNNs), including the advanced variants such as deep long short-term memory neural network (DLSTM), bidirectional long short-term memory neural network (BLSTM) \cite{lstm },\cite{dlstm},\cite{bilstm}, as well as on Transformer-based architectures \cite{cite30}.
Some studies have directly utilized the DLSTM as the filter itself, and the network output is directly used as the target state estimation. 
Pioneered by \cite{cite25, cite26}, a filter structure based on the Bayesian filtering theory has been designed, leveraging DLSTM-based models to learn the conditional and predictive probabilities under the Bayesian framework, thereby enabling the tracking of maneuvering targets.
%
%
Additionally, the BLSTM was utilized to rectify the estimation of the UKF \cite{cite28}.
Unlike the direct application of RNN for target state estimation, it was proposed to use RNNs to calculate filtering parameters in a Bayesian filtering framework, achieving more robust maneuvering target tracking \cite{cite29}. 
As a step further, a Transformer was used to design a maneuvering target tracking algorithm, reducing the estimation error caused by the strong maneuvering of the target \cite{cite31}.

While existing data-driven target tracking methods that are designed under the regression-only framework have demonstrated promising performance, they predominantly treat both the radar measurements as continuous inputs fed directly into the network architecture. 
In preprocessing, data normalization or standardization is commonly applied. 
However, for maneuvering target tracking at long ranges (e.g., hundreds of kilometers), these preprocessing methods either compress data differences to very small scales or transform data into zero-mean Gaussian distributions, potentially causing performance degradation and model instability \cite{prepocess}.
In addition to the preprocessing limitations outlined above, current data-driven target tracking methods typically employ conventional regression losses such as mean squared error (MSE) or mean absolute error (MAE), which exhibit high sensitivity to outliers and weak gradients when predictions approach target values \cite{add2}.
Furthermore, existing data-driven methods generally assume fixed measurement noise distributions. 
In practice, however, measurement noise distributions vary with the target's radar cross-section, and algorithms relying on fixed distribution assumptions may suffer performance deterioration.


\begin{table*}[!ht]
    \centering
    \renewcommand{\arraystretch}{1.8} 
    \caption{\revision{Summary of existing maneuvering target tracking methods.}}
    \label{r1tt1}
    
    \begin{tabularx}{\textwidth}{
        >{\color{myred}\centering\arraybackslash}m{2.5cm} | 
        >{\color{myred}}Y | 
        >{\color{myred}}Y | 
        >{\color{myred}}Y | 
        >{\color{myred}}Y | 
        >{\color{myred}}Y
    }
    \toprule
    \textbf{Method} & \textbf{Handling of Motion Uncertainty} & \textbf{Handling of Measurement Noise} & \textbf{Tracking Algorithm Framework} & \textbf{Advantages} & \textbf{Disadvantages}\\ \midrule
    
    \cite{cite10,cite11,cite12} 
    & Adaptively adjusting the filter gain 
    & \multirow{4}{=}{\centering Integration of measurement noise covariance into the Bayesian update to derive Kalman gain} 
    & \multirow{4}{=}{\centering Bayesian filtering framework}
    & \multirow{4}{=}{\centering High interpretability; low computational complexity}
    & \multirow{4}{=}{\centering Constrained by predefined models; sensitive to model mismatch} \\ \cline{1-2}

    \cite{cite13,cite14} 
    & Adaptively adjusting the filter structure 
    & & & 

    \\ \cline{1-2}
    
    \cite{cite15,cite16} 
    & Adaptive tuning of process noise covariance 
    & & &   

    \\ \cline{1-2}

    \cite{cite17,cite18,cite19,cite20,cite21} 
    & Probability-weighted fusion among a predefined Markov model set 
    & & & 

    \\ \hline
    
    \cite{cite25,cite26,cite28,cite29,cite31} 
    & Implicit learning of complex motion models from offline data 
    & Fixed measurement noise distribution 
    & Regression-only framework 
    & Capable of learning complex motion patterns from large-scale datasets 
    & High training difficulty due to complex mapping space; neglecting variations in measurement noise distribution 
    \\ \midrule
    
    \textbf{MUPO-TTN} 
    & \textbf{Implicit learning of complex motion models from offline data} 
    & \textbf{Using MUPO to capture the distribution of target measurement noise} 
    & \textbf{Hybrid classification and regression framework} 
    & \textbf{Leveraging measurement uncertainty; simplifying mapping space to facilitate training} 
    & \textbf{Increased computational overhead} 
    \\ 
    \bottomrule
    \end{tabularx}\label{r1tt1}
\end{table*}

To address the aforementioned issues, we propose a different target tracking framework.
Table \ref{r1tt1} provides a summary of existing maneuvering target tracking methods. 
Unlike established tracking frameworks, the proposed framework reformulates maneuvering target tracking as a hybrid task of classification and regression. Specifically, the proposed framework transforms part of the output space into a finite set, thereby reducing its complexity. 
In parallel, we design a measurement representation scheme tailored to this framework, which not only leverages the statistical characteristics of measurement noise distributions but also simplifies the complexity of the network’s input space. 
Different from the existing data-driven algorithms designed based on regression, this approach achieves higher estimation precision.
Furthermore, the proposed framework extends to multi-target tracking, demonstrating superior tracking performance for multi-target compared to conventional methods.
In summary, the main contributions of this paper are as follows:
\begin{itemize}
    \item A measurement uncertainty projection operator (MUPO), as a presentation for characterizing measurement sequences, has been proposed.
    It samples the statistical characteristics from a target state space (TSS) established based on the distribution of target measurement noise.
    Under this representation, the network is capable of leveraging the time-varying distribution information.
    \item A multi-channel MUPO has been proposed, where it integrates multiple MUPO representations derived from the original measurement sequence and its diverse algorithmic preprocessing results. 
    This design leverages complementary information across channels to guide model training and enhance performance in downstream tasks. 
    \item A framework for maneuvering target tracking is proposed, which formulates target tracking as a hybrid task of classification and regression. 
    In order to realize its functionality, a MUPO-based target tracking network (MUPO-TTN) is designed to exploit the uncertainty of radar measurements across multiple frames. 
    A domain knowledge-informed loss function is designed to assist the network training process.
\end{itemize}

The remainder of this paper is organized as follows. 
Section \ref{prob form} introduces maneuvering target tracking and the problems brought by the assumption of a fixed distribution. 
Section \ref{mupotnn} describes the specific maneuvering target tracking method proposed in this paper. 
Section \ref{ne} discusses the generation of the dataset, the validation of the network's rationality, and comparisons with other algorithms.
Finally, Section \ref{co} concludes the paper.
\section{PRELIMINARY}\label{prob form}
The mathematical models in the target tracking comprise the target state model and the measurement model. 
The target state model depicts the evolution of the target state over time, while the measurement model portrays the relationship between the target states and the sensor measurements. 
For the discrete-time tracking system, the state model and the measurement model can respectively be expressed as
\begin{equation}\label{eq1}
\mathbf{x}_k=f\left(\mathbf{x}_{k-1}, \mathbf{v}_{k-1}\right),
\end{equation}
\begin{equation}\label{eq2}
\mathbf{z}_k=h\left(\mathbf{x}_k, \mathbf{w}_k\right),
\end{equation}
where subscript $k$ represents the time index, $\mathbf{x}_k$ is the state vector, such as target position, $f(\cdot)$  is the evolution function of the target, $ \mathbf{v}_{k}$ is the process noise, $\mathbf{z}_k$ represents the measurement vector such as radial distance and azimuth of radar observation, the measurement function is denoted as $h(\cdot)$, and the measurement noise is denoted as $\mathbf{w}_k$. 
$\mathbf{v}_{k}$ and $\mathbf{w}_k$ are typically modeled as a zero-mean Gaussian white noise \cite{cite32}.

When the target is maneuvering, the target dynamic behavior changes rapidly, which means both the evolution function and the distribution of process noise may change. 
In the above case, (\ref{eq1}) can be rewritten as
\begin{equation}\label{eq3}
\mathbf{x}_k=f_k\left(\mathbf{x}_{k-1}, \mathbf{v}_{k-1}\right), f_k \in \mathcal{F},
\end{equation}
where $\mathcal{F}$ represents the class of evolution function $f_k(\cdot)$. 
$\mathbf{v}_{k}$ obeys the distribution $P_k$ and $P_k \in \mathcal{P}$, where $\mathcal{P}$ is the class of process noise's distribution.

In tracking of maneuvering targets, a key step is to match the target motion model with the target dynamic behavior. 
However, the target state evolution function is typically unknown. 
In such a case, the model matching in target tracking is more akin to solving the following optimization problem to acquire the optimal state evolution function $f_k^*$ and process noise distribution $P_k^*$:
\begin{equation}\label{eq4}
(f_k^*, P_k^*):=\arg \min _{f, P} \mathcal{L}_k(f, P),
\end{equation}
where the functional $\mathcal{L}_k(f, P): \mathcal{F} \times \mathcal{P} \mapsto \mathbb{R}$ measures the tracking performance of the tracking algorithm. 

However, in practical applications, the searching space of the model based tracking algorithm is $\mathcal{F}^{\prime} \times \mathcal{P}^{\prime}$ , where $\mathcal{F}^{\prime}, \mathcal{P}^{\prime}$ are predetermined in advance, and $\mathcal{F}^{\prime} \subseteq \mathcal{F}, \mathcal{P}^{\prime} \subseteq \mathcal{P}$. 
The tracking algorithm merely attains a local optimum within the entire searching space. 
Its tracking performance will deteriorate in the absence of a matching real-world target motion model and process noise distribution within the $\mathcal{F}^{\prime} \times \mathcal{P}^{\prime}$.
Data-driven tracking algorithms can acquire the motion model of the target from a considerable volume of data, thereby offering a more diverse searching space and boosting the probability of obtaining a matching target motion model and process noise distribution. 
Besides, for most maneuvering target tracking algorithms, the matching process of $f_k$ and $P_k$ is typically not explicitly manifested.

From the viewpoint of functional implementation, the data-driven methods endeavor to establish a mapping from the target measurements to the requisite parameters: $\Phi\left(\mathbf{z}_{1: L}\right): \mathbb{R}^{L \times N} \mapsto \mathbb{R}^M$, where $N,M$ signify the dimensions of the measurement vector and the estimated parameter vector respectively and $\mathbf{z}_{1: L}:=\left\{\mathbf{z}_i \mid i=1, \ldots, L\right\}$ stands for the measurement sequence and $L$ is the length of it.
The estimated parameters herein include the target state, the filter gain, or the residual between the reference filter and the target state. 

\subsection{Optimization Conflicts} 
\begin{figure}[tbp]
    \centering
    \includegraphics[width=1.\linewidth]{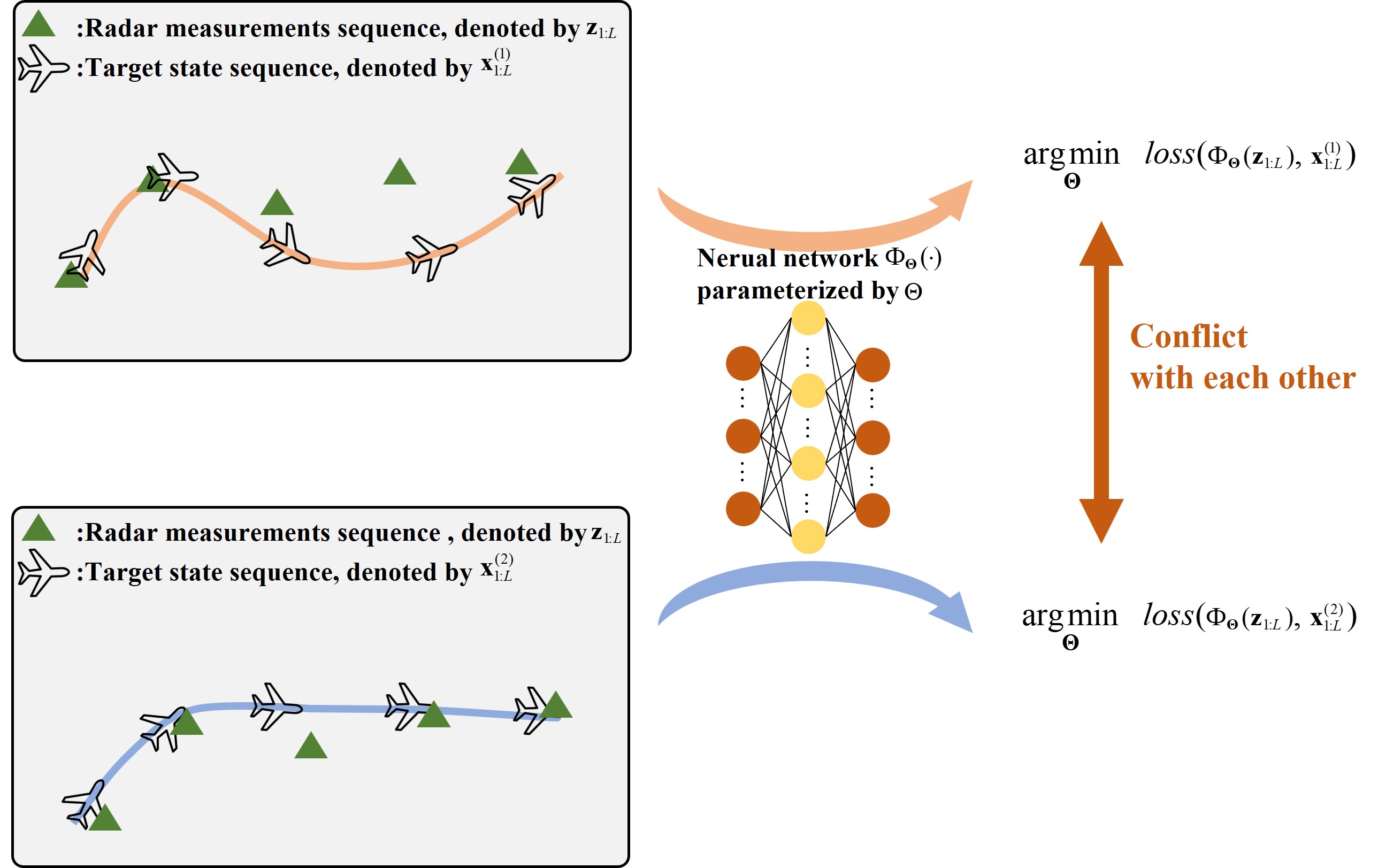}
    \caption{Illustration of optimization conflicts in existing data-driven methods arising from identical measurements under different hypotheses due to measurement noise.}
    \label{fig.1}
\end{figure}

The majority of existing data-driven approaches under the fixed distribution assumption only utilize position information from measurements, which can lead to optimization conflicts. 
As illustrated in Fig. \ref{fig.1}, two hypotheses are considered.
Under the first hypothesis, a target possesses its own states $\mathbf{x}_{1: L}^{(1)}:=\left\{\mathbf{x}_i^{(1)} \mid i=1, \ldots, L\right\}$, from which a sequence of measurements $\mathbf{z}_{1: L}^{(1)}:=\left\{\mathbf{z}_i^{(1)} \mid i=1, \ldots, L\right\}$ is obtained via radar observation.
Under the second hypothesis, the target state is assumed to be $\mathbf{x}_{1: L}^{(2)}$, which represents a dynamic behavior entirely different from that described by $\mathbf{x}_{1: L}^{(1)}$. Radar observation in this case yields a measurement sequence $\mathbf{z}_{1: L}^{(2)}$.
Due to the stochastic nature of measurement noise, there may exist scenarios in which the instantaneous measurement values under the two hypotheses are close, particularly when the measurement noise variance is large.
In such cases, $\mathbf{z}_{1: L}^{(1)}$ and $\mathbf{z}_{1: L}^{(2)}$ may take identical values, and are therefore represented by the same symbol $\mathbf{z}_{1: L}$ in Fig. \ref{fig.1}.
Moreover, common preprocessing procedures in the existing data-driven methods tend to further reduce this discrepancy.
In such a case, which is analogous to the scenario described above, the training inputs consist of two measurement sequences with highly similar values, whereas their corresponding references are different.
This inconsistency leads to conflicting parameter update directions, since the network is faced with nearly identical inputs associated with distinct targets, making it unclear toward which input–reference pair the optimization should proceed.
The primary cause of the aforementioned problem is that most existing data-driven methods for maneuvering target tracking fail to account for the confidence of individual measurements, namely the distribution of measurement noise.
\subsection{Degradation of Tracking Performance}In practical radar target tracking, the distribution of measurement noise is stochastic. 
Algorithms designed based on fixed noise distributions treat the measurement noise distribution as prior information and solely utilize target position for tracking. 
However, the disregard for the confidence information of measurements will potentially lead to the degradation of tracking performance.

To investigate the factors contributing to the degradation in tracking performance, the scenario depicted in Fig. \ref{fig.1} is considered.
Under the two hypotheses illustrated in Fig. \ref{fig.1}, the dynamic behaviors of the two targets are distinct, which can be expressed as $\sum_{i=1}^L\left\|\mathbf{x}_i^{(1)}-\mathbf{x}_i^{(2)}\right\| \gg 0$ where $\|\cdot\|$ are norm-based operations. 
However, their corresponding measurements are highly similar, particularly after certain normalization procedures, which can be expressed as $\sum_{i=1}^L\left\|\mathbf{z}_i^{'(1)}-\mathbf{z}_i^{'(2)}\right\| \approx 0$ where $\mathbf{z}_i^{'(1)},\mathbf{z}_i^{'(2)}$ denote the measurements that are preprocessed.
Since mapping $\Phi\left(\cdot\right)$ can be regarded as a continuous function \cite{cite22}, \cite{cite23}, \cite{cite33}, a slight perturbation in the input leads to a correspondingly slight variation in the output:
\begin{equation}\label{eq5}
\begin{aligned}
& \forall \epsilon>0, \exists \delta>0, \text { s.t. if }\sum_{i=1}^L\left\|\mathbf{z}_{i}-\mathbf{z}_{i}^{(0)}\right\|<\delta, \\
& \text { then }\left\|\Phi\left(\mathbf{z}_{1: L}\right)-\Phi\left(\mathbf{z}_{1: L}^{(0)}\right)\right\|<\epsilon.
\end{aligned}
\end{equation}
where $\mathbf{z}_{i}^{(0)}$ represents a variable that differs slightly from $\mathbf{z}_{i}$.
In the network’s input space, the measurements corresponding to the two targets become difficult to distinguish, resulting in final network estimates of their states that tend to be similar, as shown in (\ref{eq5}). 
In this case, higher network precision and more complex training strategies are required to achieve discrimination; otherwise, the degradation of tracking performance becomes inevitable. 
This issue can be mitigated by incorporating higher-dimensional uncertainty information, specifically the measurement noise distribution, which serves to increase the effective distance between the target measurements in the input space.

\section{MEASUREMENT UNCERTAINTY PROJECTION OPERATOR-BASED TARGET TRACKING NETWORK}\label{mupotnn}
Research has demonstrated that an appropriate task form can lead to training stability and better performance \cite{add1}.
We propose a target tracking framework, termed target state estimation via point prediction, which formulates tracking as a joint classification and regression task.
Specifically, the space surrounding the current sensor measurement is discretized into multiple sub-regions.
A classifier first identifies the sub-region containing the true target, followed by a local regression model that estimates the precise target location within it.
Compared to regression-only based data-driven approaches, the proposed hybrid formulation alleviates the difficulty of network training because classification tasks induce sharper, more discriminative internal features compared to regression \cite{add3}.
To address the issues outlined in Section \ref{prob form}, we design a preprocessing operator, MUPO, which incorporates measurement noise uncertainty information and ensures compatibility with the proposed framework.
By integrating MUPO, we design the MUPO-TTN to realize the functionality of the proposed framework. 
\begin{figure*}[!tp]
    \centering
    \includegraphics[width=0.7\linewidth]{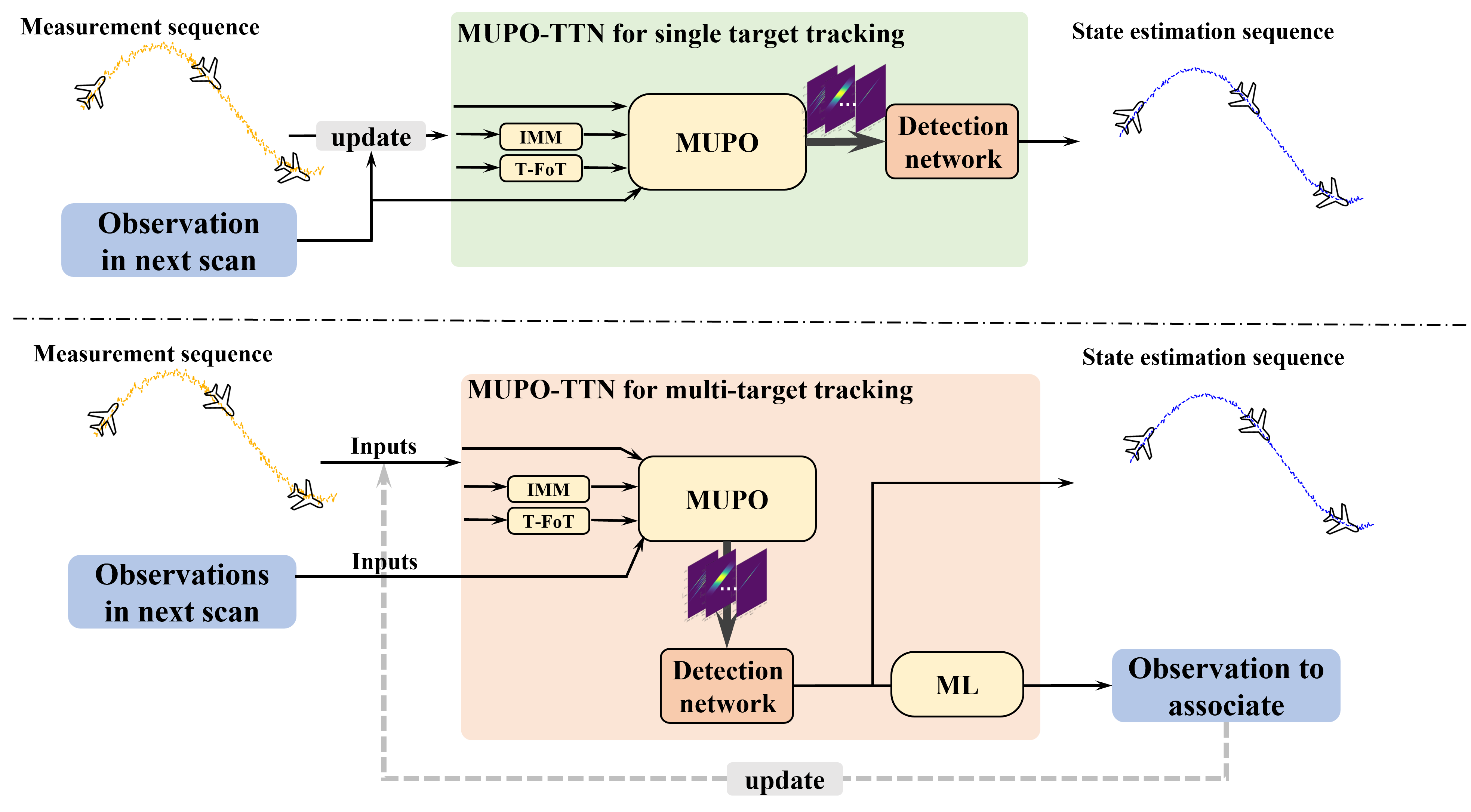}
    \caption{Diagram of MUPO-TTN.}
    \label{fig.2}
\end{figure*}

Fig. \ref{fig.2} illustrates the algorithmic workflows for single-target and multi-target tracking. 
In single-target tracking, the observation in the next frame contains only one target, and the new observation can be directly used to update the existing measurement sequence. 
The new measurement sequence, along with its results after IMM and FoT \cite{cite20}, \cite{cite42} processing, is combined into a multi-channel MUPO through MUPO fusion. 
The detection network performs target state estimation on the multi-channel MUPO.
In multi-target tracking, the observations in the next frame consist of multiple targets. 
The adopted strategy integrates data association and state estimation. 
Specifically, the observations to be associated in the current frame is passed through MUPO, and the existing measurement sequence, after being processed by IMM and FoT, is also passed through MUPO. 
Subsequently, the detection network estimates the target state, then the best-associated target in the current frame is selected using the maximum likelihood (ML) criterion.
Specifically, for multi-target scenarios, the Hungarian algorithm is employed to perform the global optimal assignment between tracks and measurements, thereby updating the sequence of target measurements.
\subsection{Measurement Uncertainty Projection Operator}\label{MUPO}
In light of the problem described in Section \ref{prob form}, the measurement noise distribution information constitutes an indispensable piece of data for attaining more precise target state estimation. 
For the sake of better exploiting this information, it is necessary to undertake an analysis of the measurement noise distribution. 
The distribution of the measurement noise is generally associated with the target SNR. 
Without loss of generality, the measurement noise is assumed to be additive. 
For example, in the context of the two-dimensional coordinates, radar measurements consist of radial distance and azimuth angle. 
The relationship between the precisions of these radar measurements and the SNR is as follows:

\begin{equation}\label{eq6}
\left\{
\begin{aligned}
& \sigma_\rho \propto \frac{c}{2 B \sqrt{2 S N R}} \\
& \sigma_\theta \propto \frac{\theta_{3 d B}}{1.6 \sqrt{2 S N R}}
\end{aligned}\right.,
\end{equation}
where $\sigma_\rho,\sigma_\theta$ represent the measurement precisions of distance and azimuth angle respectively, the speed of light is denoted as $c$, transmitted signal bandwidth as $B$, beamwidth of $3dB$ as $\theta_{3dB}$, and SNR is typically estimable throughout the detection process \cite{cite34, cite35}.

In case of measurements converted into the Cartesian coordinate system, the noise covariance matrix $\mathbf{\Sigma}$ will likewise be influenced by the radial distance of the target, which is described as
\begin{equation}\label{eq7}
\mathbf{\Sigma}=\mathbf{A}\left[\begin{array}{cc}
\sigma_\rho^2 & 0 \\
0 & \sigma_\theta^2
\end{array}\right] \mathbf{A}^{\top}, \mathbf{A}=\left[\begin{array}{cc}
\cos \theta & -\rho \sin \theta \\
\sin \theta & \rho \cos \theta
\end{array}\right],
\end{equation}
It is widely acknowledged that the measurement noise pertains to zero-mean Gaussian white noise. 
Its distribution can be comprehensively encapsulated by a covariance matrix. 
Nevertheless, the covariance matrix proves relatively abstract for neural networks, thereby rendering it challenging for the network to accurately learn the relationship between the matrix and the measurement noise distribution. 
Neural networks demonstrate outstanding performance in image processing \cite{cite36}, \cite{ cite37}, \cite{cite38}, illustrating that images represent a data format well-suited for neural network processing.
In this paper, we propose a representation of measurement uncertainty, MUPO, which converts the target measurement sequence into an image, facilitating the network to fully exploit the information of the measurement noise distribution.

\begin{figure}[tbp!]
    \centering
    \includegraphics[width=1.\linewidth]{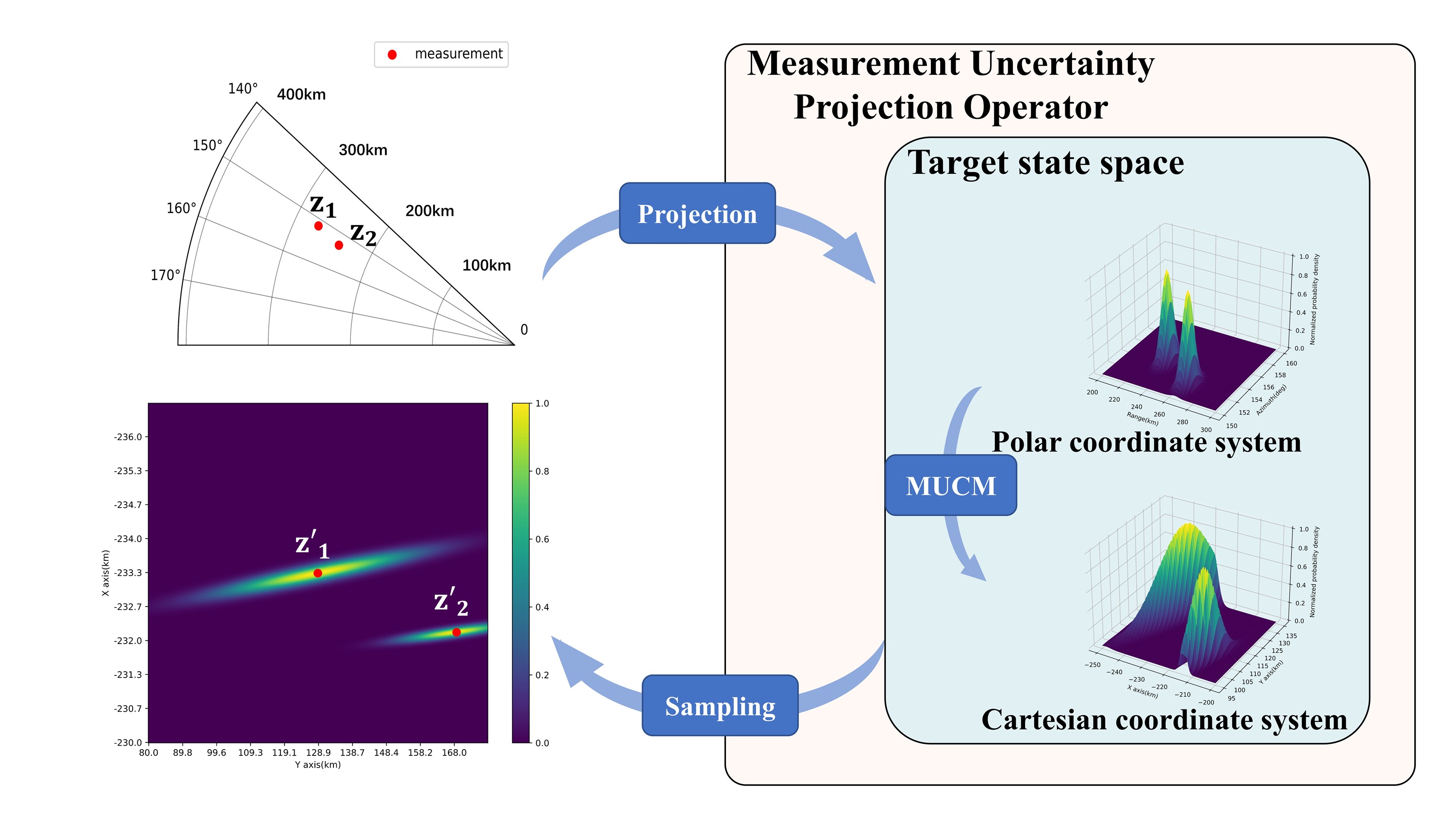}
    \caption{Flowchart of MUPO.}
    \label{fig.3}
\end{figure}

MUPO transforms measurement sequences into images primarily via two steps: \textbf{projection} and \textbf{sampling}. 
The first step is to construct the TSS and project the measurement sequence into TSS. 
In the first step, the modified unbiased conversion measurement (MUCM) \cite{cite39} is utilized in the TSS to perform the conversion from the polar coordinate system to the Cartesian coordinate system.
The second step is to sample the distribution information from the Cartesian TSS to obtain the representation in the form of an image. 
The specific flowchart of MUPO is depicted in Fig. \ref{fig.3}.

\subsubsection{\textbf{Projection}}\label{projection}
The foregoing discussion has shown that the measurement noise distribution information can be summarized as a covariance matrix under the Gaussian assumption. 
Nonetheless, the abstract nature of variance often requires the network to implicitly learn how to interpret and propagate features through its layers, which can lead to slower convergence, higher sensitivity to initialization, or suboptimal feature extraction.
One solution is to feed the complete probability distribution to the network, which is proven to improve training stability and generalization without relying on abstract variance \cite{supo1}. 
In this paper, we formulate a TSS to capture the complete distribution of the measurement sequence.

In radar data processing, measurement noise is commonly modeled as Gaussian white noise. 
The covariance matrix $\boldsymbol{\Sigma}=\operatorname{diag}\left[\sigma_\rho^2, \sigma_\theta^2\right]$ in polar coordinates can be determined via (\ref{eq6}). 
The corresponding likelihood function for one measurement $\mathbf{z}$ can be deduced as
\begin{equation}\label{eq8}
p\left(\mathbf{s}_p, \mathbf{z}, \boldsymbol{\Sigma}\right)=\frac{e^{-\frac{1}{2} \cdot\left[(\mathbf{s}_p-\mathbf{z})^{\top} \boldsymbol{\Sigma}^{-1}(\mathbf{s}_p-\mathbf{z})\right]}}{(2 \pi)|\mathbf{\Sigma}|^{\frac{1}{2}}} \quad, \mathbf{s}_p \in \mathbb{R}^N ,
\end{equation}
specifically, the potential state of the target in polar coordinates is denoted as $\mathbf{s}_p$, where $|\cdot|$ represents the determinant operation.

The conversion of the aforesaid likelihood function to a nonlinear transformation within the Cartesian coordinate will entail nonlinear errors. 
In this paper, MUCM is utilized for coordinate conversion, along with the subsequent conversion formula:
\begin{equation}\label{eq9}
\left\{\begin{array}{l}
x_{\text {MUCM}}=\lambda_\theta^{-1} \rho \cos \theta \\
y_{\text {MUCM}}=\lambda_\theta^{-1} \rho \sin \theta
\end{array}\right.,
\end{equation}
where $\lambda_\theta=e^{-\sigma_\theta^2 / 2}$. 
The corresponding measurement noise covariance matrix $\boldsymbol{\Sigma}_{\text {MUCM}}=\left\lceil\sigma_{\text {X}}^2, \sigma_{\text {XY}} ; \sigma_{\text {XY}}, \sigma_{\text {Y}}^2\right\rceil$, whose formulas for the individual elements are presented as follows:
\begin{equation}\label{eq10}
\left\{\begin{aligned}
\sigma_{\text {X}}^2= & \frac{1}{2}\left[\rho^2+\sigma_\rho^2\right] \cdot\left[1+\cos 2 \theta \cdot e^{-2 \cdot \sigma_\theta^2}\right] \\
& -e^{-\sigma_\theta^2} \cdot \rho^2 \cdot \cos ^2 \theta \\
\sigma_{\text {X}}^2= & \frac{1}{2}\left[\rho^2+\sigma_\rho^2\right] \cdot\left[1-\cos 2 \theta \cdot e^{-2 \cdot \sigma_\theta^2}\right] \\
& -e^{-\sigma_\theta^2} \cdot \rho^2 \cdot \sin ^2 \theta \\
\sigma_{\text {XY}}= & \sigma_{\text {YX}}=\frac{1}{2}\left[\rho^2+\sigma_\rho^2\right] \cdot \sin 2 \theta \cdot e^{-2 \sigma_\theta^2} \\
& -e^{-\sigma_\theta^2} \cdot \rho^2 \cdot \cos \theta \sin \theta
\end{aligned}\right.,
\end{equation}

Employing (\ref{eq9}) and \eqref{eq10}, the converted measurements and their corresponding covariance matrix can be obtained. 
By introducing the converted variables into (\ref{eq8}), the Cartesian probability function $p\left(\mathbf{s}_c, \mathbf{x}_{\text {MUCM}}, \boldsymbol{\Sigma}_{\text {MUCM}}\right)$ can be obtained, as $\mathbf{x}_{\text {MUCM}}=\left[x_{\text {MUCM}}, y_{\text {MUCM}}\right]^{\top}$ and $\mathbf{s}_c$ represent the potential state of the target in the Cartesian coordinate.

The TSS is designed to record the complete probability of the target at each observation moment.
However, the denominator term in (\ref{eq8}) will give rise to a numerical instability in the density of each measurement. 
%
In practical applications, a normalization parameter $\gamma_\cdot=(2 \pi)\left|\boldsymbol{\Sigma}_{\text {MUCM}}\right|^{\frac{1}{2}}$ is proposed for preprocessing and normalizing the probability density. 
%
%
%
In accordance with statistical knowledge, the denominator in (\ref{eq8}) is utilized to guarantee that the integral of the PDF over the domain of definition amounts to 1; however, this constraint is dispensable when constructing TSS in this paper. 
Based on (\ref{eq8}), the projection from the measurement sequence to TSS defined herein is as follows:

\begin{equation}\label{eq11}
\begin{aligned}
& \psi\left(\mathbf{s}_c, \mathbf{x}_{\mathrm{MUCM}, 1: L}, \boldsymbol{\Sigma}_{\mathrm{MUCM}, 1: L}\right):= \\
&\quad\quad \sum_{l=1}^L \gamma_l \cdot p\left(\mathbf{s}_c, \mathbf{x}_{\mathrm{MUCM}, l}, \boldsymbol{\Sigma}_{\mathrm{MUCM}, l}\right),
\end{aligned}
\end{equation}
where $\boldsymbol{\Sigma}_{\text {MUCM}, 1 \cdot L}=\left\{\boldsymbol{\Sigma}_{\text {MUCM}, l} \mid l=1, \ldots, L\right\}$ represents the covariance matrix of the corresponding measurements by  MUCM.
For notational simplicity and uniformity in subsequent derivations, it is henceforth assumed that the measurements in the track span from time instant 1 to $L$.


\subsubsection{\textbf{Sampling}}\label{sampling}
The sampling procedure discretizes the TSS and samples the embedded distribution information to generate a structured representation. 
In the projection stage, which precedes the sampling procedure, the measurement sequence is projected onto the TSS to capture its probability distribution while partially disregarding sequential information. 
To mitigate this limitation, the first measurement point in the sequence is designated as the center of the sampling region.
This establishes it as an origin, allowing subsequent target trajectories to extend outward from the center and thereby implicitly conveying the temporal ordering of the sequence to the network.

The size of the sampling region also requires careful definition.
Ideally, sampling should be performed in a manner that comprehensively encompasses all distribution information within the sampling region.
Moreover, the true position of the target should be included within this region, which ensures that it can be estimated from the sampling.
In the MUPO presented in this paper, two sampling methodologies are devised to satisfy these requirements:

\textbf{Fixed-Size Sampling}: 
\revision{
For a given target class, the maximum ground speed $v_{\max}$ is prescribed based on its physical maneuverability limits and specific mission requirements. In practice, $v_{\max}$ is strictly constrained by the target’s aerodynamic envelope. For hypersonic glide vehicles or high-performance fighters, $v_{\max}$ is derived from their peak Mach numbers at operational altitudes. Conversely, $v_{\max}$ remains significantly lower for "low-slow-small" targets and maritime vessels.
}
By combining this with the time interval $\Delta t$ between consecutive measurements, the maximum potential displacement relative to the initial measurement point $\mathbf{x}_{\text{MUCM}, k_1}$ can be computed along the x- and y-axes. 
Specifically, assuming isotropic motion constraints, the displacement bound in each dimension is given by $d = v_{\max} \cdot (L-1) \cdot \Delta t$, where $L$ denotes the length of the measurement sequence. 
This defines a fixed sampling region, such as $[-d, d] \times [-d, d]$ centered at $\mathbf{x}_{\text{MUCM}, k_1}$. 
The sampling rate, defined as the reciprocal of the uniform sampling step size applied across this region, governs the density of sampled points.
Consequently, for various measurement sequences $\mathbf{x}_{\text{MUCM}, 1: L}$, the resulting sampled image has a size that depends solely on $L$, facilitating a consistent input size for network training.
\revision{
It is important to highlight that while $v_{\max}$ directly influences the spatial extent of the sampling region and the resulting computational complexity, the input size does not necessarily scale linearly with target speed. In radar systems, high-velocity targets typically necessitate a significantly smaller $\Delta t$ to ensure track maintenance. Consequently, the reduction in $\Delta t$ often compensates for the increase in $v_{\max}$, effectively regularizing the product $v_{\max} \cdot \Delta t$. 
}

\textbf{Flexible-Size Sampling}: 
In contrast to fixed-size sampling, which defines the sampling region based on the principle of maximum ground speed and employs a relatively low sampling rate, flexible-size sampling adopts a higher sampling rate while dynamically adapting the sampling region. 
Specifically, the boundaries of the sampling region are established using the 3-sigma rule applied to the Gaussian distribution centered at the last measurement point $\mathbf{x}_{\text{MUCM}, L}$ in the track. 
To account for potential off-diagonal elements in the covariance matrix $\boldsymbol{\Sigma}_{\text{MUCM},L}$, an eigenvalue decomposition is performed: $\boldsymbol{\Sigma}_{\text{MUCM},L} = \mathbf{U} \boldsymbol{\Lambda} \mathbf{U}^T$, where $\mathbf{U}$ contains the eigenvectors (principal axes) and $\boldsymbol{\Lambda}$ is the diagonal matrix of eigenvalues $\lambda_1$ and $\lambda_2$. 
The 3-sigma bounds are then computed along each principal axis, yielding four endpoints: $\mathbf{x}_{\text{MUCM}, L} \pm 3\sqrt{\lambda_i} \mathbf{u}_i$ for $i=1,2$, where $\mathbf{u}_i$ is the $i$-th eigenvector. 
These endpoints define the enclosing rectangular boundary for the sampling region. 
The sampling rate, similarly defined as the reciprocal of the uniform sampling step size across this adaptive region, ensures denser point sampling to capture finer details in the distribution. 
As a result, the size of the generated image depends not only on the sequence length $L$ but also on the $\boldsymbol{\Sigma}_{\text{MUCM},L}$ and $\mathbf{x}_{\text{MUCM}, L}$.

\begin{figure}[tb]
    \centering
    \includegraphics[width=1.\linewidth]{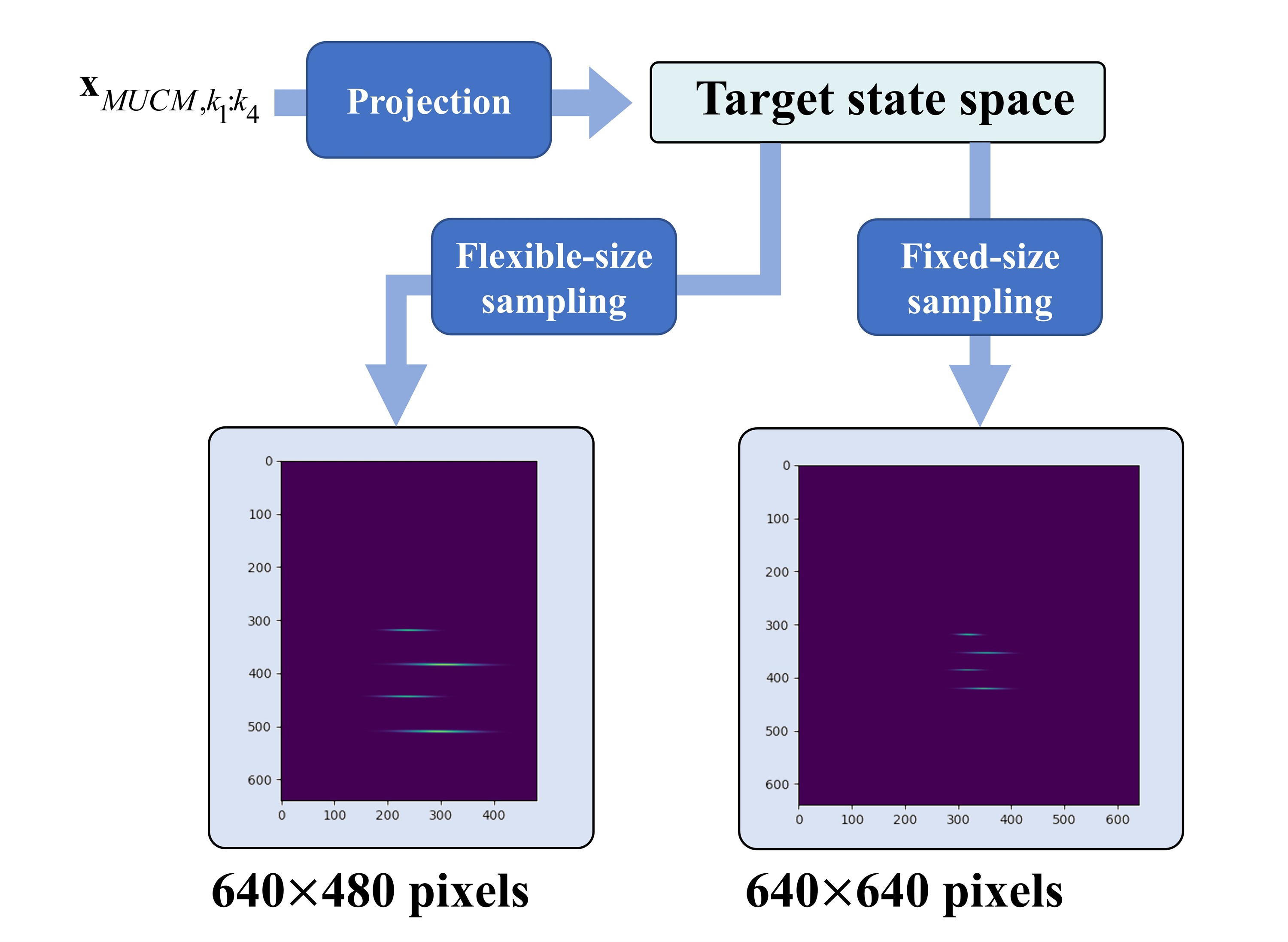}
    \caption{Fixed-size sampling and flexible-size sampling in MUPO in the case of measurements sequence's length is 4.}
    \label{fig:5}
\end{figure}

The comparison of the aforementioned two sampling methods is depicted in Fig. \ref{fig:5}.
The advantage of fixed-size sampling lies in its straightforward computation and fixed sampling size, which facilitates network training.
Flexible-size sampling, however, results in a variable sampling region, which can significantly increase network training time\cite{cite40}.
Nevertheless, in fixed-size sampling, the relatively low sampling rate results in a coarser discretization of the region, yielding images with lower resolution.
In contrast, flexible-size sampling, with its higher sampling rate, yields denser and higher-resolution results.
Additionally, the sampling rate must be considered.
A low sampling rate may result in insufficient utilization of information in the TSS, whereas an excessively high sampling rate introduces larger data volumes, imposing a computational burden on subsequent processing steps.
For convenience in subsequent discussions, the aforementioned two methods will be referred to as fixed-sampling and flexible-sampling.

\subsection{Multi-channel MUPO}\label{mcmupo}
The studies in \cite{ass1,ass2} employed Bayesian statistics and Kalman filtering techniques to guide neural network optimization, accelerating convergence and boosting overall model performance.
Inspired by this concept, two channels are designed as auxiliary ones to direct the network's training.

%
One of the channels within the auxiliary channel is configured as the estimation of coarse-tuning IMM. 
On one hand, when model matching, the results of IMM can assist in locating the target’s true position; on the other hand, in cases of model mismatch, they serve as negative samples to help the network refine its decision boundary.
The projection to the corresponding TSS is
\begin{equation}\label{eq12}
\begin{aligned}
& \psi_{I M M}\left(\mathbf{s}_c, \hat{\mathbf{x}}_{\text{IMM},1: L}, \boldsymbol{\Sigma}_{1: L}^*\right)= \\
&\quad\quad \sum_{l=1}^L e^{-\frac{1}{2} \cdot\left[\left(\mathbf{s}-\hat{\mathbf{x}}_{l}\right)^{\top} \boldsymbol{\Sigma}_{l}^{*-1}\left(\mathbf{s}-\hat{\mathbf{x}}_{l}\right)\right]},
\end{aligned}
\end{equation}
where $\hat{\mathbf{x}}_{\text{IMM},1: L}$ represents the estimations of IMM. 
The covariance matrix $\Sigma^*$ mentioned in (\ref{eq12}) is computed equivalently using the first measurement point in the track. 
Given the SNR of the first measurement point, the expectancy of SNRs can be estimated from the radar equation  below: $S N R_{l}=S N R_{1} \cdot \rho_{1}^4 / \rho_{l}^4, l=1, \ldots, L$, where $\rho$ denotes the radial distance. 
And $\Sigma^*$ denotes the corresponding covariance, which is calculated by (\ref{eq6}), (\ref{eq9}) and (\ref{eq10}). 

Discrete observations provide measurements only at specific instants, rendering the target state unobservable during intervening periods. 
Another channel attempts to incorporate predictions of the measurement sequence trend, thereby compensating for partially unobservable states.
To address this, the T-FoT \cite{cite42} method constructs a continuous trajectory describing the target's state evolution, while enhancing the incorporation of temporal information.

Specifically, the T-FoT algorithm is intended to acquire an estimation of $F(t)$, designated as $\widehat{F}(t ; A)$, and presumes that $\widehat{F}(t ; A)$ can be depicted as a linear combination of a set of orthogonal basis vectors:
\begin{equation}\label{eq13}
\widehat{F}(t ; A)=a_0 \phi_0(t)+a_1 \phi_1(t)+\cdots+a_\gamma \phi_\gamma(t),
\end{equation}
$\left\{\phi_i(t)\right\}_{i=0,1, \ldots, \gamma}$ constitutes a set of orthogonal basis. 
For the sake of convenience, the power series is often used as the orthogonal basis. 
$A:=\left\{a_i\right\}_{i=0,1}$ denotes the coefficients of each orthogonal basis, which are derived by resolving the following optimization problem:
\begin{equation}\label{eq14}
\underset{A}{\arg \min } \sum_{t=t_{\mathrm{1}}}^{t_{L}}\left\|\mathbf{x}_t-h\left(\widehat{F}(t ; A), \mathbf{m}_t\right)\right\|+\lambda \Omega(t, A),
\end{equation}
for $\Omega(t, A):=\left\|\widehat{F}(t ; A)-\mathbf{x}_t\right\|$, the regulation parameter is denoted as $\lambda$.

Given an estimated T-FoT $\widehat{F}(t ; A)$, $N_s$ sampling points are uniformly placed in intervals $\left[t_{1}, t_{L}\right]$, where $N_s$ is typically much larger than $L$. 
The sequence that consists of T-FoT's sampling values $\widehat{F}(t_{k_{l}} ; A)$ can be represented via MUPO, and the corresponding projection to TSS is
\begin{equation}\label{eq15}
\begin{aligned}
& \psi_{F o T}\left(\mathbf{s}_c, \hat{F}(t ; A), \mathbf{\Sigma}_{1: N_s}^*\right)= \\
&\quad\quad \sum_{l=1}^{N_s} e^{-\frac{1}{2}\left[\left(\mathbf{s}_c-\hat{F}\left(t_l ; A\right)\right)^{\top} \mathbf{\Sigma}_l^{*-1}\left(\mathbf{s}_c-\hat{F}\left(t_l ; A\right)\right)\right]},
\end{aligned}
\end{equation}
$\Sigma^*$ is computed in the same manner as $\Sigma^*$ in (\ref{eq12}), wherein the estimated values used in the covariance calculation are replaced by $\hat{F}\left(t_l ; A\right)$.

Meanwhile, in our preliminary experiments, we found that incorporating the MUPO corresponding to the last point in the track can effectively improve the network training speed.
The corresponding projection is
$
\psi\left(\mathbf{s}_c, \mathbf{x}_{\text {MUCM}, L}, \mathbf{\Sigma}_{\text {MUCM}, L}\right)
$.
In terms of single target tracking, the multi-channel MUPO is constructed as 
\begin{equation}\label{r1eq1}
\begin{aligned}
\mathbf{T} & =\mathbf{e}_1 \otimes \psi\left(\mathbf{s}_c, \mathbf{x}_{\mathrm{MUCM}, 1: L}, \boldsymbol{\Sigma}_{\mathrm{MUCM}, 1: L}\right) \\
& +\mathbf{e}_2 \otimes \psi_{I M M}\left(\mathbf{s}_c, \hat{\mathbf{x}}_{\mathrm{MUCM}, 1: L}, \boldsymbol{\Sigma}_{\mathrm{MUCM}, 1: L}^*\right) \\
& +\mathbf{e}_3 \otimes \psi_{F o T}\left(\mathbf{s}_c, \hat{F}(t ; A), \boldsymbol{\Sigma}_{1: N_s}^*\right) \\
& +\mathbf{e}_4 \otimes \psi\left(\mathbf{s}_c, \mathbf{x}_{\mathrm{MUCM}, L}, \boldsymbol{\Sigma}_{\mathrm{MUCM}, L}\right),
\end{aligned}
\end{equation}
where $\mathbf{T}$ represents the four-channel MUPO tensor. 
$\mathbf{e}_k$ is a standard orthonormal basis vector, $(\mathbf{e}_k)_m = 1$ only when $k=m$; otherwise, when $k \neq m$, $(\mathbf{e}_k)_m = 0$. 
$\otimes$ denotes a tensor product operation that maps the output of $\psi$ functions to the $k$-th channel of $\mathbf{T}$.
For multi-target tracking, the corresponding multi-channel MUPO is represented as
\begin{equation}\label{r1eq2}
\begin{aligned}
\mathbf{T} & =\mathbf{e}_1 \otimes \psi\left(\mathbf{s}_c, \mathbf{x}_{\mathrm{MUCM}, 1: L_t}, \boldsymbol{\Sigma}_{\mathrm{MUCM}, 1: L_t}\right) \\
& +\mathbf{e}_2 \otimes \psi_{I M M}\left(\mathbf{s}_c, \hat{\mathbf{x}}_{\mathrm{MUCM}, 1: L_t}, \boldsymbol{\Sigma}_{\mathrm{MUCM}, 1: L_t}^*\right) \\
& +\mathbf{e}_3 \otimes \psi_{F o T}\left(\mathbf{s}_c, \hat{F}(t ; A), \boldsymbol{\Sigma}_{1: N_s}^*\right) \\
& +\mathbf{e}_4 \otimes \psi\left(\mathbf{s}_c, \mathbf{x}_{\mathrm{MUCM}, L+1:L_t}, \boldsymbol{\Sigma}_{\mathrm{MUCM}, L+1:L_t}\right),
\end{aligned}
\end{equation}
where $L_t$ is the length of the existing measurement sequence, and $\mathbf{x}_{\mathrm{MUCM}, L+1:L_t}$ and $\boldsymbol{\Sigma}_{\mathrm{MUCM}, L+1:L_t}$ represent the observations and the corresponding covariance matrix within the latest frame.

\subsection{Target State Estimation via Point Prediction}\label{detectionnet}
\begin{figure}[tbp!]
    \centering
    \includegraphics[width=0.7\linewidth]{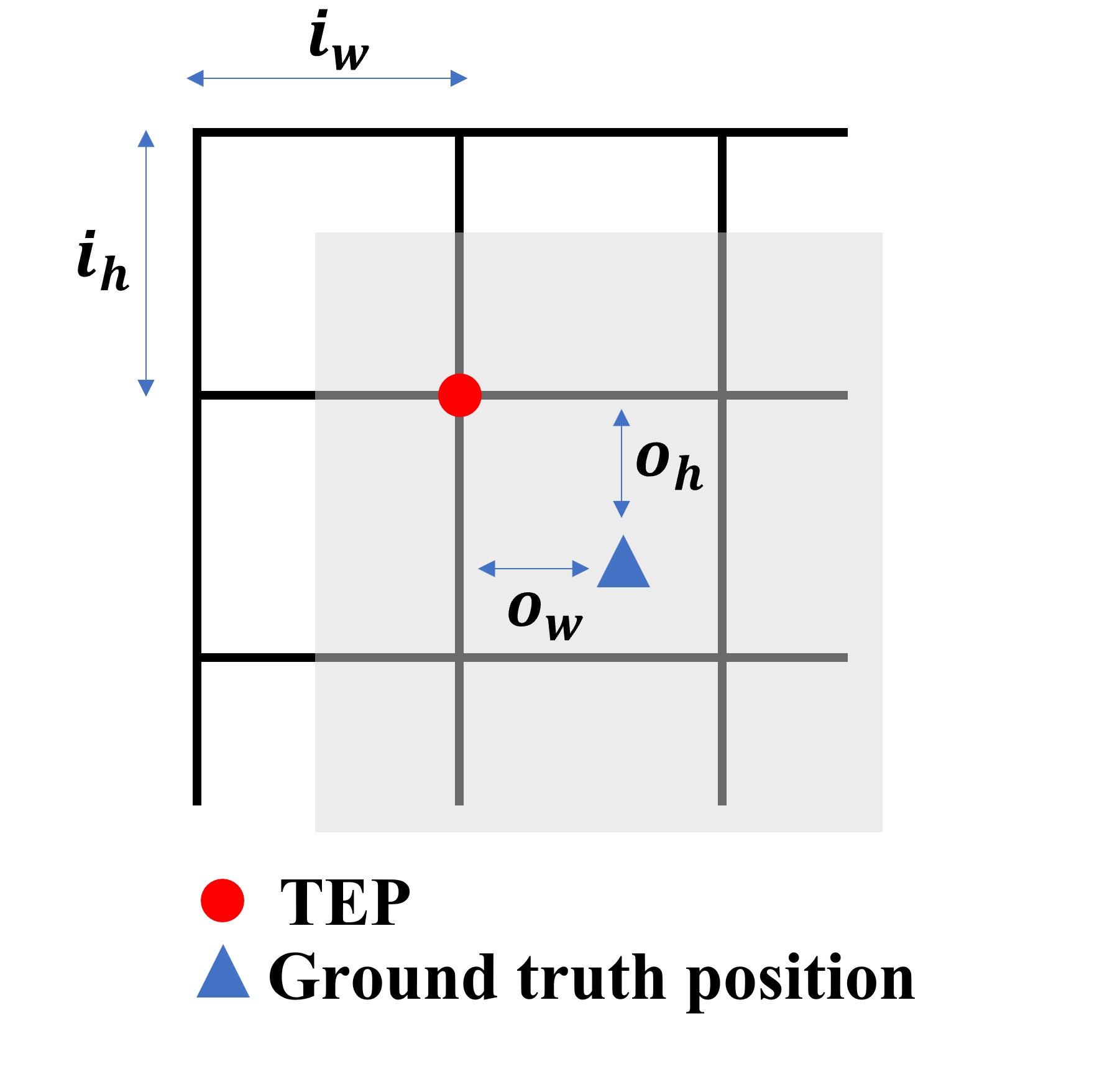}
    \caption{TEP prediction in precise estimation. (The gray shaded area represents the responsibility area for this TEP.)}
    \label{fig:add1}
\end{figure}
In the proposed framework, we partition multi-channel MUPO uniformly with a certain density and take the top-left corner of each partitioned region as the target state estimation point (TEP).
The network predicts 3 features for each TEP, $o_p,o_w,o_h$, representing existence probability, offsets along the images horizontally and vertically.
If the ground truth position is offset from the TEP by ($i_w, i_h$), then the prediction of the position is $\mathbf{i}=[i_w + o_w,i_h+o_h]\top$, as illustrated in Fig. \ref{fig:add1}.
The rectangular domain that $\mathbf{i}$ is constrained within can be considered as the responsibility area for this TEP.
If the center coordinate is $\mathbf{c}=[c_w,c_h]\top$, then the exact target state estimation corresponds to
\begin{equation}\label{eqadd1}
\hat{\mathbf{x}}=\eta \cdot(\mathbf{i}-\mathbf{c}),
\end{equation}
where $\eta$ denotes the scaling parameter between the partitioned region and the physical space, $\hat{\mathbf{x}}_{\text{IMM}}$ represents the corresponding result of IMM.
\revision{
For the multi-channel MUPO input, the fusion of channel-wise information is adaptively performed through the learned weights of the convolutional kernels. Unlike static equal-weighting, this mechanism allows the network to capture complex inter-channel dependencies effectively.
}

\begin{figure}[!tbp]
    \centering
    \includegraphics[width=1.\linewidth]{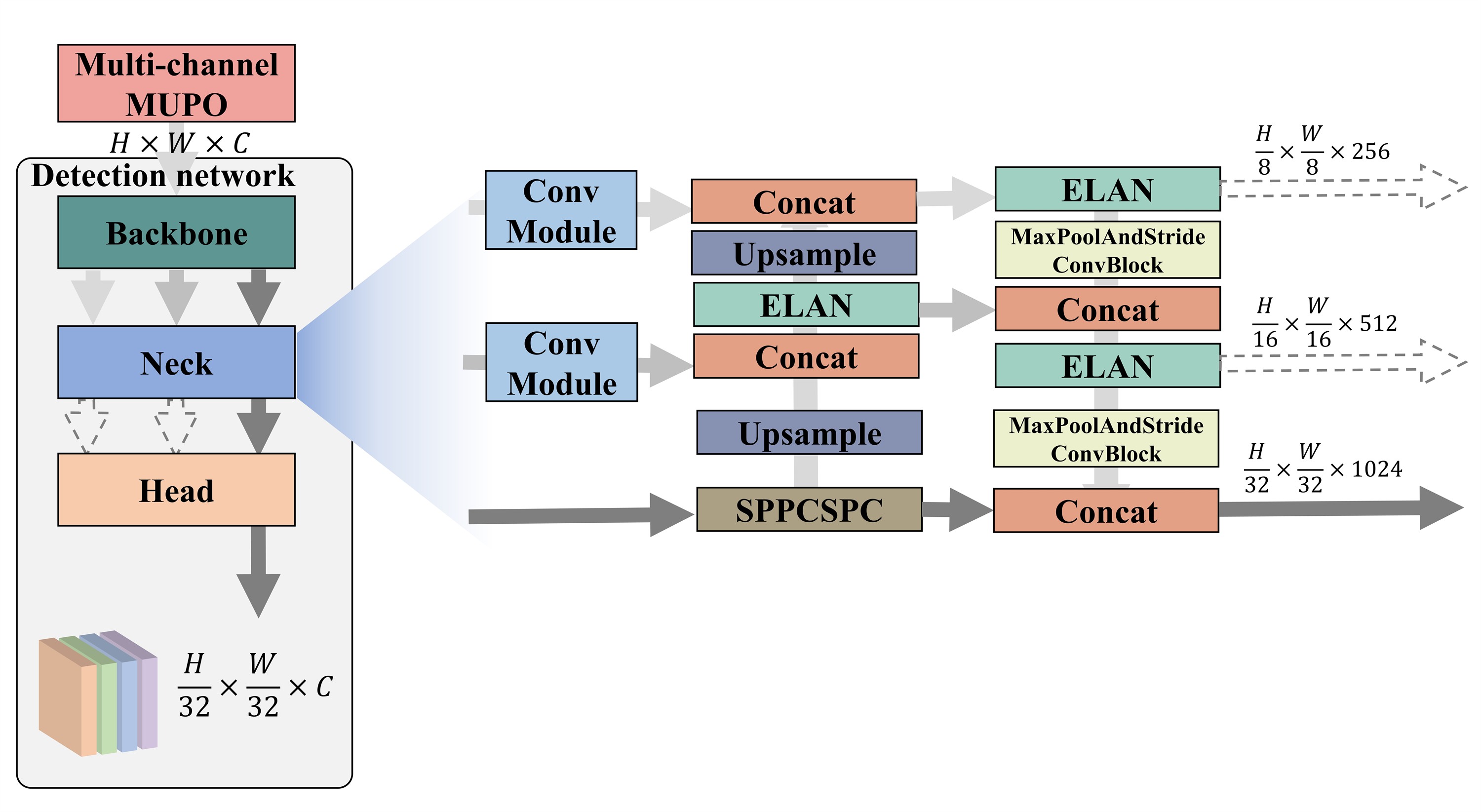}
    \caption{Detection network framework. Specific design of each module can be found in \cite{cite47}.}
    \label{fig:7}
\end{figure}

Maneuvering targets with various velocities and motions may lead to multi-scale multi-channel MUPO.
Additionally, the image projected to TSS varies because of the complex target motions.
We leverage a feature pyramid network \cite{cite48} and a path aggregation network \cite{cite49} to effectively detect targets under complex scenes. 
As illustrated in Fig. \ref{fig:7}, the detection network we applied is divided into three key components: the backbone, neck, and head. 
The backbone is Darknet-53, a 53-layer convolutional neural network designed for efficient feature extraction \cite{cite47}. 
It incorporates residual connections to facilitate gradient flow, enabling deeper networks with improved performance. 
The neck enhances feature propagation by aggregating features across different scales and promoting better feature fusion. 
Finally, the head generates the final output through one detection layer, which predicts TEP features. 
\subsection{Network Training Loss}\label{loss}
Multi-channel MUPO can be denoted as $\mathbf{T}_j \in \mathbb{R}^{C \times H_j \times W_j}$, with $j$ representing the sequence serial number, $C$ representing the number of channels, and $H_j, W_j$ representing height and width, respectively. 
As depicted in Section \ref{mupotnn}, the detection network is designed to accomplish three-stage tasks: classification, regression, and fusion. 
To optimally guide the network training and attain the desired performance, the loss function devised in this paper is partitioned into four components: classification, confidence, regression, and constraint.

Within the TEP responsibility area, we assign labels to be 1 if they encapsulate the targets to detect, and 0 otherwise. 
The design of the loss function for the classification component assumes the form of cross-entropy:
\begin{equation}\label{eq16}
\begin{gathered}
 \mathcal{J}_{\text {classification }}:=-\sum_{j=1}^{N_D} \sum_{m=1}^{H_j^{\prime} \times W_j^{\prime}} \left[ o_p^{(m, j)} \log \hat{o}_p^{(m, j)}+ \right.\\
\quad\quad\left. \left(1-o_p^{(m, j)}\right) \log \left(1-\hat{o}_p^{(m, j)}\right) \right],
\end{gathered}
\end{equation}
for $H^{\prime}=H_j \cdot r, W^{\prime}=W_j \cdot r$, where $r=1/32$ represents the ratio of the partitioned region's size to the image pixel, then $H_j^{\prime} \times W_j^{\prime}$ is the total number of the TEPs in one sample. 
The number of training samples is denoted as $N_{\mathcal{D}}$, $\hat{o}_p^{(m, j)}$ is the existence probability prediction of the $m^{th}$ TEP in $j^{th}$ sample. 


The target state can be estimated based on (\ref{eqadd1}). 
Meanwhile, we assume that the TEP whose responsibility area contains the true target position possesses the capability to accurately estimate the target state. 
This assumption increases the number of positive samples, and the regression loss is defined as follows:
\begin{equation}\label{eq18}
 \mathcal{J}_{\text{regression}}:=\sum_{j=1}^{N_{\mathcal{D}}} \sum_{m \in \mathcal{A} j}\big(\mathbf{x}^{(j)}-\hat{\mathbf{x}}^{(m, j)}\big)^{\top}\big(\mathbf{x}^{(j)}-\hat{\mathbf{x}}^{(m, j)}\big),
\end{equation}

In addition, we define a loss function to obtain the desired prediction results, and the constraint loss can be described as
\begin{equation}\label{eq20}
\begin{gathered}
    \mathcal{J}_{\text {constraint}}:=\sum_{j=1}^{N_D}\Bigg[\underbrace{\lambda_e\left(-\sum_{m \notin C_j} \hat{o}_p^{(m, j)} \log \hat{o}_p^{(m, j)}\right)}_{\text {entropy }}+ \\
    \quad\quad \underbrace{\lambda_p\left(\prod_{m \in C_j} \hat{o}_p^{(m, j)}\right)^{-\frac{1}{\left|c_j\right|}}}_{\text {perplexity }}\Bigg],
\end{gathered}
\end{equation}
for $
\mathcal{C}_j=\left\{m\mid\left\|\left[ i_w^{(i)}, i_h^{(i)}\right]^{\top}-\mathbf{i}_\text{obj}^{(j)}\right\|_2<d\right\}
$, where $|\cdot|$ is the operation for determining the cardinality, $\mathbf{i}_{\text{obj}}$ is the target position coordinates, $d$ determines the range of the constraint region and $\lambda_e, \lambda_p$ are the weighting coefficients. 
The constrained loss is bifurcated into two components. 
The entropy part constrains the distribution of the predictions in the absence of a target.
The perplexity part functions as an evaluation metric to assess how well the network predicts the existence of targets.

Conclusively, the loss function within the training procedure of the network is 
\begin{equation}\label{eq21}
\begin{aligned}
\mathcal{J}= & \beta_{\mathrm{d}} \cdot \mathcal{J}_{\text {detection }}+\beta_{\mathrm{r}} \cdot \mathcal{J}_{\text {regression }} +\beta_{\mathrm{c}} \cdot \mathcal{J}_{\text {constraint }},
\end{aligned}
\end{equation}
where $\beta_{\mathrm{d}}, \beta_{\mathrm{r}}, \beta_{\mathrm{c}}$ are the corresponding weighting coefficients for different losses.
\subsection{Data Association Criterion}
In multi-target tracking, after the network provides target state estimates $\hat{\mathbf{x}}$, an association step is performed to select the most suitable measurement in the latest frame for existing measurement sequence. 
In this step, a maximum likelihood criterion is employed to compute the likelihood between the network-outputted estimated states and the candidate measurements, thereby associating the measurement with the maximum likelihood.
The log-likelihood function for $\hat{\mathbf{x}}$ is given by
\begin{equation}\label{r1eq3}
\begin{aligned}
& \ln p\left(\hat{\mathbf{x}} \mid \mathbf{x}_{\mathrm{MUCM}, k}, \boldsymbol{\Sigma}_{\mathrm{MUCM}, k}\right) \\
& =-\frac{d}{2} \ln (2 \pi)-\frac{1}{2} \ln \left|\boldsymbol{\Sigma}_{\mathrm{MUCM}, k}\right| \\
& -\frac{1}{2}\left(\hat{\mathbf{x}}-\mathbf{x}_{\mathrm{MUCM}, k}\right)^\top \boldsymbol{\Sigma}_{\mathrm{MUCM}, k}^{-1}\left(\hat{\mathbf{x}}-\mathbf{x}_{\mathrm{MUCM}, k}\right),
\end{aligned}
\end{equation}
where $k=L+1,...,L_t$.

To maximize $\ln p\left(\hat{\mathbf{x}} \mid \mathbf{x}_{\mathrm{MUCM}, k}, \boldsymbol{\Sigma}_{\mathrm{MUCM}, k}\right)$, is equivalent to minimizing the expression:  
\begin{equation}\label{r1eq5}
\begin{aligned}
&B\left(\hat{\mathbf{x}} , \mathbf{x}_{\mathrm{MUCM}, k}, \boldsymbol{\Sigma}_{\mathrm{MUCM}, k}\right)=\ln \left|\boldsymbol{\Sigma}_{\mathrm{MUCM}, k}\right|\\
&+\left(\hat{\mathbf{x}}-\mathbf{x}_{\mathrm{MUCM}, k}\right)^\top \boldsymbol{\Sigma}_{\mathrm{MUCM}, k}^{-1}\left(\hat{\mathbf{x}}-\mathbf{x}_{\mathrm{MUCM}, k}\right).
\end{aligned}
\end{equation}

Therefore, measurement data association is achieved by minimizing this objective function:
\begin{equation}\label{r1eq4}
\hat{k}=\arg \min _k B\left(\hat{\mathbf{x}} , \mathbf{x}_{\mathrm{MUCM}, k}, \boldsymbol{\Sigma}_{\mathrm{MUCM}, k}\right)
\end{equation}

\section{NUMERICAL RESULTS}\label{ne}
\subsection{Maneuvering Target Dataset}\label{simulation}
The target motion model presented in this paper encompasses the constant velocity (CV) model, the constant acceleration (CA) model, the Jerk model, the Singer model, the current statistics (CS) model, the Coordinated Turn (CT) model, and the CT model with unknown turn rates \cite{cite50}. 
These models are capable of comprehensively describing the diverse dynamic behavior of the target. 

The target tracking addressed in this paper takes place within the wide-field radar monitoring scenarios, and the generation of maneuvering targets is likewise conducted in this setting. 
The initial target state is uniformly sampled within a prescribed range, as is indicated in Table \ref{tab:1}, assuming the radar is at the origin of coordinates and all angles are measured from the positive x-axis, taken as 0 degrees, increasing counterclockwise.
\begin{table}[!tbp]
    \centering
    \renewcommand{\arraystretch}{1.5}\caption{The sampling interval for initial target state.}
    \begin{tabular}{cc}
        \toprule
        Radial distance (km) & $[150, 400]$ \\
        \midrule
        Azimuth ($\deg$) & $[-180, 180]$\\
        \midrule
        Ground speed (m/s) & $[200, 220]$\\
        \midrule
        Course ($\deg$) & $[-180, 180]$\\
        \bottomrule
    \end{tabular}
    \label{tab:1}
\end{table}

When a fighter aircraft detects that it is being illuminated or tracked by an enemy radar system, it typically performs evasive maneuvers to degrade or break the radar tracking capability.
We model this process as two distinct parts: the time the target switches its dynamic behavior and how it switches.

\textbf{The time target switches dynamic behaviors:} In this paper, the switching of the target dynamic behavior is characterized as a random event, and a Poisson process with parameter $\lambda_{\text {switch }}$ is utilized to model the count $N(t)$ of switches, where $1/\lambda_{\text {switch }}$ represents the average time interval between two random events.

\textbf{The process target switches dynamic behaviors:} In this paper, the target dynamic behavior is described as a random variable $M$, whose possible value is one of the aforementioned motion models. 
When the target is going to switch motion models, the motion model to switch is related to the current target motion model. 
Consequently, the process of model switching is constituted as a homogeneous Markov process. 
Let $\mathbf{P}$ represent the corresponding state transition matrix of the Markov process, and the transition probability $P_{i j}=P\{M(t+1)=j \mid M(t)=i\}$ is defined as the switching probability between model $i \rightarrow j$. 
It is noteworthy that the adjustment of matrix $\mathbf{P}$ can indirectly govern the strategy of the target motion model switching.

In actuality, the distribution of the radar measurement errors is frequently variable.
According to the mechanism of measurement error generation, it is evident that, under the registration of system bias, the distribution of measurement error is predominantly related to the target SNR \cite{cite35}. 
We utilize the radar equation and the Swerling model \cite{cite34, cite52} to simulate variations in the target SNR, thereby obtaining measurement noise with a varying distribution.

In this paper, the target state $\mathbf{x}_k=\left[x_k, \dot{x}_k, y_k, \dot{y}_k\right]^{\top}$ is postulated to represent the target position and velocity within the Cartesian coordinate system, while the target measurement $\mathbf{z}_k=\left[\rho_k, \theta_k\right]$ lies in the polar coordinate system. 
The measurement function $h(\cdot)$ is nonlinear and can be described as
\begin{equation}\label{22}
\underbrace{\left[\begin{array}{l}
\rho_k \\
\theta_k
\end{array}\right]}_{\mathbf{z}_k}=\left[\begin{array}{l}
\arctan \frac{y_k}{x_k} \\
\sqrt{x_k^2+y_k^2}
\end{array}\right]+\underbrace{\left[\begin{array}{l}
m_\rho \\
m_\theta
\end{array}\right]}_{\mathbf{m}_k},
\end{equation}
where $\mathbf{m}_k=\left[m_\rho, m_\theta\right]^{\top}$ is the radar measurement error, whose covariance matrix can be estimated through target SNR \cite{cite35}.

\begin{figure}[!tp]
    \centering
    \includegraphics[width=1.\linewidth]{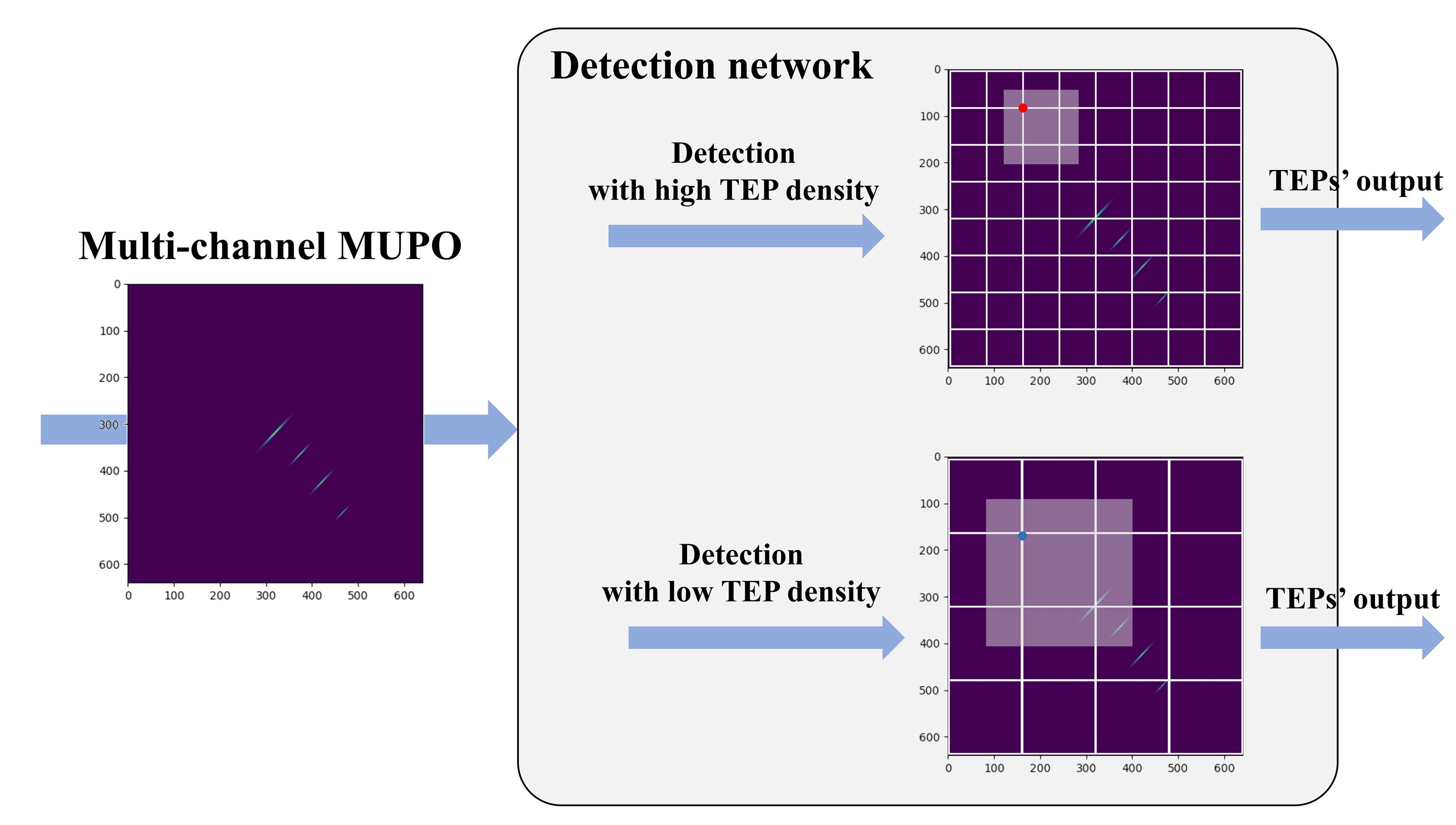}
    \caption{Prediction with different TEP densities in the case of multi-channel MUPO's channel number is 1.}
    \label{fig:6}
\end{figure}
\subsection{Validation Experiments for the Rationality of MUPO-TTN}
The multi-channel MUPO proposed in Subsection \hyperref[mcmupo]{III.B} integrates multiple MUPO representations derived from the original measurement sequence and its diverse algorithmic preprocessing results. 
We designed a comparative experiment to evaluate the impact of the multiple channels on tracking performance. 
In the experiment, four networks were trained using a single-channel MUPO, MUPO with IMM, MUPO with T-FoT, and the complete multi-channel MUPO, all under the same training dataset and update steps. 
The tracking performance of each network was tested on the same validation set, and the detailed results are presented in Table \ref{tab:abl}.
\begin{table}[!tbp]
    \centering
    \renewcommand{\arraystretch}{1.5}\caption{Tracking performance on validation set.}
    \begin{tabular}{cc}
        \toprule
        MUPO type & average MSE (m) \\
        \midrule
        signal channel & 365.95\\
        \midrule
        signal channel with IMM& 99.78\\
        \midrule
        signal channel with T-FoT& 106.17\\
        \midrule
        complete multi-channel & 86.26\\
        \bottomrule
    \end{tabular}
    \label{tab:abl}
\end{table}

The tracking performance on the validation set indicates that both MUPO with IMM and MUPO with T-FoT can improve the network’s ability to estimate target states. 
Among the four configurations, the proposed multi-channel MUPO achieves the best estimation precision under the proposed target tracking framework.
It suggests that both MUPO with IMM and MUPO with T-FoT provide additional target information compared to the single-channel MUPO. 
Furthermore, the superior performance of multi-channel MUPO demonstrates that combining MUPO with IMM and MUPO with T-FoT can further exploit incremental information, thereby enhancing tracking performance.

Due to the use of a feature pyramid architecture, as shown in Fig. \ref{fig:7}, three feature fusion results are produced. 
The feature map from different layers corresponds to different TEP densities when predicting the TEP features.
Fig. \ref{fig:6} illustrates the TEP features prediction with different TEP densities.

\begin{figure}[!tbp]
    \centering
    \includegraphics[width=0.8\linewidth]{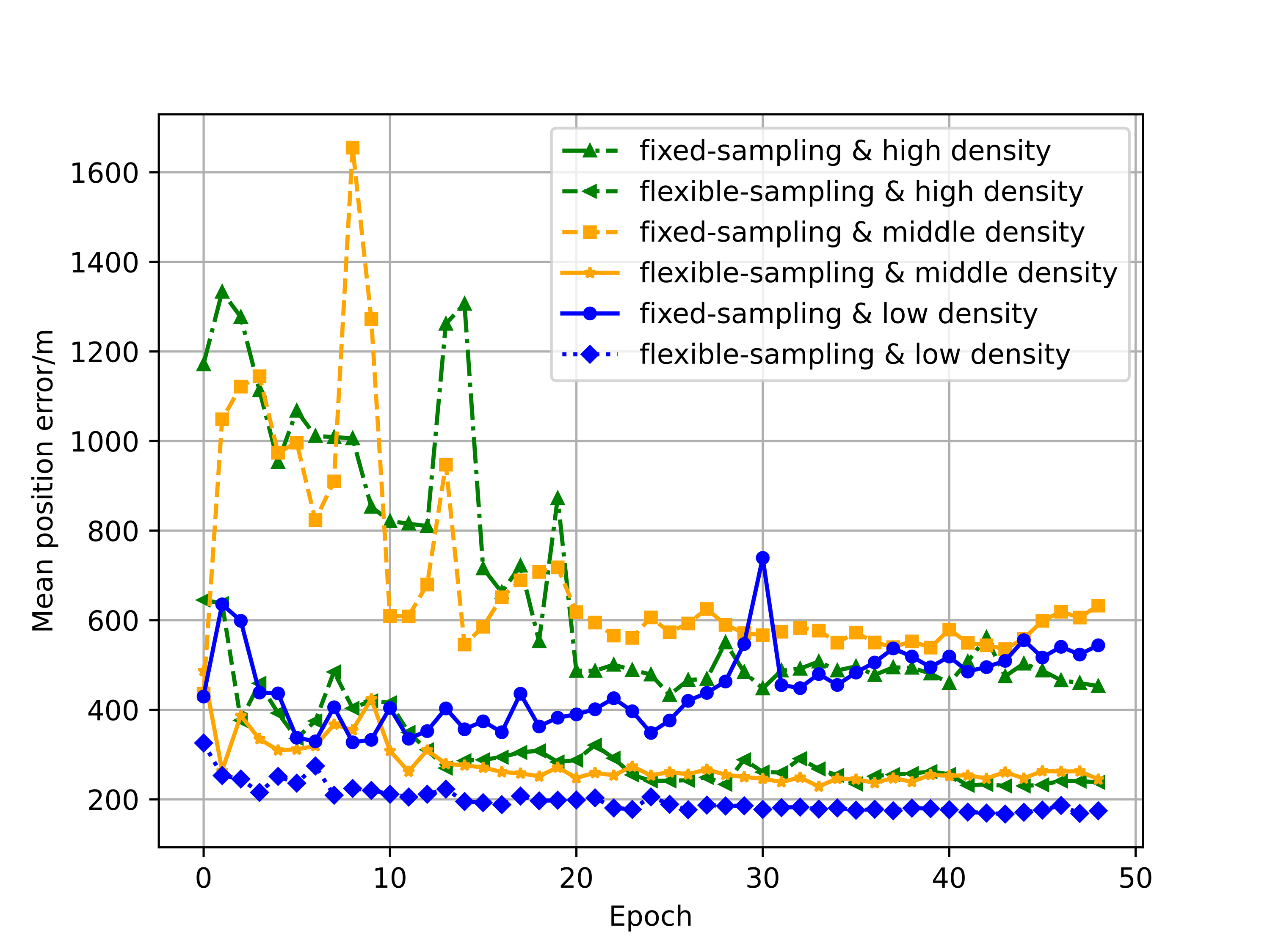}
    \caption{Comparative experiments conducted to evaluate the network performance of detection networks employing different sampling methods and TEP densities.}
    \label{fig:8}
\end{figure}

In the validation study, we present a performance comparison across different TEP densities and sampling methods.
\revision{
The parameter $v_{\max}$ for the fixed-sampling scheme is determined by the target of interest and the operational requirements of maneuvering target tracking. In this study, $v_{\max}$ is set to $500 \text{ m/s}$, which represents the maximum operational speed of a typical fighter aircraft.
}
Fig. \ref{fig:8} illustrates the training process, with the x-axis representing the number of training epochs and the y-axis showing the network performance.
Under the identical sampling method, the estimation precision of low TEP density attains the highest level. 
Low TEP density indicates a more complex network structure that can extract senior information from the image for target state estimation. 
The core difference between the two sampling methods resides in the resolution, whereby a higher resolution facilitates the capture of more detailed information within the image. 
Under the same TEP density, flexible-sampling demonstrates superior performance since it possesses a higher resolution compared to fixed-sampling, which contains more distribution information within the image. 

\revision{
Table \ref{tab:2} summarizes the training cost and complexity indicators of various models. Notably, since the input dimensions are variable under the flexible-sampling strategy, we only report the average single-step FLOPs specifically for the fixed-sampling configuration with a size of $C=4, H=320, W=320$ to provide a consistent baseline for comparison.
}
The flexible-sampling method inhibits effective parallel computation at scale for network training, leading to a marked increase in training time.
As presented in Table \ref{tab:2}, under the identical hardware circumstances, the average training time per epoch of MUPO-TTN adopting flexible-sampling is conspicuously higher than that of MUPO-TTN employing fixed-sampling. 
Simultaneously, for target state estimation, the mean inference time of MUPO-TTN utilizing flexible-sampling is longer than that using fixed-sampling.

\begin{table}[!tb]
    \centering
    \setlength{\tabcolsep}{4pt}
    \caption{\revision{Training cost and complexity indicators of different MUPO-TTNs.}}
    \renewcommand{\arraystretch}{1.2}
    \begin{tabular}{>{\centering\arraybackslash}p{2.2cm} >{\centering\arraybackslash}p{1.2cm} >{\centering\arraybackslash}p{1.2cm} >{\centering\arraybackslash}p{1.2cm}}
        \toprule
        \textbf{MUPO-TTN \quad\quad(sampling type \& density)} & \textbf{Mean training time} & \textbf{Mean inference time}& \revision{\textbf{GFLOPs}} \\
        \midrule
        fixed \& low  & 460s & 11.88ms & \revision{10.93}\\
        fixed \& middle  & 462s & 13.39ms & \revision{11.53}\\
        fixed \& high  & 459s & 15.10ms & \revision{12.13}\\
        flexible \& low  & 3512s & 12.40ms & -\\
        flexible \& middle  & 3479s & 13.54ms & -\\
        flexible \& high  & 3423s & 15.23ms & -\\
        \bottomrule
    \end{tabular}
    \label{tab:2}
\end{table}

\begin{figure*}[htbp]
\centering
\subfloat[Maneuvering target generation.]{
		\includegraphics[scale=0.45]{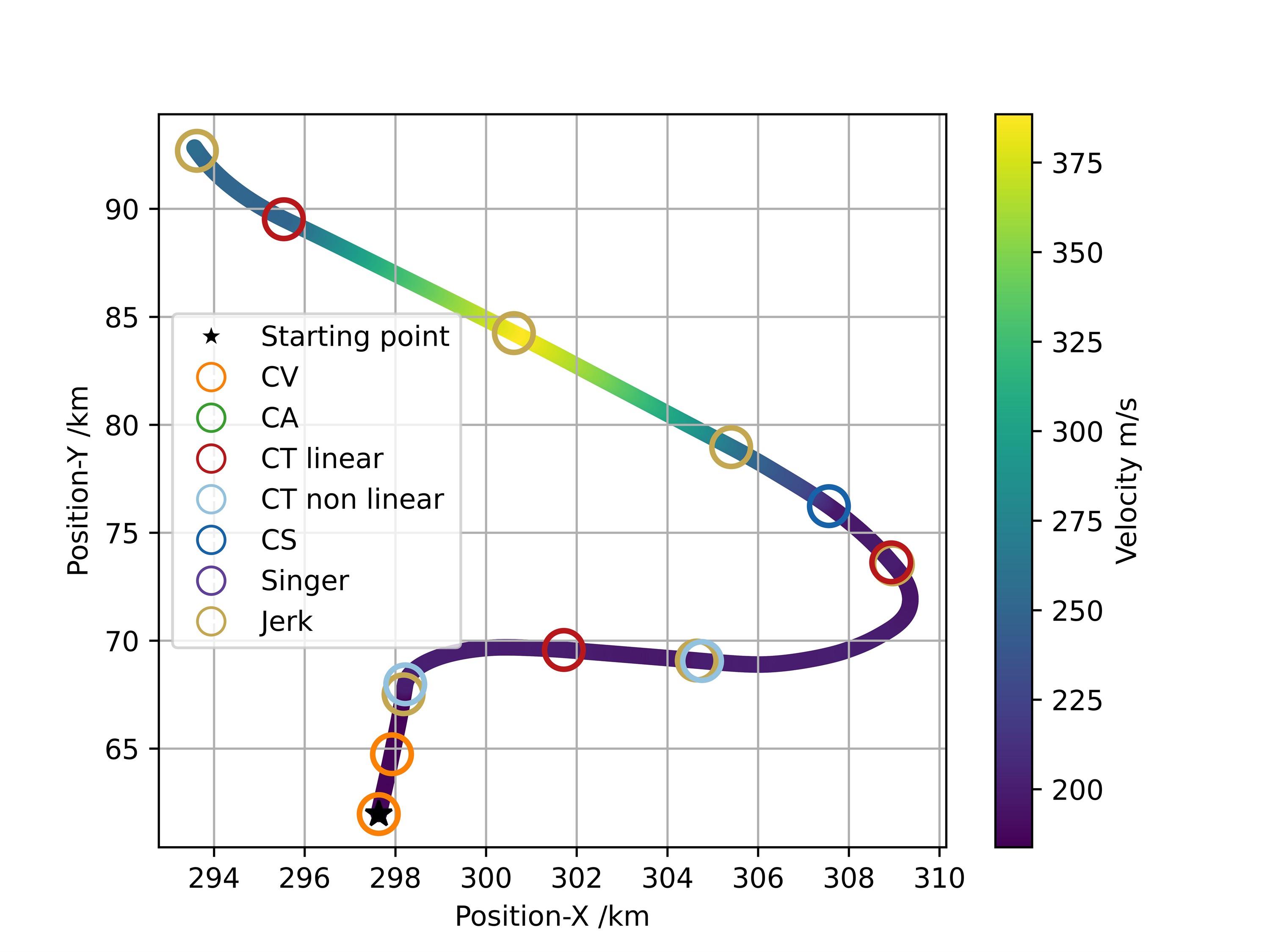}\label{fig:9a}}
\subfloat[Radar measurements generation.]{
		\includegraphics[scale=0.45]{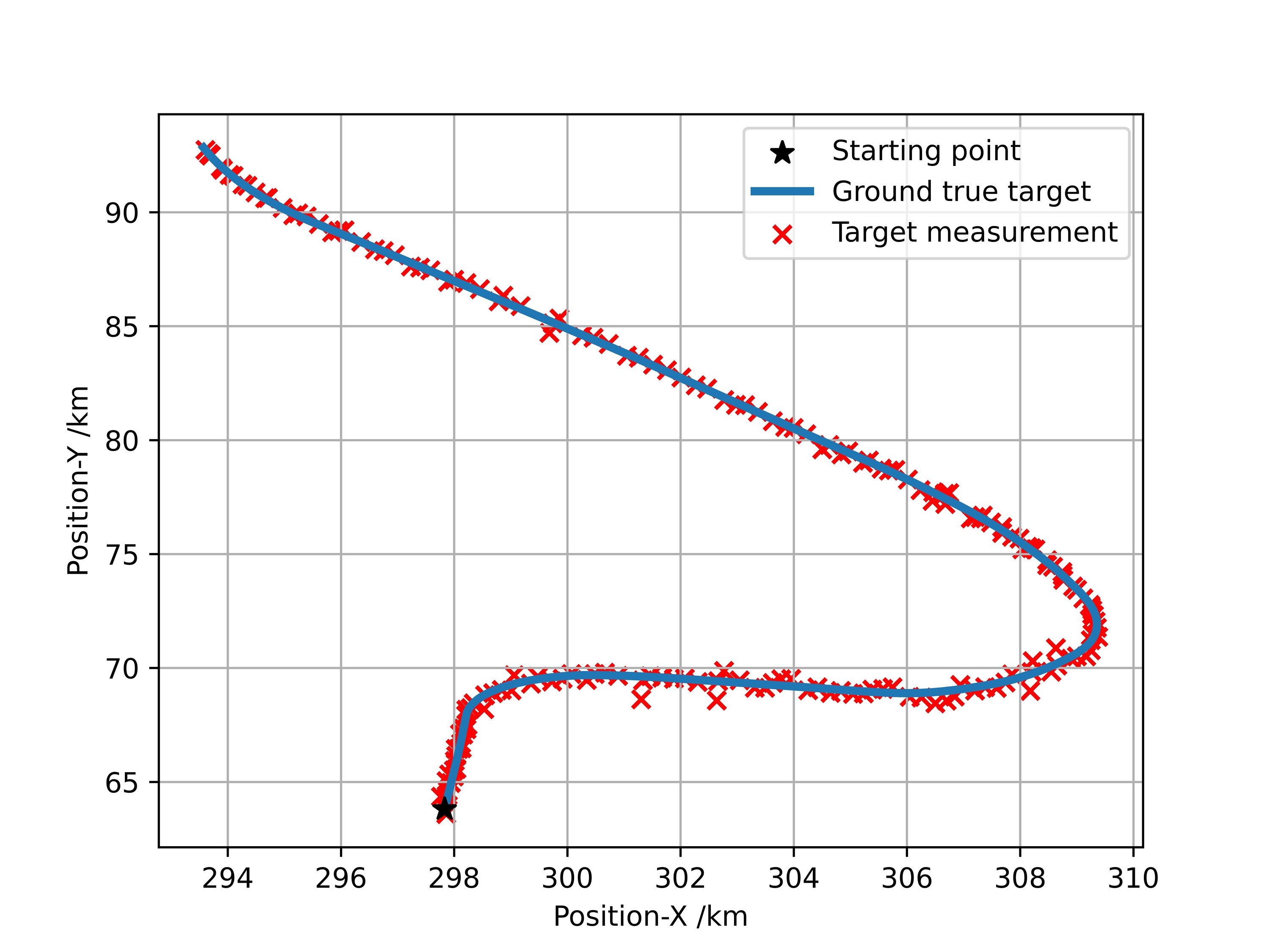}\label{fig:9b}}
\caption{Maneuvering target measurement data generation scene 1.}
\label{fig:9}
\end{figure*}
\begin{figure*}[htbp]
\centering
\subfloat[Maneuvering target generation.]{
		\includegraphics[scale=0.45]{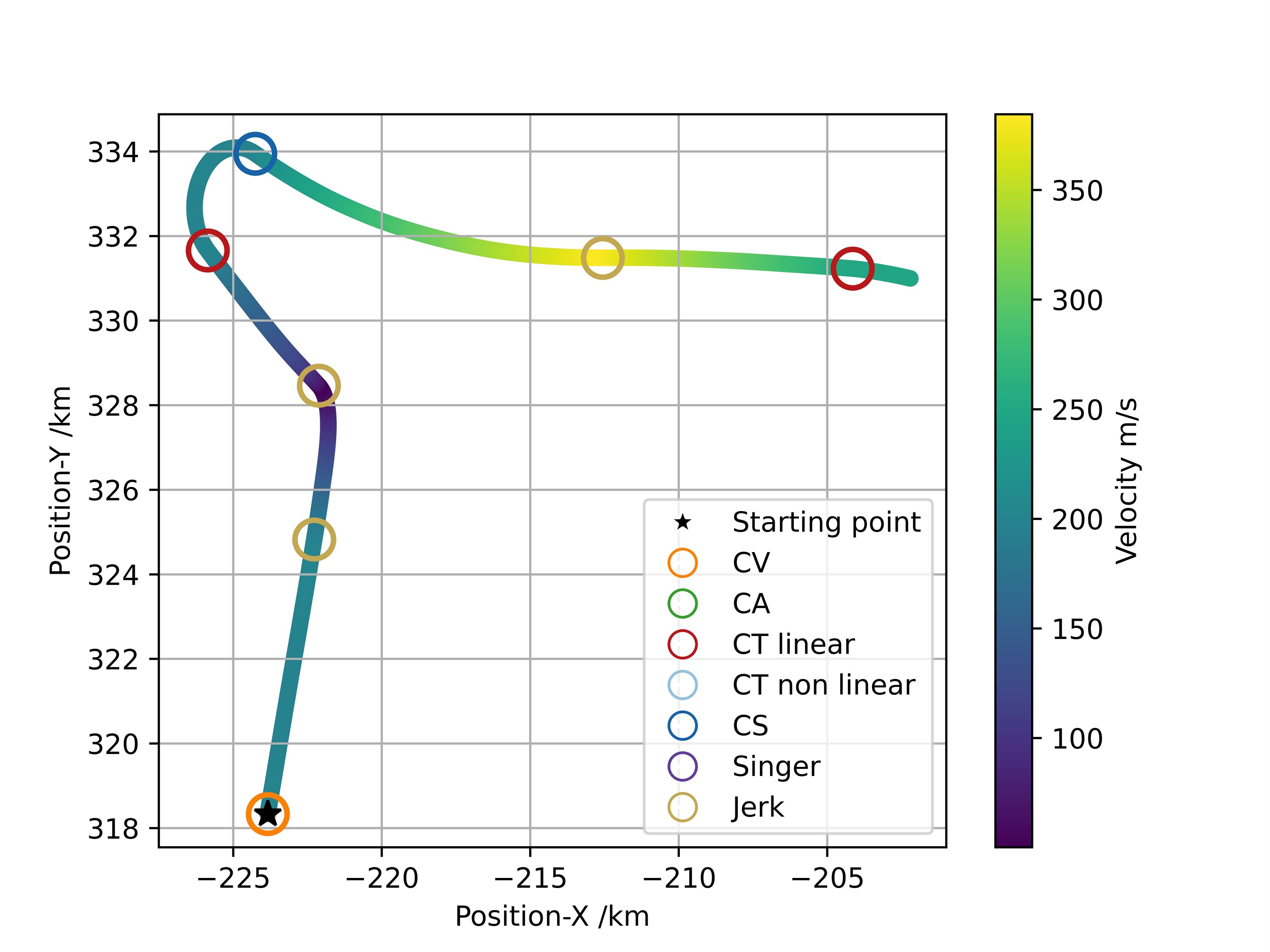}}\label{fig:10a}
\subfloat[Radar measurements generation.]{
		\includegraphics[scale=0.45]{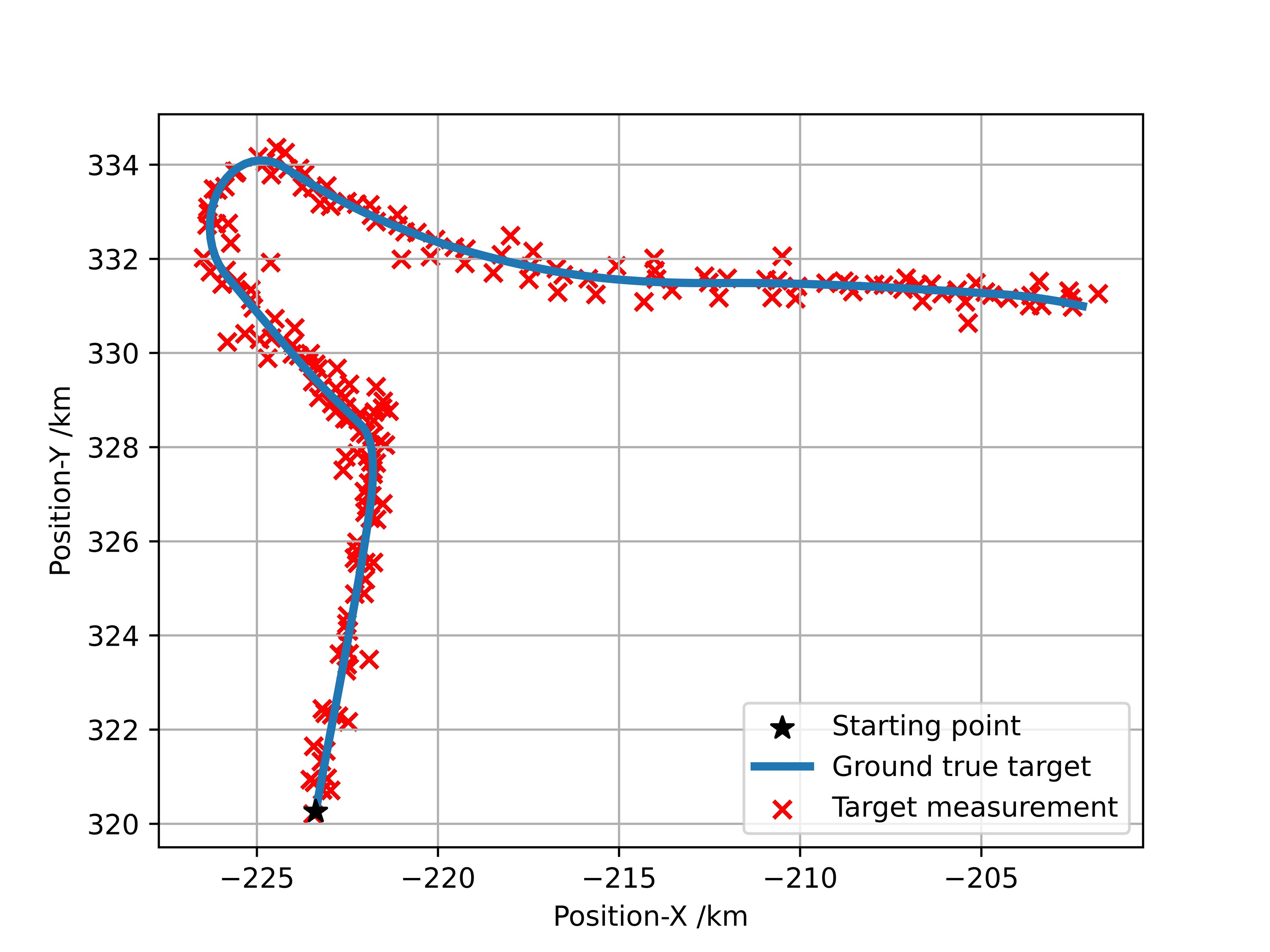}\label{fig:10b}}
\caption{Maneuvering target measurement data generation scene 2.}
\label{fig:10}
\end{figure*}
\subsection{Single-target Tracking}
\begin{figure*}[!t]
\centering
\subfloat[Position RMSE of 100 MC tracking for scene 1.]{
		\includegraphics[width=13cm]{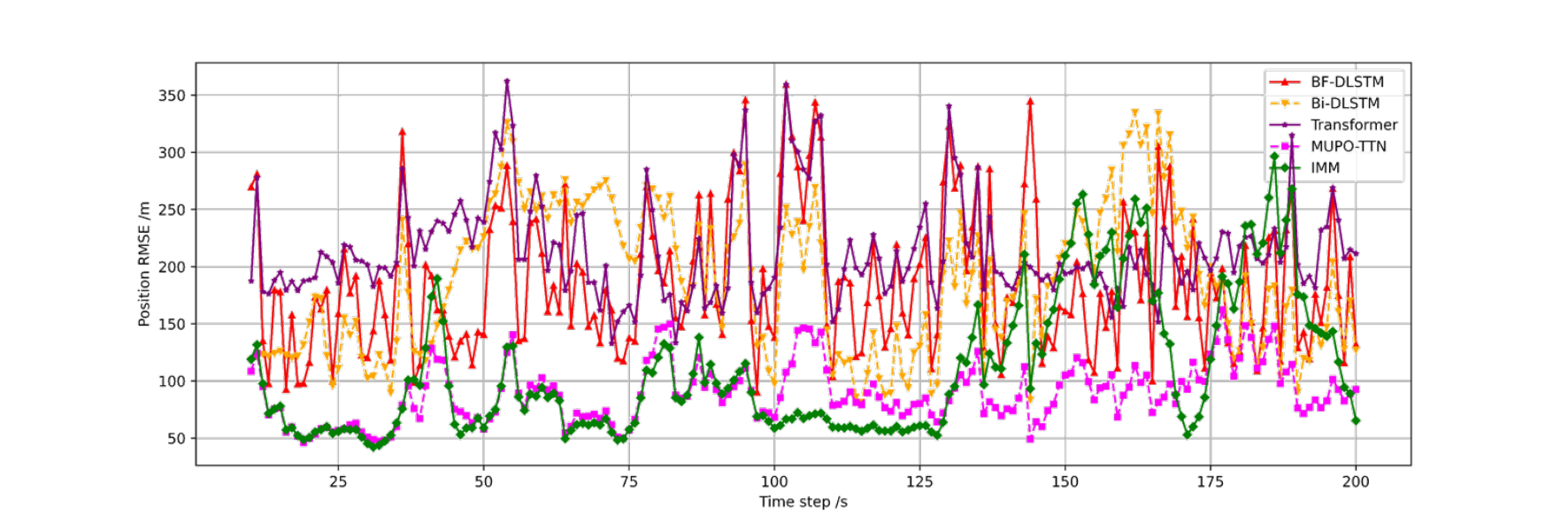}}\\
\subfloat[Position RMSE of 100 MC tracking for scene 2.]{
		\includegraphics[width=13cm]{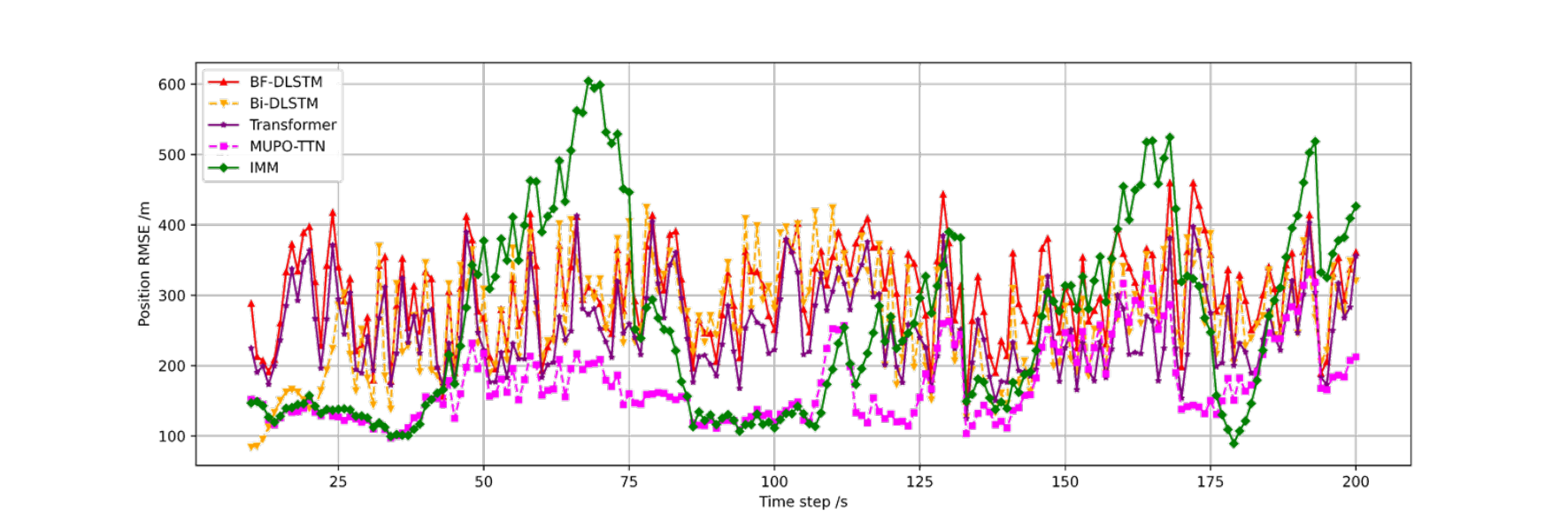}}
\caption{Tracking performance of different tracking methods.}
\label{fig:11}
\end{figure*}
The challenge in maneuvering target tracking lies in whether the tracking algorithm can promptly match the target motion model when the target dynamic behavior changes abruptly. 
In this subsection, a comparative experiment is devised to compare the performance disparities between the proposed MUPO-TTN and other maneuvering target tracking algorithms. 
MUPO-TTN with flexible-sampling and low density is utilized to track a maneuvering target in this subsection. 
The other maneuvering target tracking algorithms are classified into model-based and data-driven. 
Among the model-based algorithms, the classic IMM method is elected as the comparison algorithm. 
Within the IMM method, the target motion model encompasses the CV models, CA models which incorporate diverse process noises, and CT models with different turning rates. 

In terms of data-driven methods, several deep neural networks proficient in handling sequential tasks, such as BF-DLSTM \cite{cite25}, Bi-DLSTM \cite{cite28}, and Transformer \cite{cite31}, are selected as the comparison methods. 

The BF-DLSTM, Bi-DLSTM, and Transformer networks are configured with the following key parameters for comparison: the BF-DLSTM model uses multiple stacked LSTM layers, with 6 LSTM layers and a hidden unit size of 128. 
The Bi-DLSTM model processes sequences in both directions, with 6 bidirectional LSTM layers and a hidden unit size of 128. 
The Transformer model features 4 attention heads, 6 encoder and decoder layers, a hidden size of 128.
In order to demonstrate the impact of using MUPO representation on the tracking performance, we also input the measurement noise distribution information into the other comparison methods. 
Specifically, the IMM method uses $\boldsymbol{\Sigma}_{\text {MUCM}}$ obtained from (\ref{eq10}) during state update, while the data-driven methods incorporate the target SNR and $\boldsymbol{\Sigma}_{\text {MUCM}}$ as inputs during training.

\begin{table}[!t]
    \centering
    \renewcommand{\arraystretch}{1.5}\caption{Target dynamic model in scene 1.}
    \begin{tabular}{cc}
        \toprule
        Time range (time interval is 0.1 s) & Target motion model \\
        \midrule
        0s-33.3s & CV\\
        33.4s-103.3s & Jerk\\
        103.4s-123.3s & CT\\
        123.4s-164.8s & CS\\
        164.9s-191.9s & Jerk\\
        192.0s-200.0s & CT\\
        \bottomrule
    \end{tabular}
    \label{tab:3}
\end{table}
Assume that the radar is situated at the origin of the two-dimensional Cartesian coordinate system. 
The initial target motion model is the CV model, and the initial state of the target is acquired as the way mentioned in Subsection \hyperref[simulation]{IV.A}. 
The target trajectory duration is set as 200 s. 
In a single experiment, the types of the target motion model are presented in Table \ref{tab:3}, and the target trajectory is depicted in Fig. \ref{fig:9a}. 
When generating the target measurement sequence, the target measurement noise is devised based on SNR, and the measurement interval is set to be 1 s. 
The diagram of the target measurement sequence is shown in Fig. \ref{fig:9b}. 
To substantiate that the tracking performance of the method proposed in this paper is not confined to one single track, another maneuvering target track is concurrently generated in accordance with the aforementioned steps, as depicted in Fig. \ref{fig:10}.
\begin{table}[!t]
    \centering
    \caption{ARMSE(m) of 100 MC tracking for extra 10 targets.}
    \begin{tabular}{c>{\centering\arraybackslash}p{1cm}>{\centering\arraybackslash}p{1cm}>{\centering\arraybackslash}p{1.3cm}>{\centering\arraybackslash}p{1cm}>{\centering\arraybackslash}p{1cm}}
        \toprule
         & BF-DLSTM & Bi-DLSTM & Transformer & MUPO-TTN & IMM \\
        \midrule
         1 & 247.53 & 303.44 & 219.43 & \textbf{131.62} & 148.98 \\
         2 & 232.75 & 201.41 & 211.73 & \textbf{133.25} & 174.18 \\
         3 & 218.00 & 202.75 & 200.78 & \textbf{127.30} & 146.64 \\
         4 & 161.25 & 138.68 & 178.44 & \textbf{99.18} & 145.02 \\
         5 & 132.34 & 143.29 & 145.63 & \textbf{79.40} & 105.83 \\
         6 & 159.43 & 169.47 & 202.61 & \textbf{101.47} & 138.22 \\
         7 & 174.47 & 190.26 & 281.50 & \textbf{83.73} & 87.63 \\
         8 & 187.40 & 174.14 & 222.59 & \textbf{70.18} & 92.35 \\
         9 & 293.11 & 324.88 & 275.20 & \textbf{153.23} & 217.05 \\
         10 & 242.46 & 198.25 & 203.35 & \textbf{111.86} & 138.14 \\
        \bottomrule
    \end{tabular}
    \label{tab:4}
\end{table}


For the purpose of comparing the tracking performance of diverse methods under steady tracking circumstances, the Monte Carlo (MC) simulations of maneuvering target tracking in distinct scenarios commence from 10 s. 
As illustrated in Fig. \ref{fig:11}, when the target motion model matches, the IMM method can manifest high tracking performance. 
Nonetheless, when the target motion model undergoes a strong maneuver, its tracking performance declines conspicuously. 
Other data-driven methods are less susceptible to target maneuvers compared with the IMM method, but their tracking accuracy during steady tracking remains low. 
While other data-driven methods also exploit noise distribution information, they do not significantly improve tracking performance over our proposed approach.
This also suggests that regression-only based data-driven methods struggle to train an ideal network to accurately fit the exact map. 
%
%

To further illustrate the tracking performance of the proposed approach, an extra 10-target scene is generated.
The tracking performance of these 10 targets and detailed motion model information will be presented in \hyperref[app]{Appendix}.
The average root-mean-square error (ARMSE) of the proposed approach and other methods in 100 MC is tabulated in Table \ref{tab:4}.
The data in Table \ref{tab:4} can also demonstrate that the proposed method exhibits superior tracking performance compared to the other approaches.
\revision{
To evaluate the statistical significance of the performance improvements, a one-sided paired $t$-test was conducted on the RMSE sequences across all experimental scenarios. The results indicate that the $t$-statistics, adjusted for multiple comparisons, consistently exceed the critical threshold at a significance level of $p < 0.05$. These statistical metrics, derived from the RMSE differences between the comparison algorithms and the proposed method, are detailed in the Table \ref{app_tab1}. Based on these findings, it can be concluded that the observed tracking performance gains are not attributable to random variation.
}
\begin{table}[!t]
    \centering
    \caption{\revision{$t$-statistic of 100 MC tracking for extra 10 targets.}}
    \revision{
    \begin{tabular}{c>{\centering\arraybackslash}p{1.3cm}>{\centering\arraybackslash}p{1.3cm}>{\centering\arraybackslash}p{1.3cm}>{\centering\arraybackslash}p{1.3cm}>{\centering\arraybackslash}p{1.3cm}}
        \toprule
         & BF-DLSTM & Bi-DLSTM & Transformer & IMM \\
        \midrule
         1 & 25.28 & 39.75 & 24.17 & 4.35 \\
         2 & 21.91 & 13.41 & 18.55 & 9.61 \\
         3 & 24.70 & 16.56 & 23.31 & 6.74 \\
         4 & 24.08 & 14.65 & 31.17 & 13.07 \\
         5 & 21.77 & 16.24 & 24.10 & 9.48 \\
         6 & 21.63 & 19.27 & 42.10 & 12.98 \\
         7 & 25.48 & 25.43 & 58.67 & 2.19 \\
         8 & 38.40 & 17.55 & 45.29 & 9.23 \\
         9 & 28.36 & 21.15 & 24.48 & 11.86 \\
         10 & 39.55 & 27.95 & 29.99 & 13.18 \\
        \bottomrule
    \end{tabular}
    }
    \label{app_tab1}
\end{table}
\subsection{Single-target Tracking in 3D Space}

\begin{figure*}[htbp]
\centering
\subfloat[Radar measurements generation in scene 3.\label{r1fig91}]{
		\includegraphics[scale=0.32]{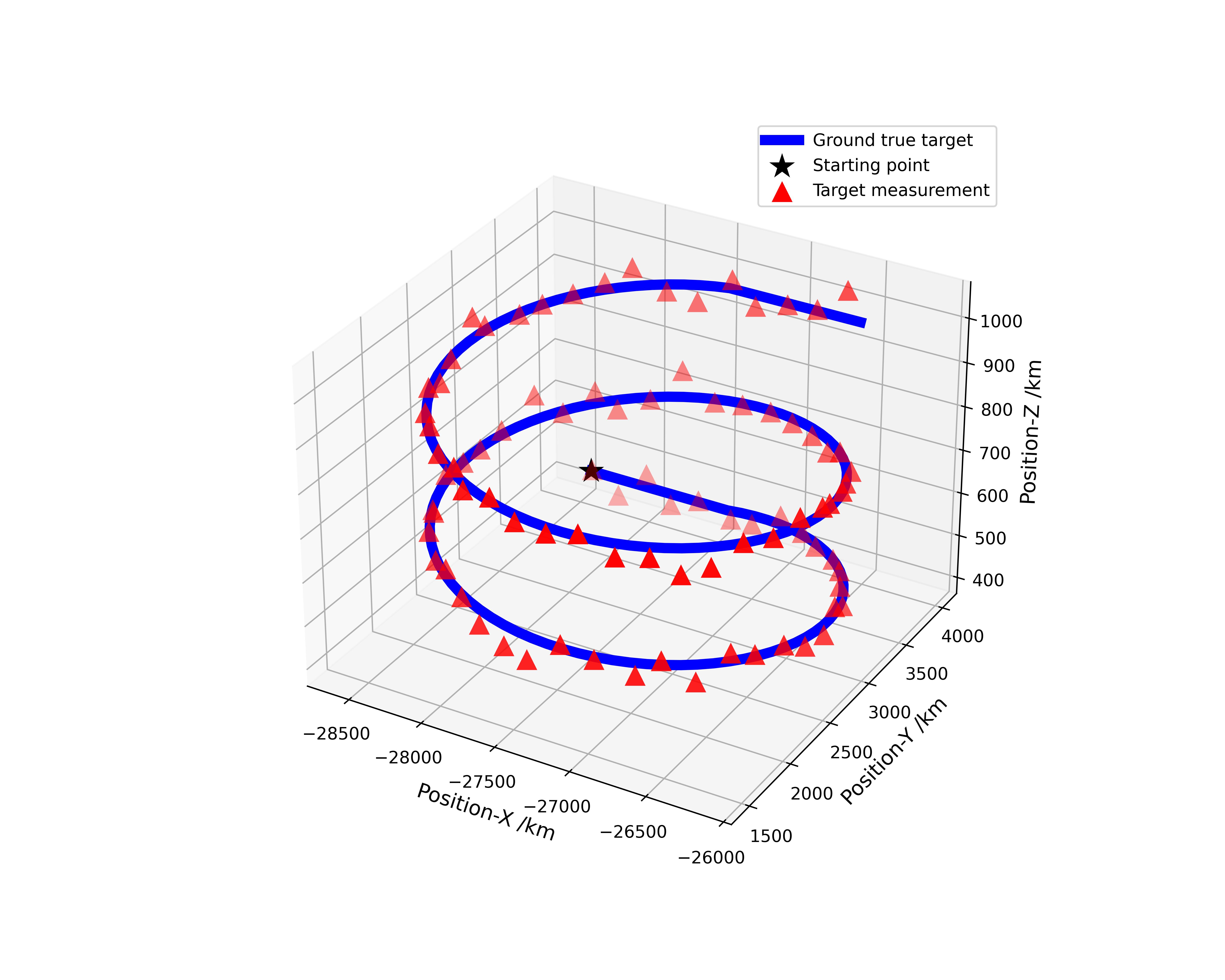}}
\subfloat[Position RMSE of 100 MC tracking for scene 3.\label{r1fig92}]{
		\includegraphics[scale=0.52]{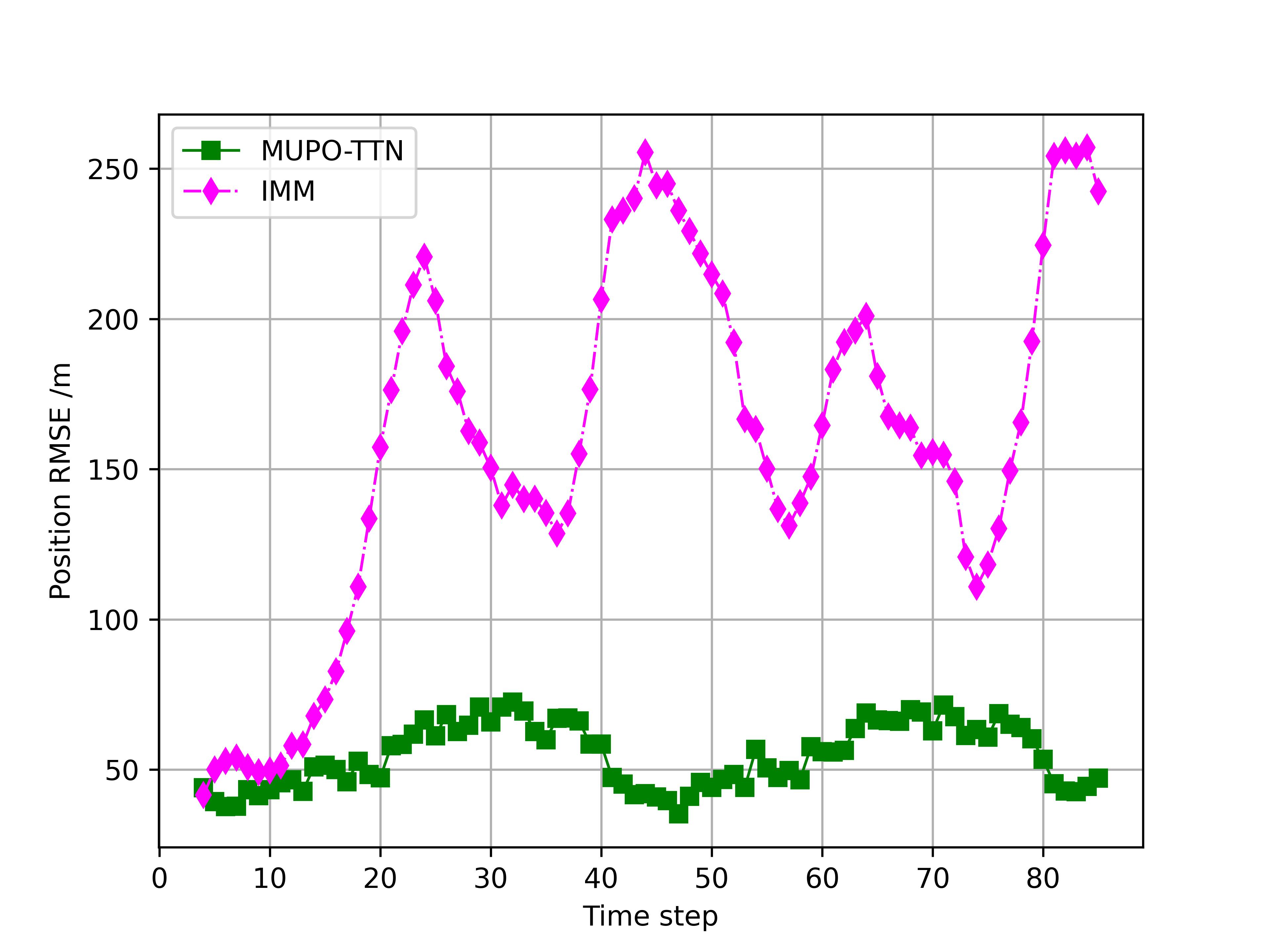}}
\caption{Tracking performance of different tracking methods.}
\label{r1fig99}
\end{figure*}
To demonstrate the proposed framework's scalability to 3D space, comparative tracking experiments were conducted, with simulation settings mirroring Subsection \hyperref[simulation]{IV.A}. 
In 3D space, the multi-channel MUPO was extended by concatenating two independent four-channel MUPOs: one for radial distance and azimuth, another for radial distance and elevation. 
This resulted in an eight-channel tensor, which fed the detection network. 
The network, utilizing multiple prediction heads, outputted target x and y coordinates $[\hat{x}_x, \hat{x}_y]$ from one head, and radial distance $\hat{r}_{xy} = \sqrt{x^2_x+x^2_y}$ along with z coordinates $\hat{x}_z$ from another, where $\hat{\mathbf{x}} = [\hat{x}_x, \hat{x}_y, \hat{x}_z]$ denotes the estimation of target state. 
This dual-head strategy supported domain knowledge-informed loss functions mentioned in Subsection \hyperref[loss]{III.D} for enhanced supervision. 
During inference, the x and y output was magnitude-scaled using the directly predicted $\hat{r}_{xy}$ to resolve inconsistencies, thereby forming the final consistent 3D target state $\hat{\mathbf{x}} = [\kappa\hat{x}_x, \kappa\hat{x}_y, \hat{x}_z]$, where $\kappa=\hat{r}_{xy} / \sqrt{\hat{x}^2_x+\hat{x}^2_y}$.

The tracking scenario is established within a distance range of 20 to 30 km with the initial ground speed of the target set at 195 m/s. 
A typical 3D spiral maneuver is employed as the test case where the target first performs a period of level flight before entering the spiral maneuver stage and subsequently returning to level flight. 
The comparison is conducted using the IMM method with a library consisting of 16 motion models. 
This library includes CV and CA models with various process noise power levels along with coordinated turn models featuring different angular velocities. 
A total of 100 MC simulations are performed under the aforementioned conditions. Fig. \ref{r1fig91} illustrates the trajectory results for a single Monte Carlo run and Fig. \ref{r1fig92} presents the statistical tracking performance results obtained from the 100 MC simulations.

The experimental results demonstrate that the IMM method encounters model mismatch during the spiral maneuver of the target. 
It indicates that traditional approaches face significant challenges in modeling complex target motions within 3D space. 
In contrast, the proposed method maintains high tracking performance throughout the maneuver of the target, which validates the scalability of the proposed framework in 3D maneuvering target tracking scenarios.

\subsection{Multi-target Tracking}\label{multtt}
\begin{figure*}[htbp]
\centering
\subfloat[Maneuvering target generation in scene 4.]{
		\includegraphics[scale=0.27]{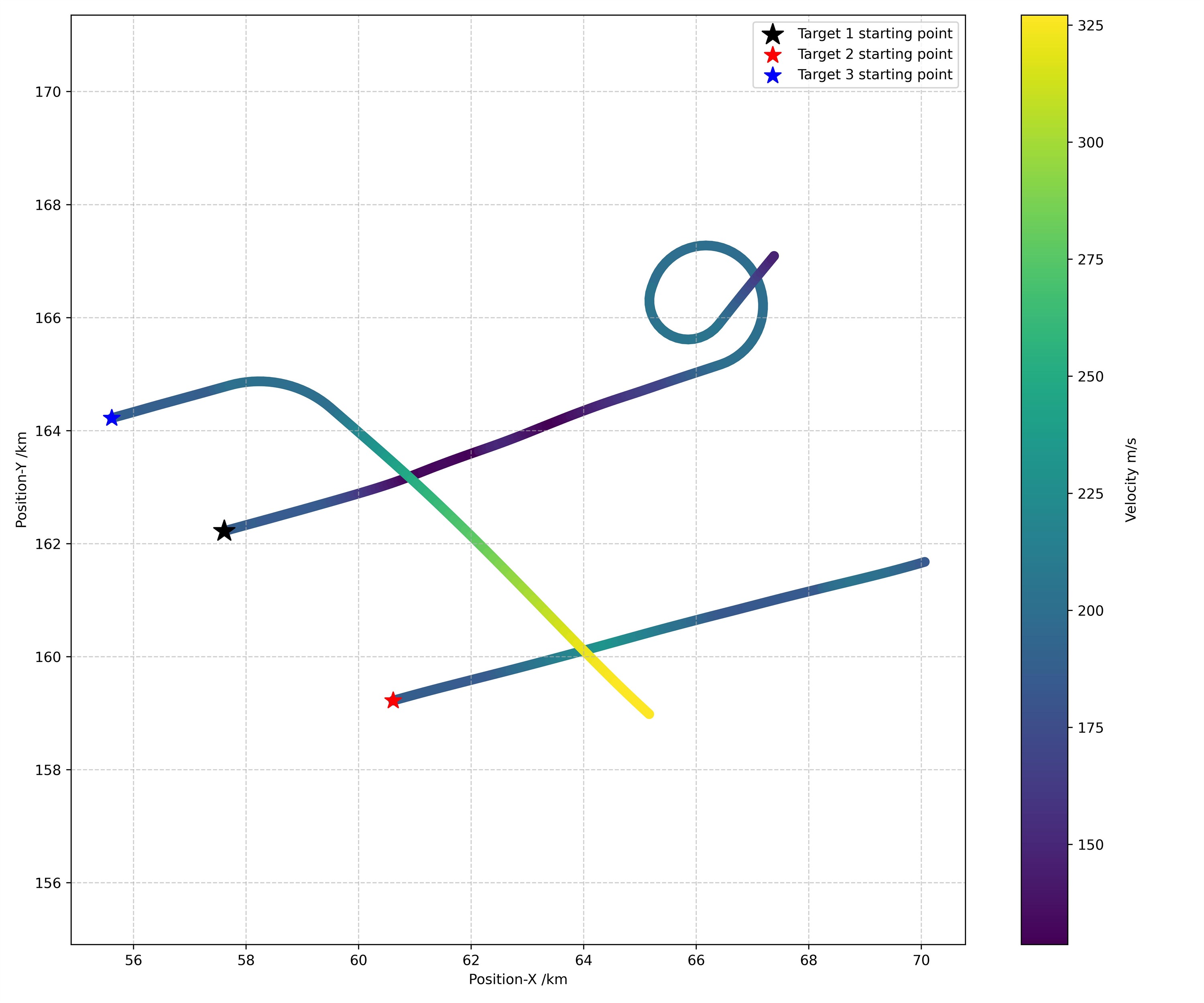}}
\subfloat[Maneuvering target generation in scene 5.]{
		\includegraphics[scale=0.27]{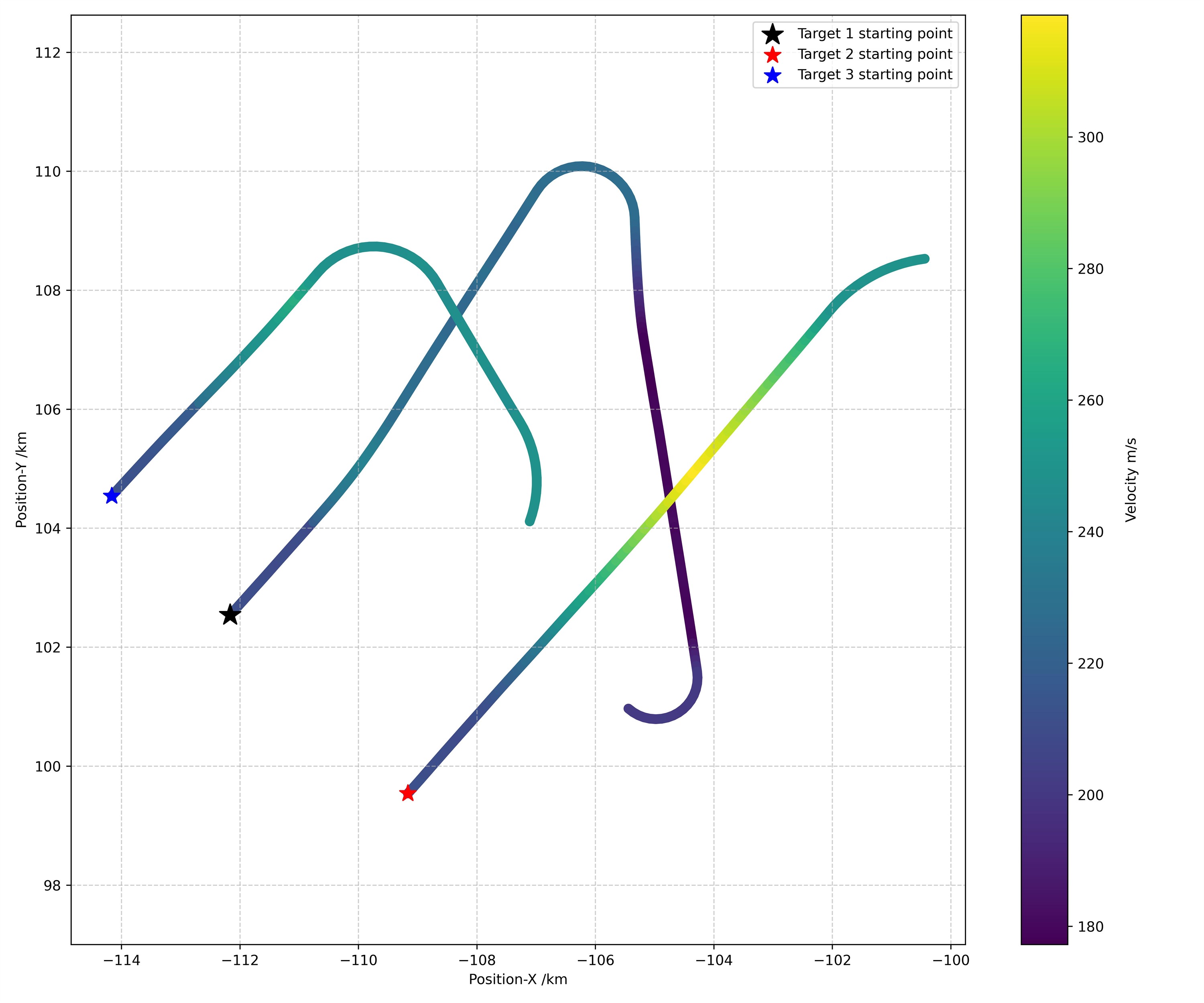}}
\caption{Maneuvering target trajectory generation.}
\label{r1fig1}
\end{figure*}
\begin{figure*}[htbp]
\centering
\subfloat[Radar measurements generation in scene 4.]{
		\includegraphics[scale=0.27]{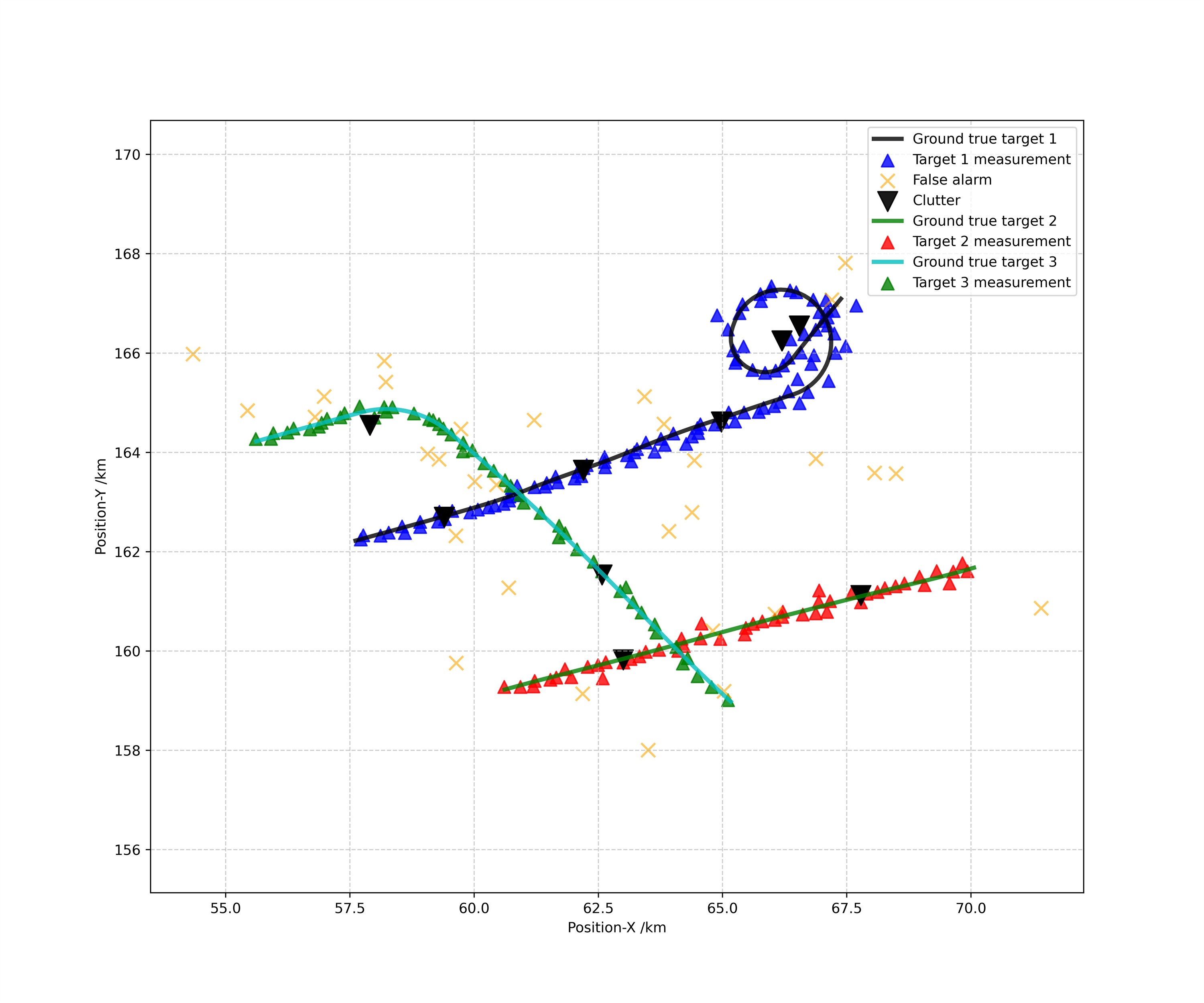}}
\subfloat[Radar measurements generation in scene 5.]{
		\includegraphics[scale=0.27]{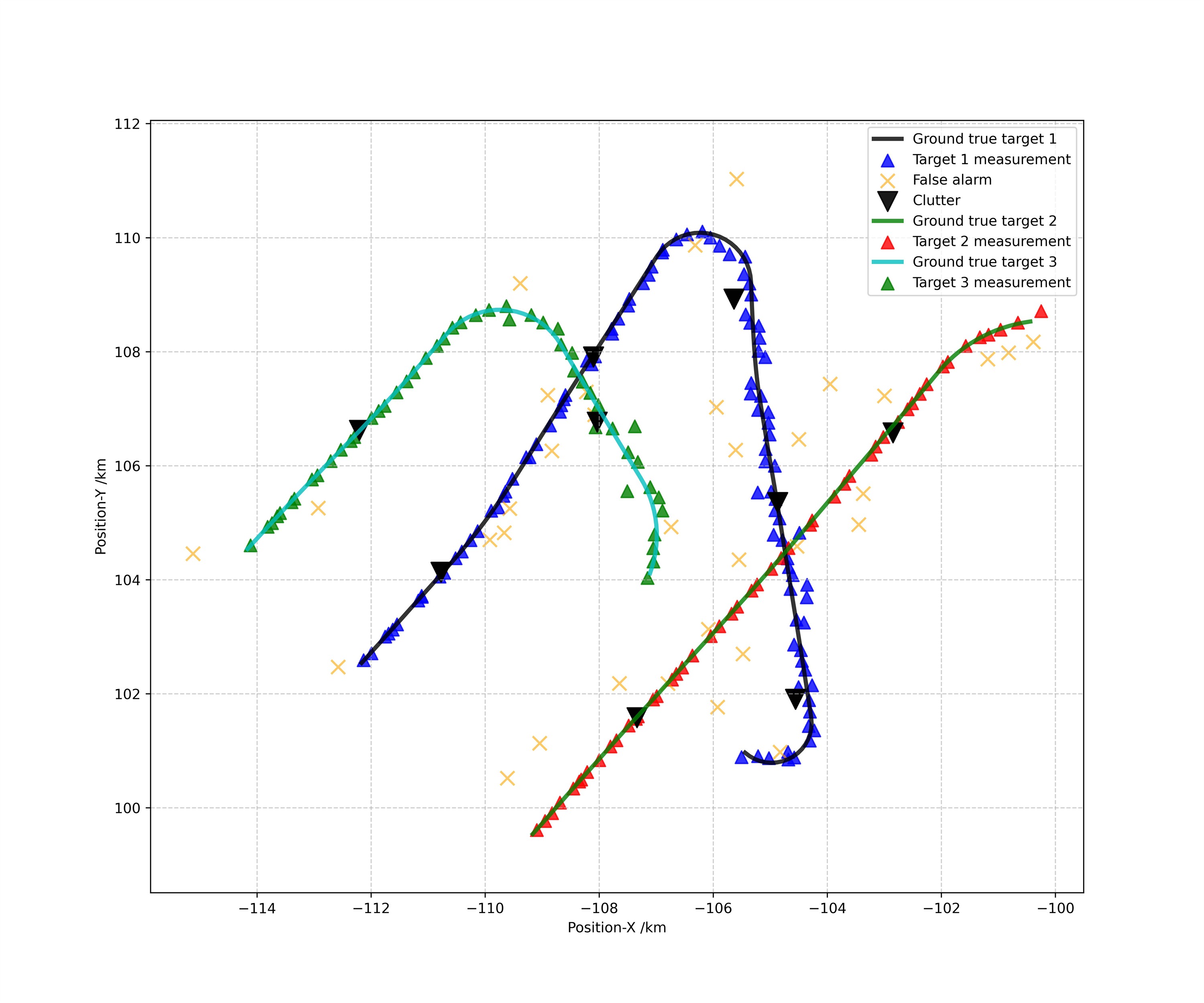}}
\caption{Maneuvering target measurement data generation in one MC simulation.}
\label{r1fig2}
\end{figure*}

\begin{figure*}[htbp]
\centering
\subfloat[Maneuvering target generation in scene 6.]{
		\includegraphics[scale=0.27]{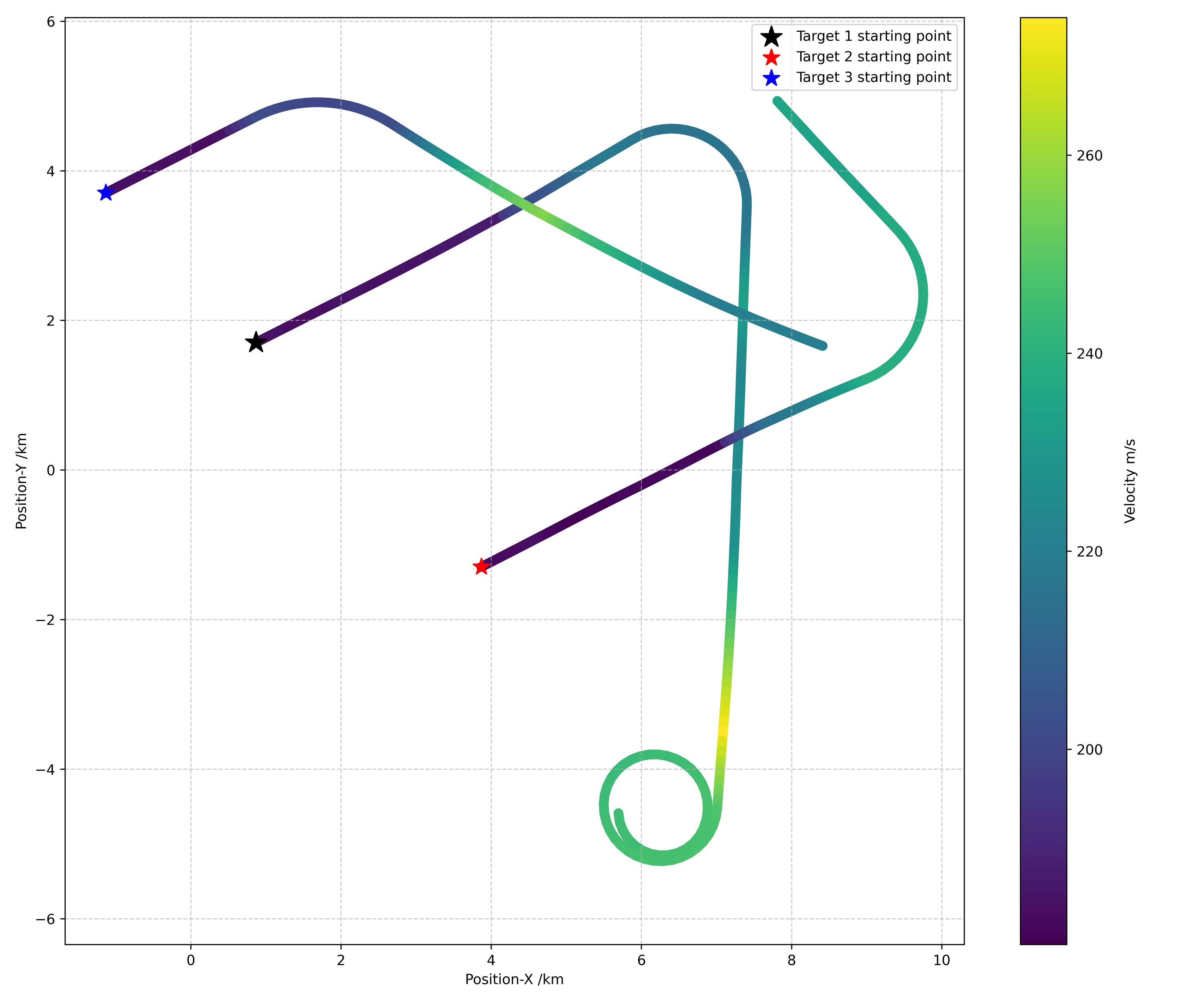}}
\subfloat[Maneuvering target generation in scene 7.]{
		\includegraphics[scale=0.27]{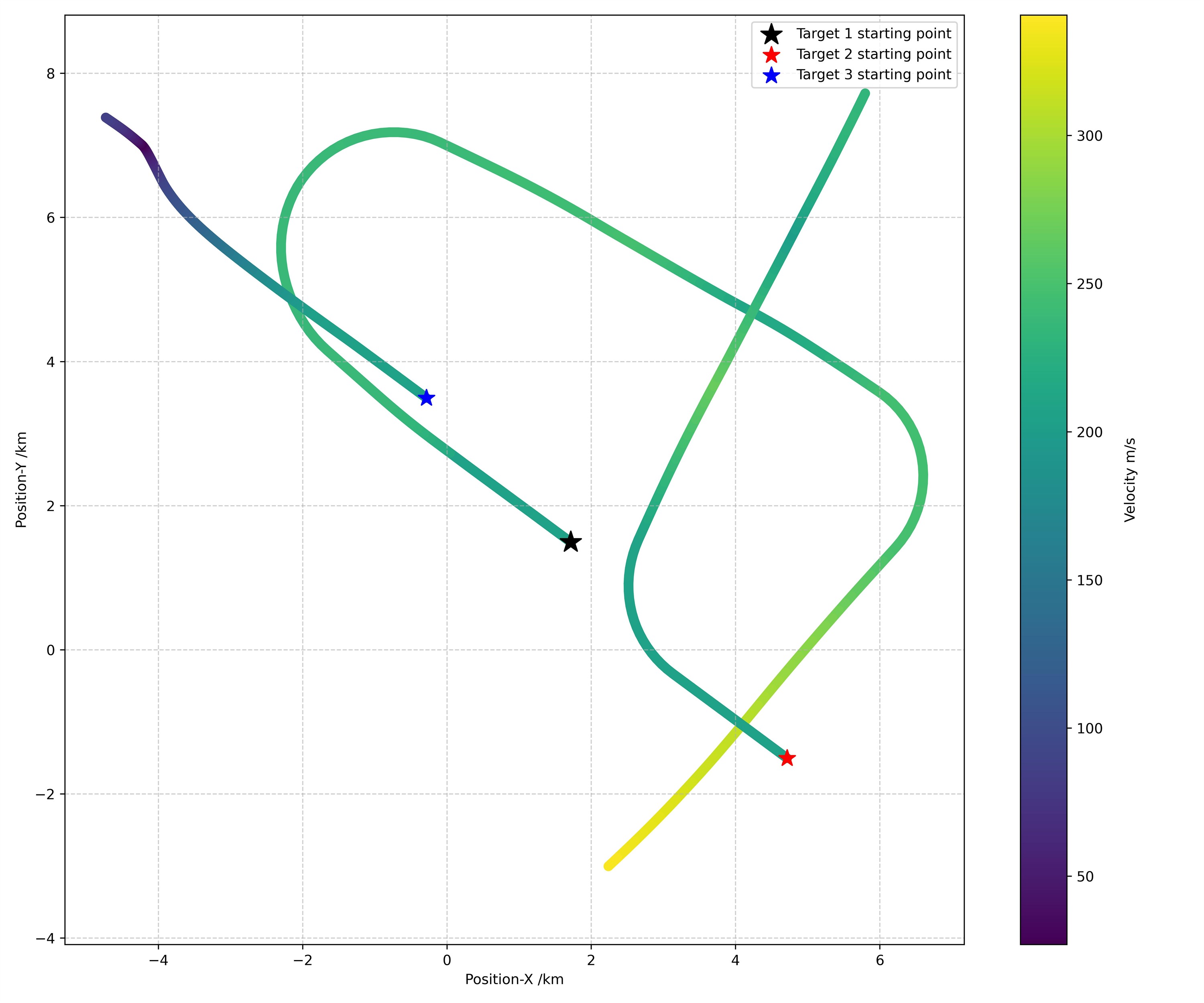}}
\caption{Maneuvering target trajectory generation.}
\label{r1fig3}
\end{figure*}
\begin{figure*}[htbp]
\centering
\subfloat[Radar measurements generation in scene 6.]{
		\includegraphics[scale=0.27]{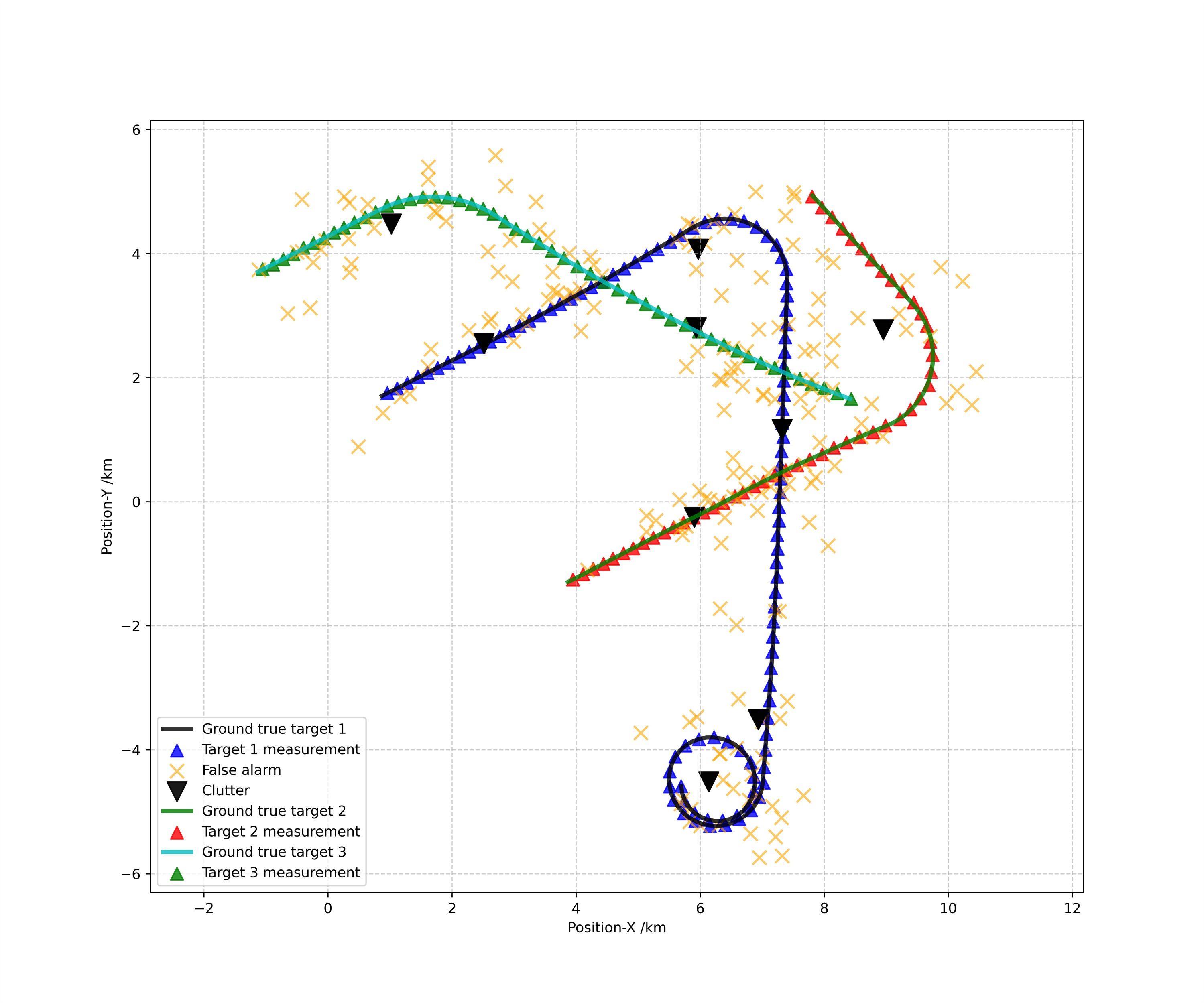}}
\subfloat[Radar measurements generation in scene 7.]{
		\includegraphics[scale=0.27]{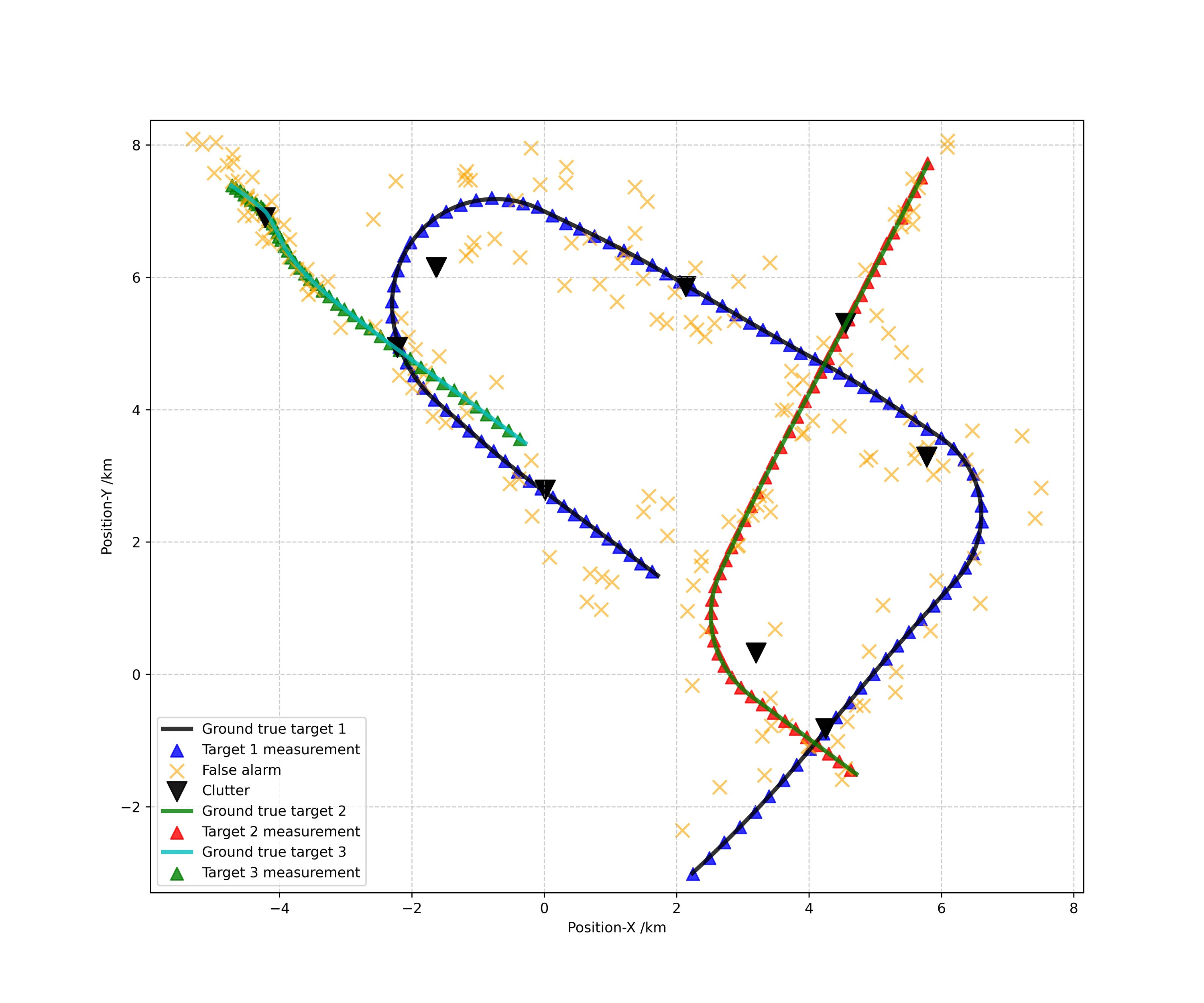}}
\caption{Maneuvering target measurement data generation in one MC simulation.}
\label{r1fig4}
\end{figure*}

In complex environments, maneuvering target tracking is inevitably challenged by false alarms, clutter and multiple overlapping targets, which lead to variations in the observed number of measurements. 
To validate the tracking performance of the proposed algorithm under such challenging scenarios, corresponding experiments are designed in this subsection. 
The simulation model for target trajectories remains consistent with the description in Subsection \hyperref[simulation]{IV.A}.

To simulate a clutter environment populated by buildings or terrain features, nine stationary targets were introduced into the scenario. 
It is noteworthy that the measurements for these stationary targets were generated using the identical observation model and noise characteristics as those applied to the targets of interest.
Furthermore, to faithfully represent a realistic clutter environment, randomly distributed false alarms are superimposed on the measurement set.The number of false alarms $J$ per frame follows a Poisson distribution with parameter $\lambda_c$. 
Specifically, the value of $J$ is determined by the given parameter $\lambda_c$ and a uniformly distributed random variable $r$ in the interval $[0, 1]$, according to the following equation:
\begin{equation}\label{r1eq5}
e^{-\lambda_c} \sum_{j=0}^{J-1} \frac{\lambda_c^j}{j!}<r \leq e^{-\lambda_c} \sum_{j=0}^J \frac{\lambda_c^j}{j!}, J=1,2, \ldots
\end{equation}

The parameter $\lambda_c$ is established based on a predefined false alarm rate $r_{fa}=1\times10^{-4}$ per second. A region of interest (ROI) is delineated within the radar's surveillance area, bounded by the minimum and maximum azimuthal and radial $\theta_{max} ,\theta_{min},\rho_{max},\rho_{min}$.
This ROI is subsequently discretized into grid cells, each characterized by dimensions $d_\theta=0.2^{\circ}$ and $d_\rho=20m$ in the azimuthal and radial directions, respectively. 
The total count of these grid cells within the defined ROI is then calculated by $N_{cell}=\left(\theta_{max} -\theta_{min}\right)\times\left(\rho_{max} - \rho_{min}\right)/\left(d_\theta\cdot d_\rho\right)$. 
Consequently, $\lambda_c$ is computed by $N_{cell}\times r_{fa}$. 
For each frame, false alarms are uniformly distributed across the delineated ROI.

In order to establish a valid benchmark, the IMM algorithm coupled with Nearest Neighbor (NN) data association is adopted as the comparative method, where the nearest neighbor is determined based on the Mahalanobis distance.
Considering that the number of false alarms varies significantly between near-field and far-field scenarios due to grid discretization in the polar coordinate system.
To address this, we trained separate network models with the same architecture for each scenario.
The experiments were categorized into near-field (1–10 km) and far-field (150–200 km) maneuvering target tracking tasks.
Each scenario comprised three distinct targets, with their trajectories exhibiting potential pairwise overlap.

Given the requirements of multi-target tracking, an algorithm must not only achieve precise state estimation but also ensure correct measurement-to-track association to maintain valid tracks. 
Therefore, we utilize the correct association probability (CAP) to assess association reliability and the Optimal Subpattern Assignment (OSPA) distance to measure overall tracking accuracy \cite{r1ref1, r1ref2}. 
The CAP is defined as the ratio of the number of correctly associated frames to the total duration of the final estimated tracks, while the OSPA distance is computed between the final estimated tracks and the ground truth. 
Regarding track maintenance during multi-target tracking, a track should be terminated if no valid measurement is detected within the predicted validation gate.

\textbf{Far-field Scenario:} Fig. \ref{r1fig1} and Fig. \ref{r1fig2} display the target trajectory and the measurements from one specific MC simulation, respectively. 
The quantitative performance averaged over 100 MC simulations is presented in Table \ref{r1t1}.
\textbf{Near-field Scenario:} Fig. \ref{r1fig3} and Fig. \ref{r1fig4} display the target trajectory and the measurements from one specific MC simulation, respectively. 
The quantitative performance averaged over 100 MC simulations is presented in Table \ref{r1t2}.
For the sake of clarity, only false alarm points within the tracking gate are plotted in Fig. \ref{r1fig2} and Fig. \ref{r1fig4}.

\begin{table}[!tbp]
\centering
\renewcommand{\arraystretch}{1.3}
\caption{Tracking performance in far-field scenarios.}
\label{r1t1}
\begin{tabularx}{\columnwidth}{>{\centering\arraybackslash}m{1.5cm} >{\centering\arraybackslash}m{1.2cm} >{\centering\arraybackslash}m{1.5cm} Y Y}
\toprule
\textbf{Scene} & \textbf{Target} & \textbf{Method} & \textbf{CAP} & \textbf{OSPA} \\ \midrule

\multirow{6}{*}{Scene 4} & \multirow{2}{*}{Target 1} & MUPO-TTN & \textbf{92.50\%} & \textbf{177.81} \\
                         &                           & IMM w NN & 89.03\% & 288.28 \\ \cline{2-5} 
                         & \multirow{2}{*}{Target 2} & MUPO-TTN & \textbf{94.10\%} & \textbf{135.64} \\
                         &                           & IMM w NN & 72.32\% & 277.90 \\ \cline{2-5} 
                         & \multirow{2}{*}{Target 3} & MUPO-TTN & \textbf{98.42\%} & \textbf{154.73} \\
                         &                           & IMM w NN & 92.62\% & 228.47 \\ \hline

\multirow{6}{*}{Scene 5} & \multirow{2}{*}{Target 1} & MUPO-TTN & \textbf{90.92\%} & \textbf{241.92} \\
                         &                           & IMM w NN & 78.55\% & 289.22 \\ \cline{2-5} 
                         & \multirow{2}{*}{Target 2} & MUPO-TTN & \textbf{93.86\%} & \textbf{219.28} \\
                         &                           & IMM w NN & 75.54\% & 299.56 \\ \cline{2-5} 
                         & \multirow{2}{*}{Target 3} & MUPO-TTN & \textbf{92.92\%} & \textbf{213.91} \\
                         &                           & IMM w NN & 70.62\% & 239.80 \\ 
\bottomrule
\end{tabularx}\label{r1t1}
\end{table}

\begin{table}[!tbp]
\centering
\renewcommand{\arraystretch}{1.3}
\caption{Tracking performance in near-field scenarios.}
\label{r1t2}
\begin{tabularx}{\columnwidth}{>{\centering\arraybackslash}m{1.5cm} >{\centering\arraybackslash}m{1.2cm} >{\centering\arraybackslash}m{1.5cm} Y Y}
\toprule
\textbf{Scene} & \textbf{Target} & \textbf{Method} & \textbf{CAP} & \textbf{OSPA} \\ \midrule

\multirow{6}{*}{Scene 6} & \multirow{2}{*}{Target 1} & MUPO-TTN & \textbf{97.87\%} & \textbf{143.50} \\
                         &                           & IMM w NN & 95.97\% & 164.03 \\ \cline{2-5} 
                         & \multirow{2}{*}{Target 2} & MUPO-TTN & \textbf{97.03\%} & \textbf{145.76} \\
                         &                           & IMM w NN & 96.75\% & 190.83 \\ \cline{2-5} 
                         & \multirow{2}{*}{Target 3} & MUPO-TTN & \textbf{88.94\%} & \textbf{202.80} \\
                         &                           & IMM w NN & 61.48\% & 293.75 \\ \hline

\multirow{6}{*}{Scene 7} & \multirow{2}{*}{Target 1} & MUPO-TTN & \textbf{92.18\%} & \textbf{152.45} \\
                         &                           & IMM w NN & 80.93\% & 263.22 \\ \cline{2-5} 
                         & \multirow{2}{*}{Target 2} & MUPO-TTN & \textbf{89.72\%} & \textbf{205.53} \\
                         &                           & IMM w NN & 73.72\% & 269.39 \\ \cline{2-5} 
                         & \multirow{2}{*}{Target 3} & MUPO-TTN & \textbf{77.72\%} & \textbf{259.60} \\
                         &                           & IMM w NN & 67.59\% & 272.18 \\ 
\bottomrule
\end{tabularx}\label{r1t2}
\end{table}



Experimental results demonstrate that the proposed method outperforms traditional approaches in terms of OSPA distance, although IMM method may achieve a higher CAP in certain scenarios.
This phenomenon can be attributed to the conservative nature of traditional algorithms. 
While they perform accurate data association when the motion model matches, they are prone to immediate track termination under target maneuvers or clutter interference due to model mismatch.
Consequently, traditional methods avoid the challenging association phases in dense false alarm and clutter, resulting in a high CAP calculated over the relatively short, surviving track segments. 
However, this premature termination incurs a severe penalty on the OSPA distance.
In scene with adjacent target trajectories and a high number of false alarms, the proposed method demonstrates superior tracking performance of multi-target compared to model-based IMM method.

\section{\revision{DISCUSSION}}
\revision{
Based on the numerical results and the architectural design of MUPO-TTN, this section provides an analysis of its operational constraints and practical deployment considerations.
}

\subsection{\revision{Operational Constraints and SNR Characteristics}}
\revision{
In practical radar operations, the tracking process is fundamentally preceded by target detection and track initiation. Once a track is successfully established, the system transitions into a dedicated tracking mode. In this phase, phased array radar systems typically optimize resource allocation to enhance the SNR of target returns by narrowing the receive beamwidth and extending dwell time.}

\revision{
For targets maintained at a high SNR, these operational adjustments yield superior measurement precision, as the reduced measurement variance directly facilitates more accurate state estimation within the MUPO framework.
Conversely, for low-SNR targets whose echo energy falls below the detection threshold due to stealth or rapid attitude maneuvers, the measurement sequence is characterized by non-uniform or irregular time intervals. In such scenarios, prediction-based interpolation can be employed to compensate for missing data points and maintain track continuity. However, in extreme cases where the target remains undetected over multiple consecutive frames, a track loss or interruption occurs, eventually necessitating a full track re-initiation process to recover the target.
}

\subsection{\revision{Computational Complexity and Practical Deployment}}
\revision{
From a theoretical perspective, the proposed MUPO-TTN exhibits suitability for real-time radar data processing. Given that the radar scanning period is set to 1 s, the inference time of the proposed method satisfies the response requirements for real-time target tracking, regardless of whether flexible-sampling or fixed-sampling is adopted.
It is worth noting that the scales of the proposed multi-channel MUPO depend on the spatial distribution of target measurements. 
Consequently, a shorter scanning period leads to a smaller measurement distribution range and MUPO size, which inherently reduces the computational cost. 
Additionally, inference speed can be further improved through model compression strategies, such as knowledge distillation \cite{r1dis} and structural pruning \cite{r1pru}. These methodologies provide a theoretical pathway for scaling the algorithm to accommodate diverse radar hardware specifications.
}

\revision{
Regarding practical hardware deployment, MUPO-TTN requires a computational budget ranging from 10.93 to 12.13 GFLOPs (Table \ref{tab:2}). Mainstream embedded edge-computing GPUs typically deliver peak throughputs at the TFLOPS scale, which easily covers this requirement. Beyond raw throughput, the deployment of MUPO-TTN on customized computing chips depends significantly on memory efficiency. The memory-efficient backbone framework \cite{cite47} ensures a controlled Video RAM footprint, facilitating its integration into resource-constrained edge devices. Concurrently, the fixed-sampling scheme produces spatially consistent input tensors, which simplifies data orchestration and maximizes parallel acceleration on hardware like FPGAs \cite{r1fpga}. These architectural advantages collectively demonstrate the practical viability of implementing MUPO-TTN in real-time radar data processors.
}

\section{CONCLUSION}\label{co}
Most data-driven methods are designed based on a regression-only framework under the assumption of a fixed measurement noise distribution. 
Such designs often introduce optimization conflicts in the tracking algorithm, increasing the complexity of network training and degrading tracking performance. 
In contrast, the target tracking framework proposed in this paper reformulates the maneuvering target tracking problem as a hybrid task of classification and regression. 
The classification task is used to coarsely identify the region containing the target, followed by a regression task for higher precision estimation.
The proposed target tracking framework demonstrates capability for multi-target tracking by incorporating simple association criteria such as maximum likelihood. 
Furthermore, the framework is not limited to 2D maneuvers but can be extended to 3D cases. 
Experimental evaluations conducted in far-field scenarios with significant measurement uncertainty and complex target maneuverability indicate that the proposed method achieves superior tracking precision compared to existing regression-only data-driven methods and IMM method. 
The effectiveness of the framework is further validated through a 3D maneuvering experiment, where it maintains high estimation precision relative to the IMM method, which suffers from model mismatch, thereby proving its dimensional scalability. 
In addition, multi-target tracking experiments conducted in environments with clutter and false alarms demonstrate that the proposed method achieves better tracking performance of multi-target in both far-field and near-field.

In the scenarios investigated within this study, the statistical characteristics of the measurement noise are inherently coupled with the fluctuations in the SNR. 
Although the RCS fluctuation process is currently simulated using the Swerling I model, specific operational conditions such as stealth modes or significant attitude changes relative to the radar line of sight can induce wide range variations in the SNR. 
Achieving robust maneuvering target tracking under such highly dynamic measurement noise distributions remains a critical unresolved challenge and will constitute the primary focus of our future research endeavors.

\section*{APPENDIX}\label{app}
\subsection{Experimental Supplements}
\begin{figure}[!hbp]
    \centering
    \begin{minipage}{0.47\textwidth}
        \centering
        \includegraphics[width=\linewidth]{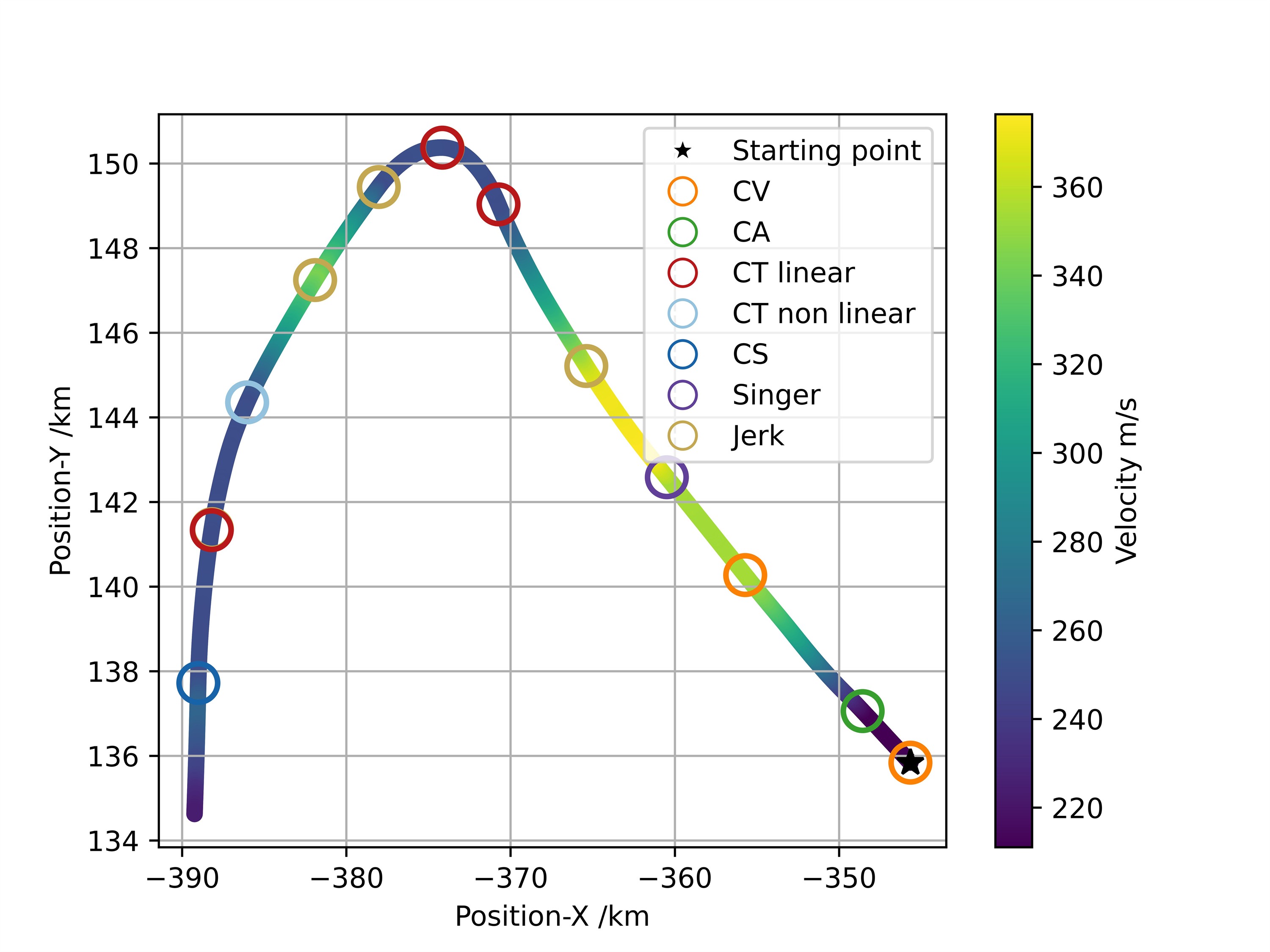}
        \caption{Trajectory for target 1.}\label{fig:13}
    \end{minipage}%

    \begin{minipage}{0.47\textwidth}
        \centering
        \includegraphics[width=\linewidth]{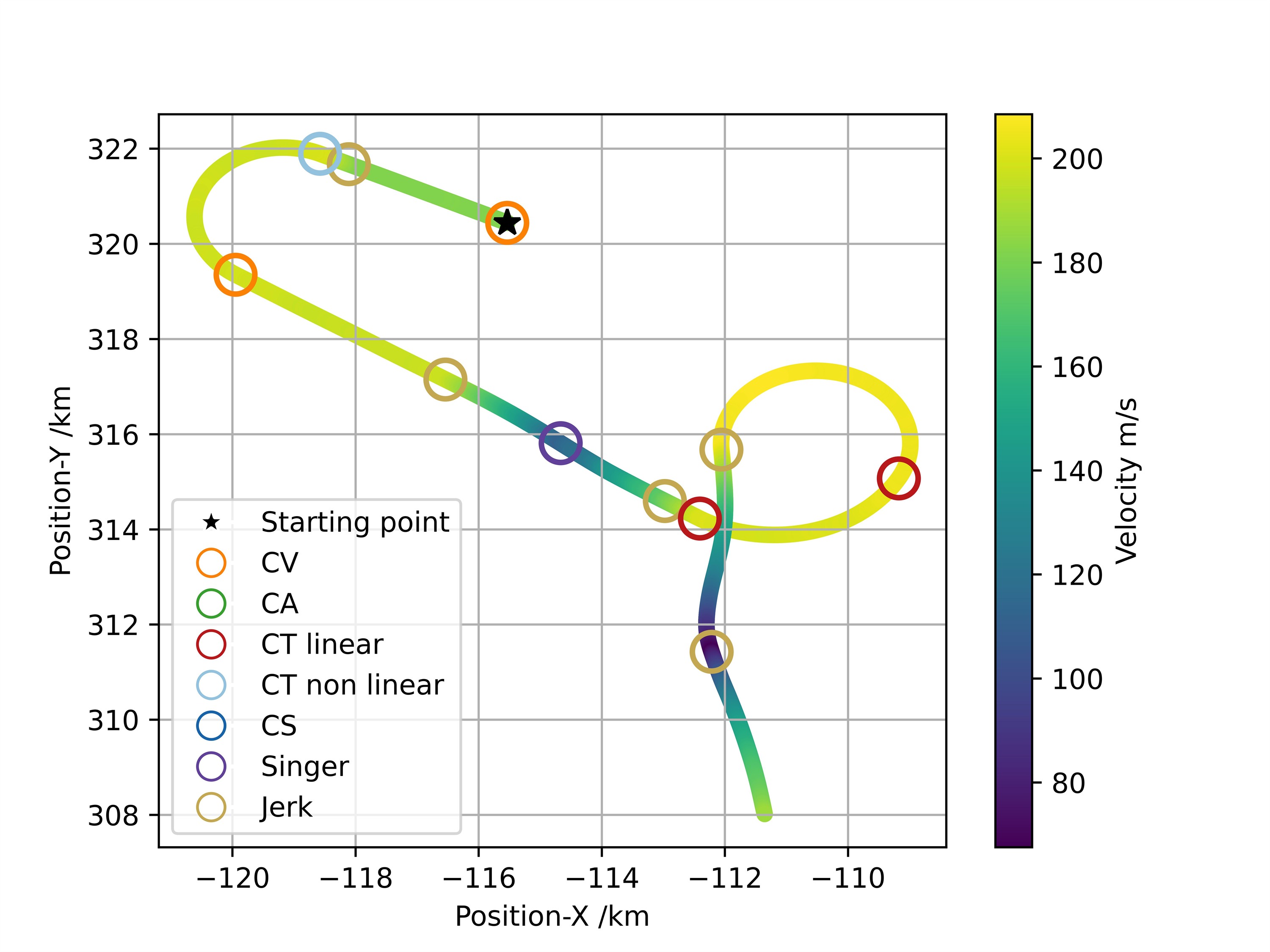}
        \caption{Trajectory for target 2.}
    \end{minipage}
    \begin{minipage}{0.47\textwidth}
        \centering
        \includegraphics[width=\linewidth]{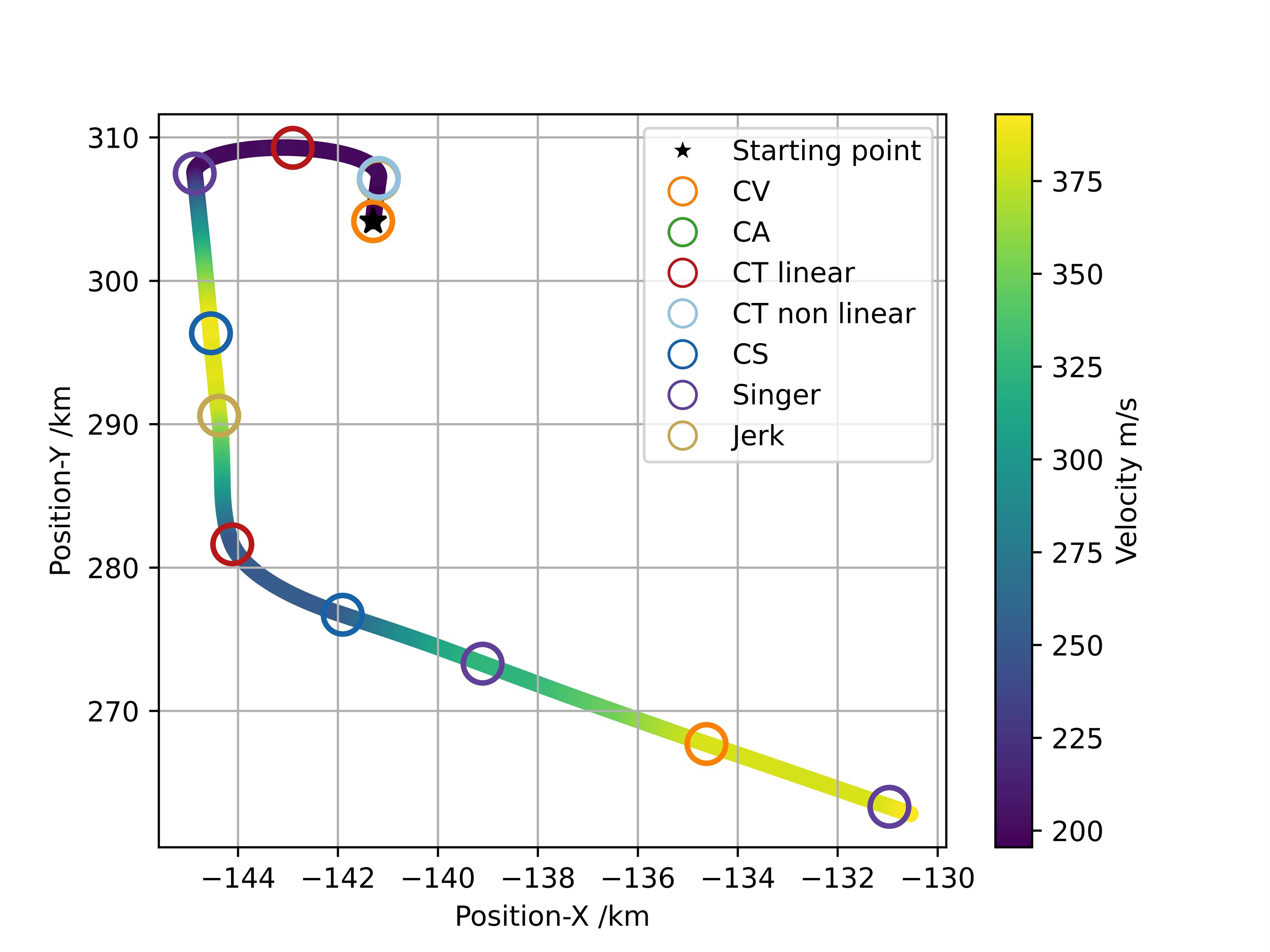}
        \caption{Trajectory for target 3.}
    \end{minipage}

\end{figure}

Suppose the radar is located at the origin of the coordinate system. 
\revision{Figs. \ref{fig:13}-\ref{fig:22}}  display the corresponding target motion model.
Fig. \ref{fig:23} shows the trajectories of these targets. 
\revision{Figs. \ref{fig:24}-\ref{fig:28}} present the tracking performance of various methods in the Monte Carlo simulations. 
Across 10 groups of 100 Monte Carlo simulation runs each, the results show that existing regression-only based data-driven methods, while resistant to significant accuracy degradation from model mismatch, consistently exhibit limited tracking accuracy.
The IMM method achieves high tracking accuracy with matched models but suffers substantial degradation under mismatch.
The proposed framework tolerates maneuver-induced tracking degradation to some extent while maintaining high tracking accuracy during steady tracking.
%


\begin{figure}[htbp]
    \centering
    \begin{minipage}{0.47\textwidth}
        \centering
        \includegraphics[width=\linewidth]{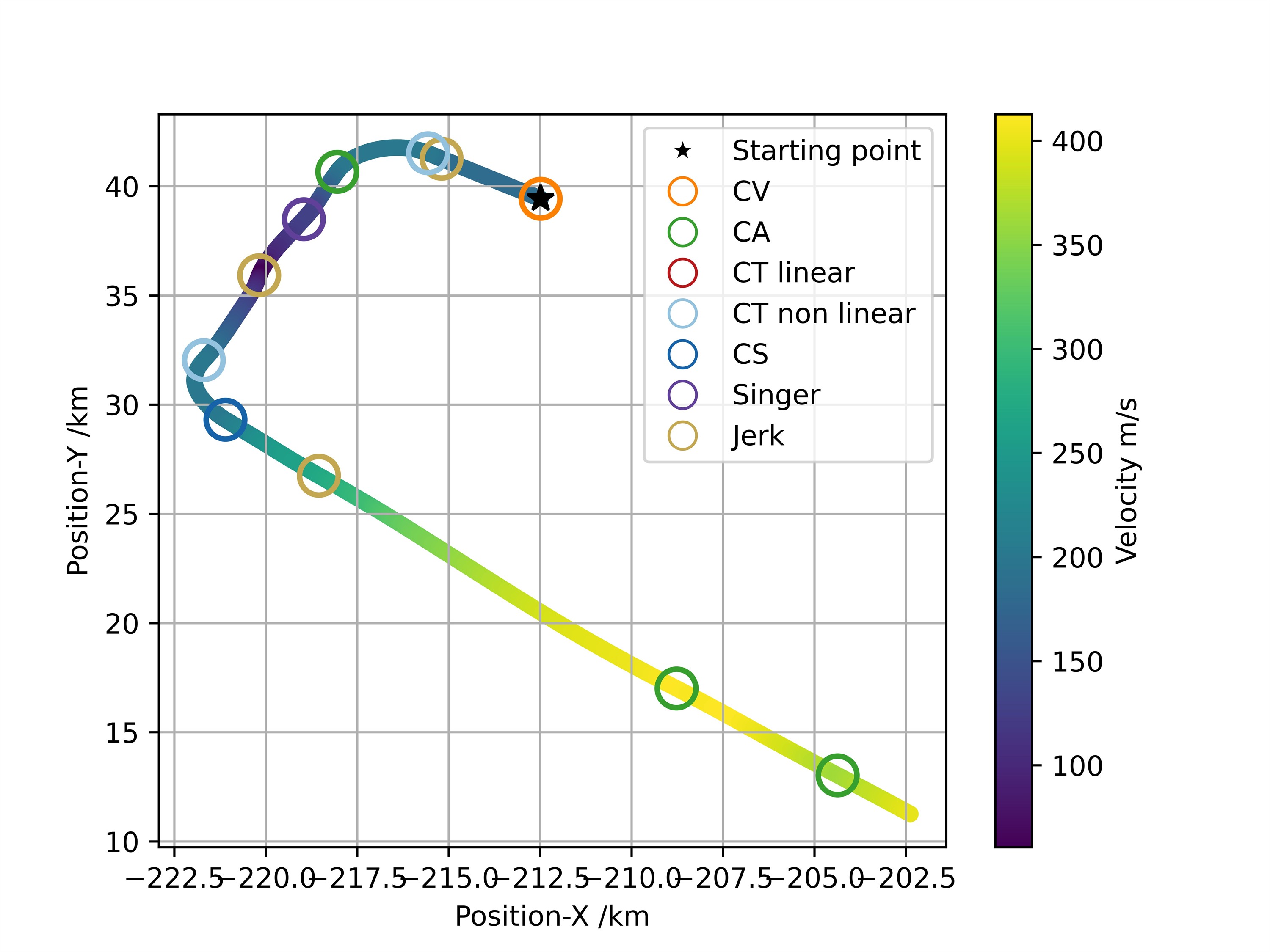}
        \caption{Trajectory for target 4.}
    \end{minipage}%
    \hfill
    \begin{minipage}{0.47\textwidth}
        \centering
        \includegraphics[width=\linewidth]{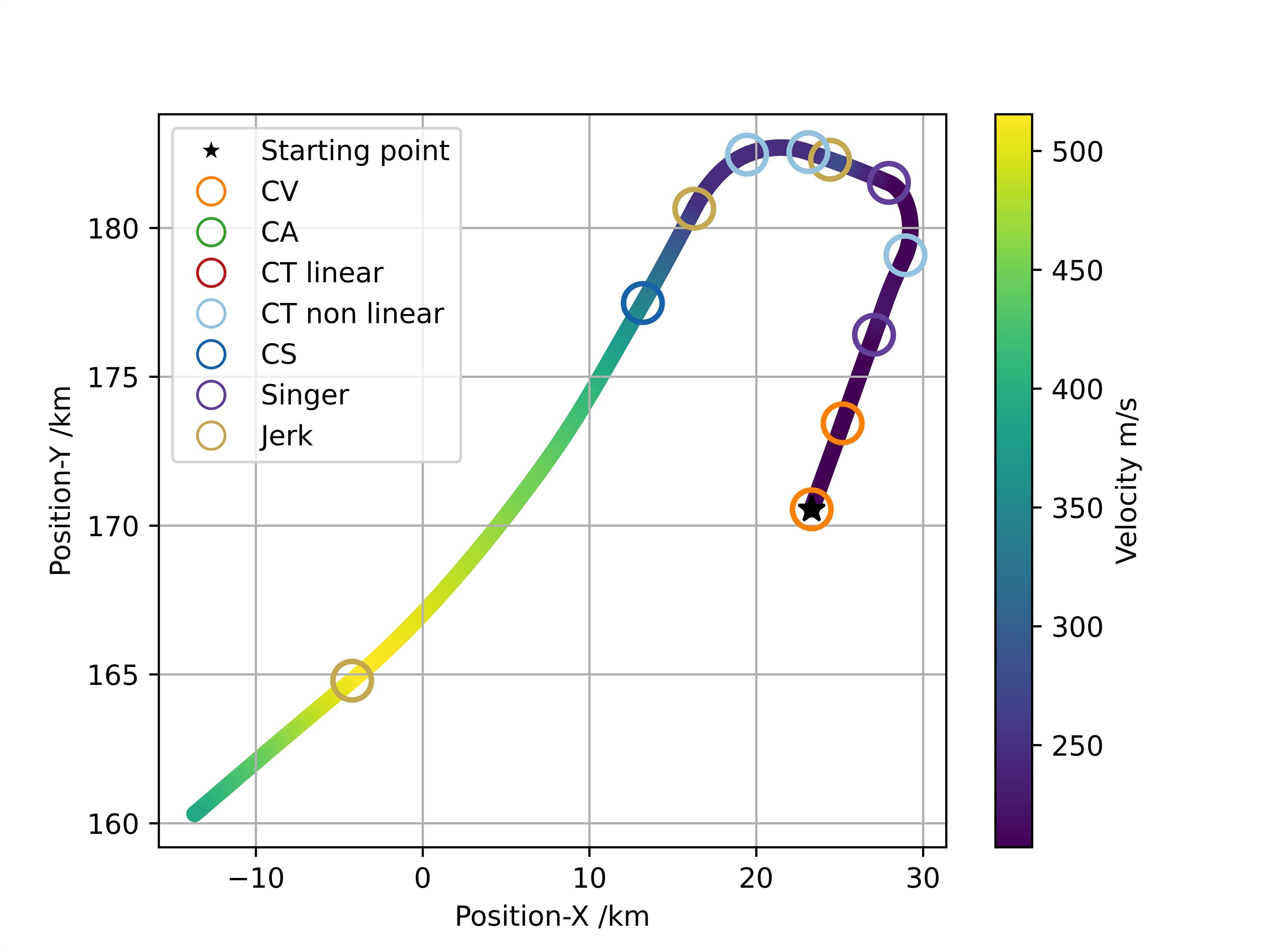}
        \caption{Trajectory for target 5.}
    \end{minipage}%
    \hfill
    \begin{minipage}{0.47\textwidth}
        \centering
        \includegraphics[width=\linewidth]{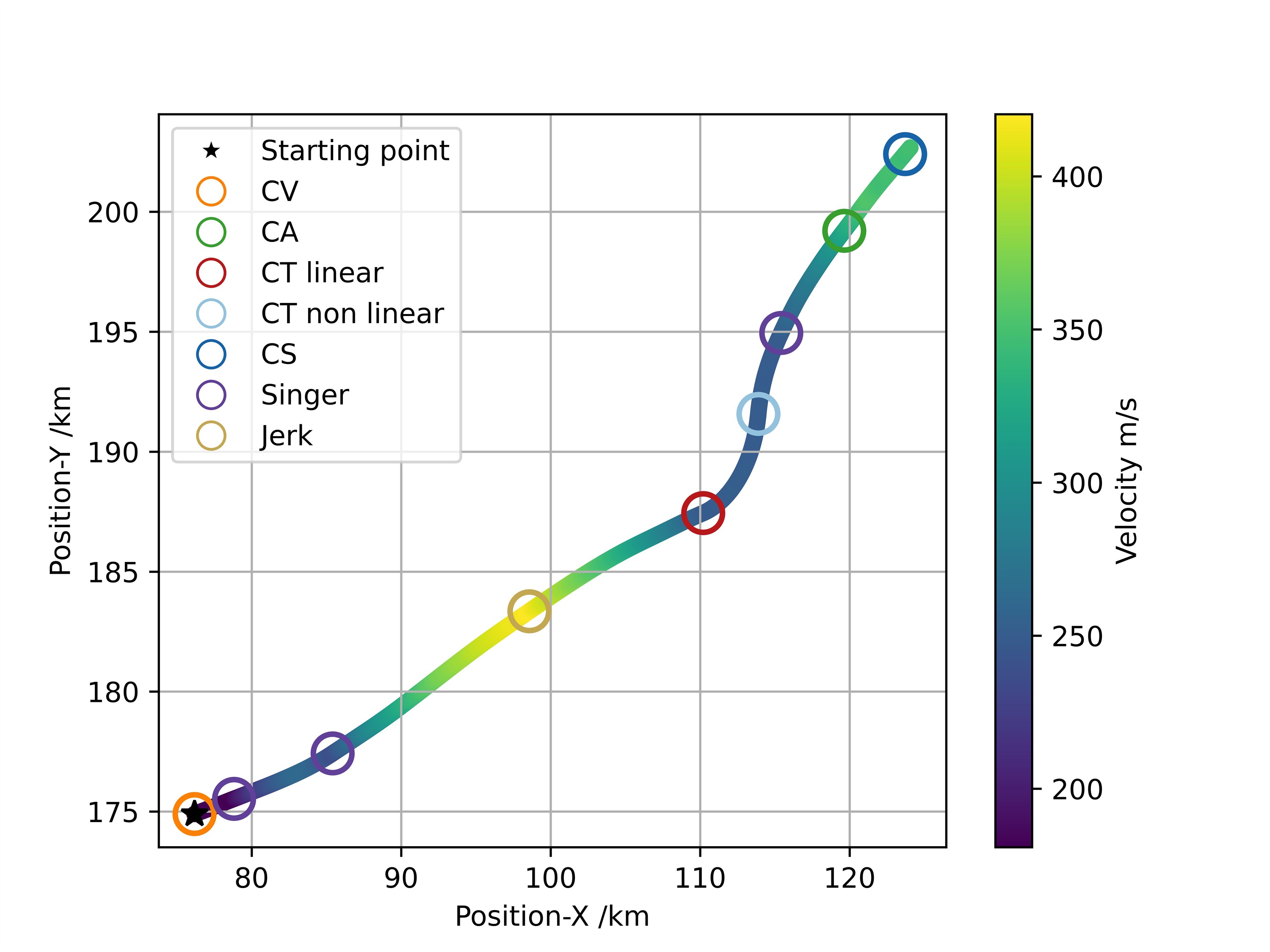}
        \caption{Trajectory for target 6.}
    \end{minipage}
\end{figure}

\begin{figure}[htbp]
    \centering
    \begin{minipage}{0.47\textwidth}
        \centering
        \includegraphics[width=\linewidth]{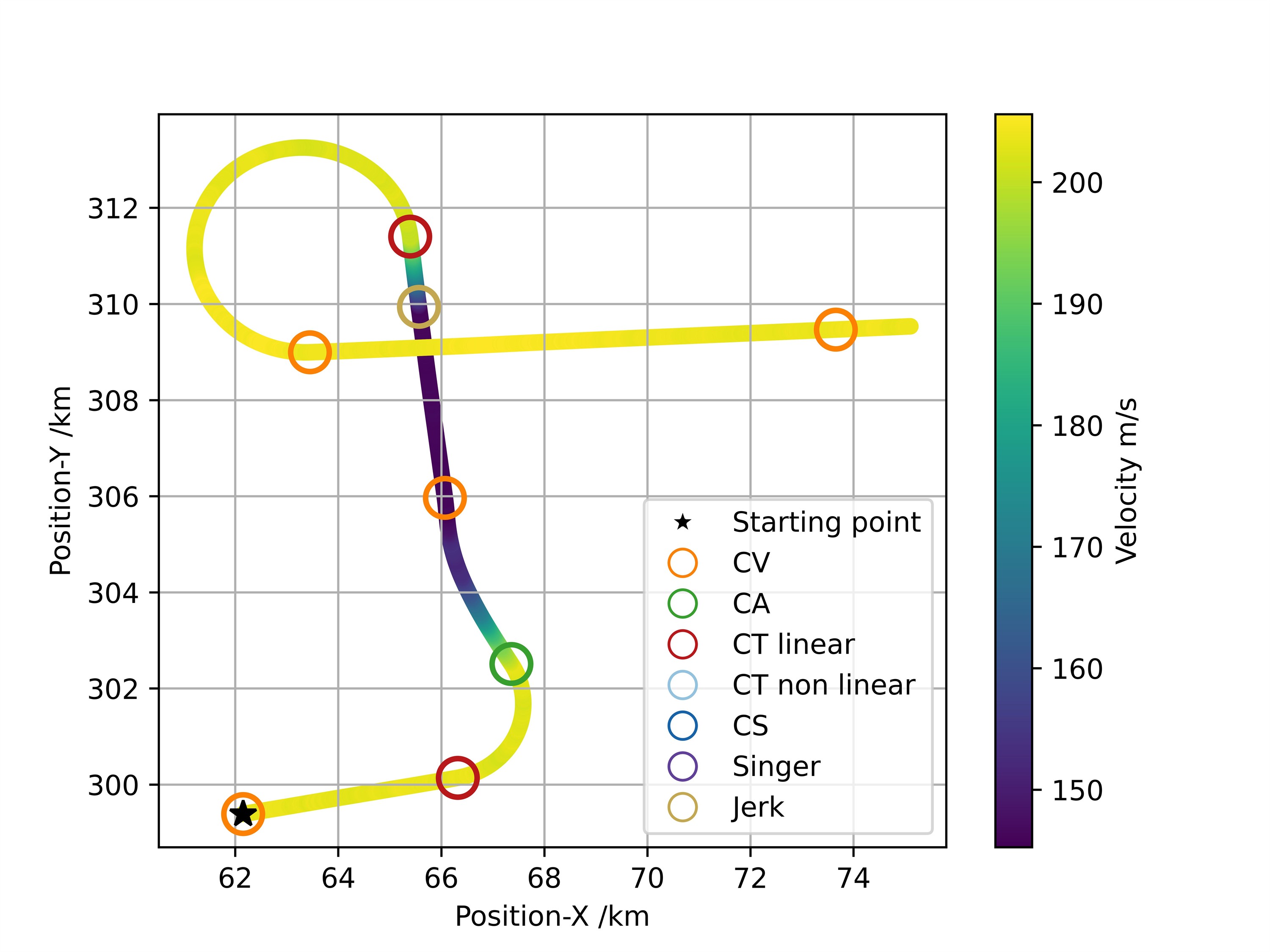}
        \caption{Trajectory for target 7.}
    \end{minipage}%
    \hfill
    \begin{minipage}{0.47\textwidth}
        \centering
        \includegraphics[width=\linewidth]{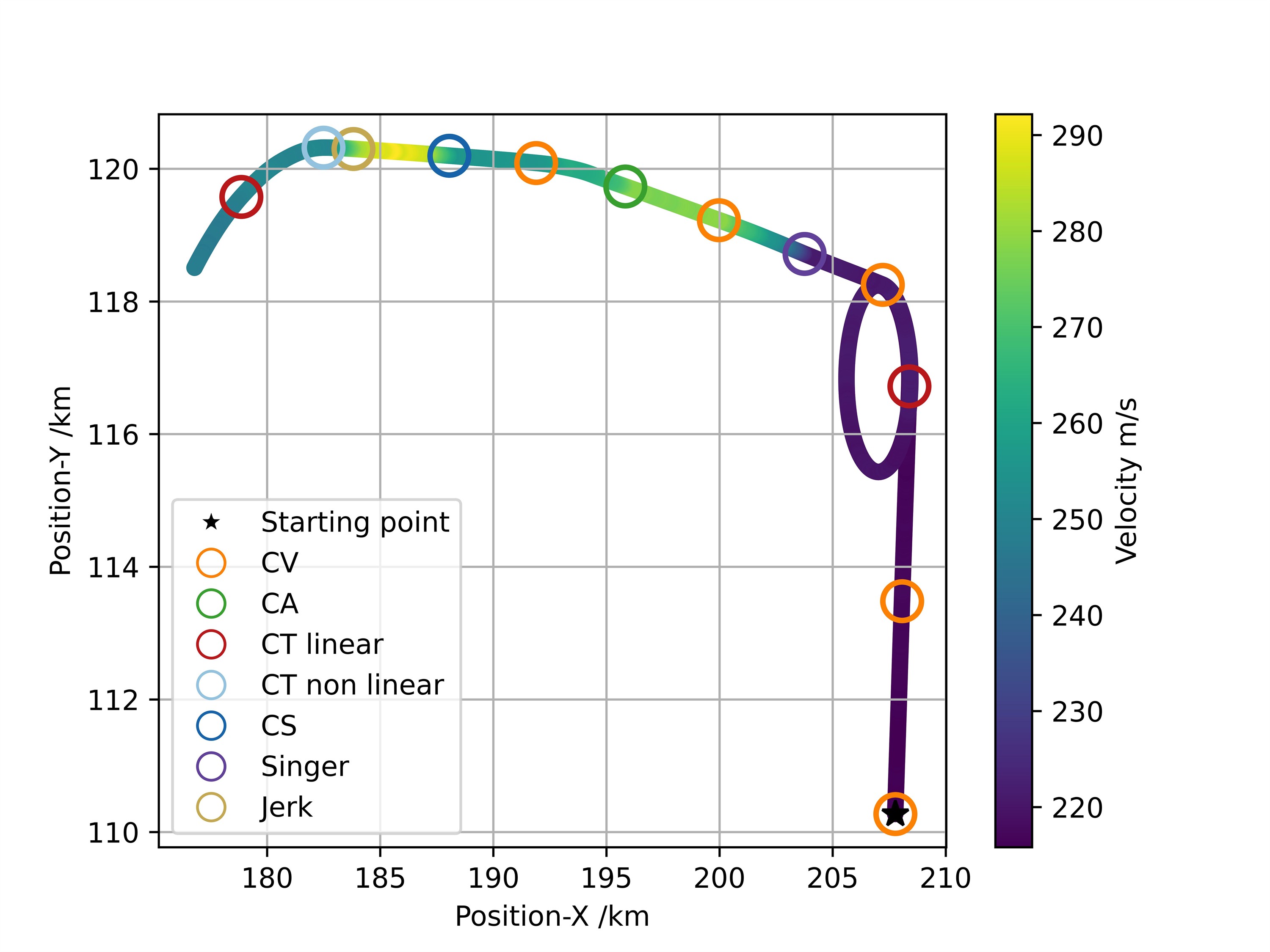}
        \caption{Trajectory for target 8.}
    \end{minipage}%
    \hfill
    \begin{minipage}{0.47\textwidth}
        \centering
        \includegraphics[width=\linewidth]{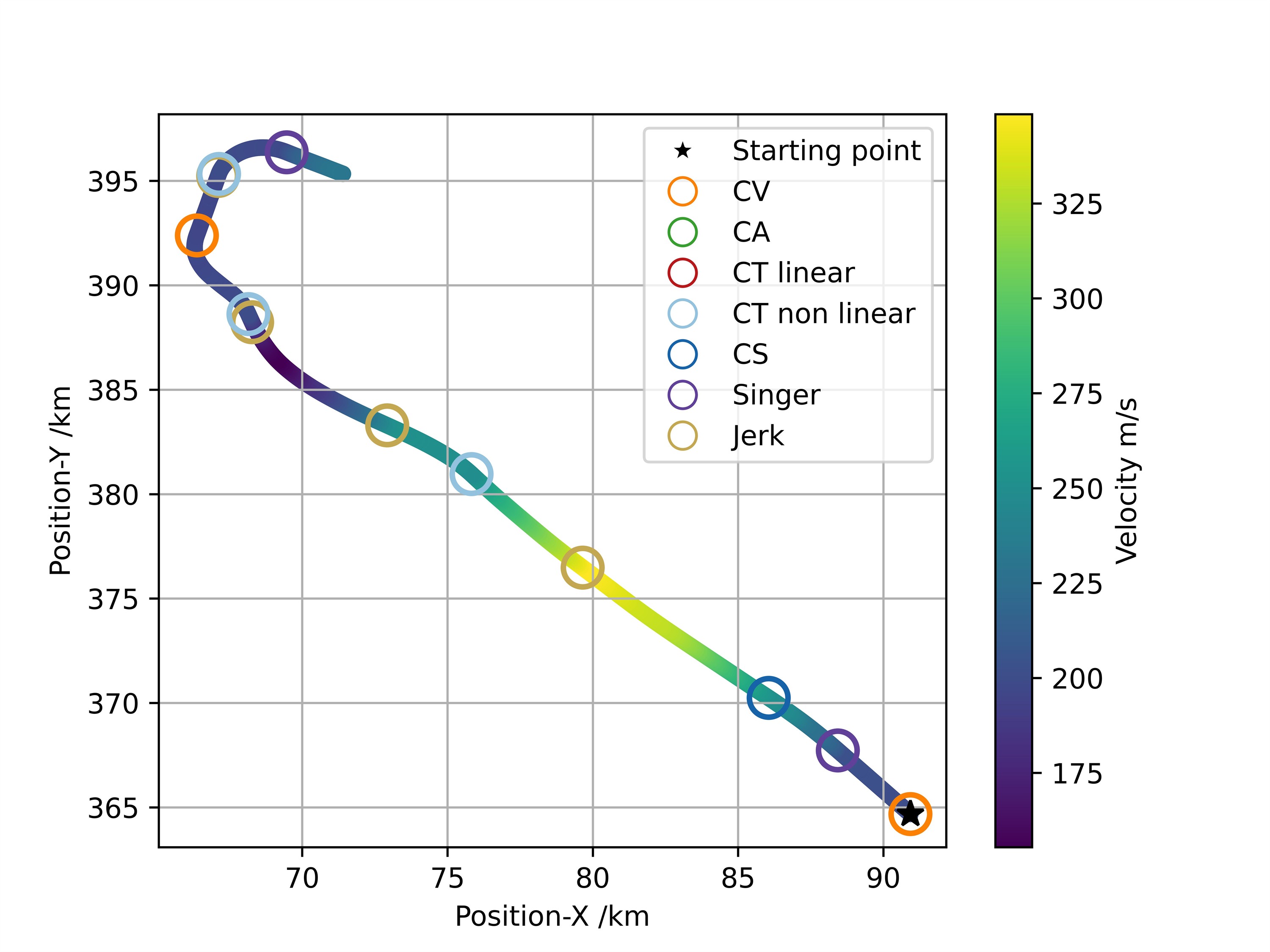}
        \caption{Trajectory for target 9.}
    \end{minipage}
\end{figure}


\begin{figure*}[htbp]
\centering
\begin{minipage}{0.45\textwidth}
    \centering
    \includegraphics[scale=0.5]{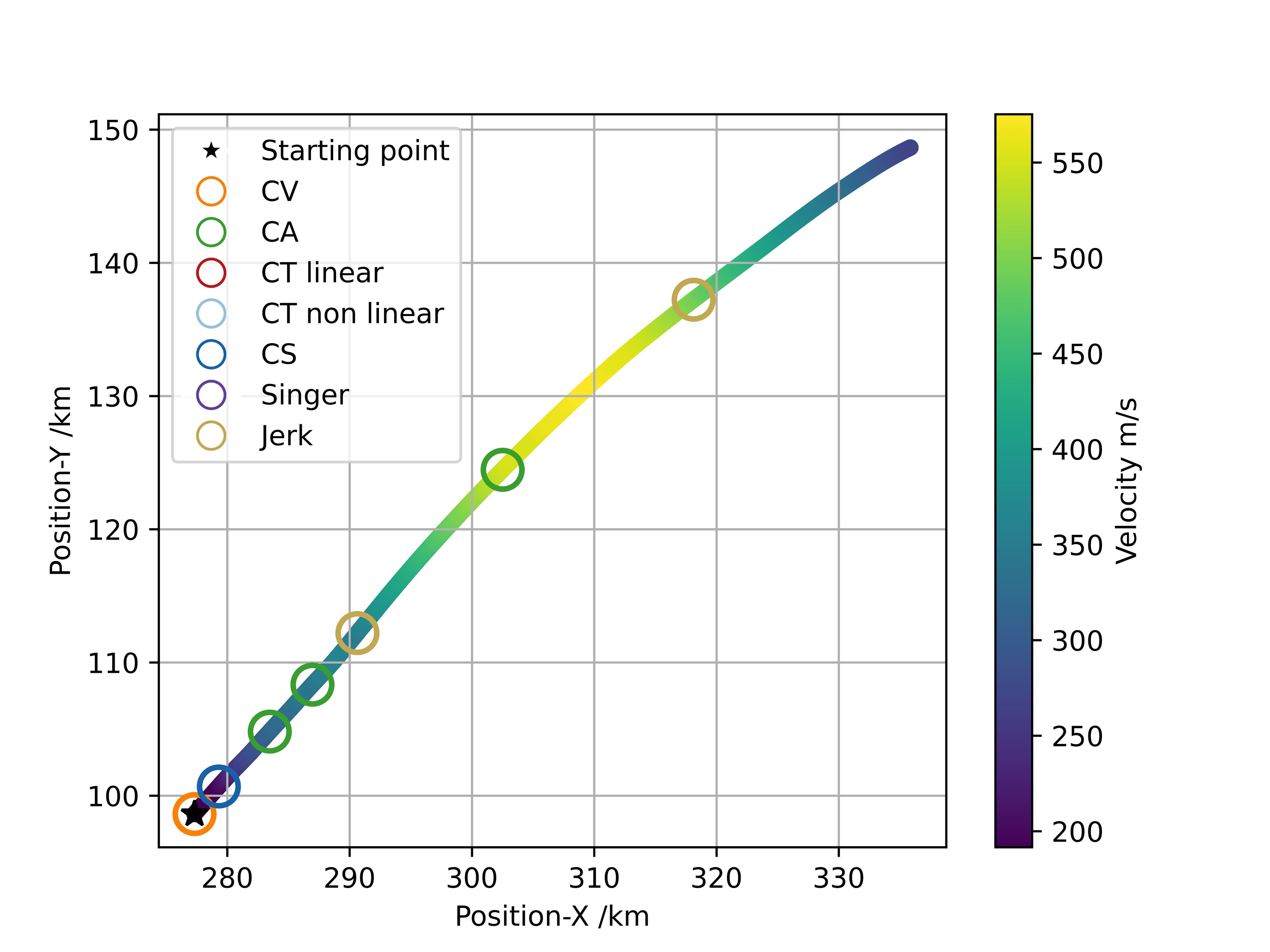}
    \captionof{figure}{Trajectory for target 10.}
    \label{fig:22}
\end{minipage}
\hspace{0.7cm} 
\begin{minipage}{0.45\textwidth}
    \centering
    \includegraphics[scale=0.5]{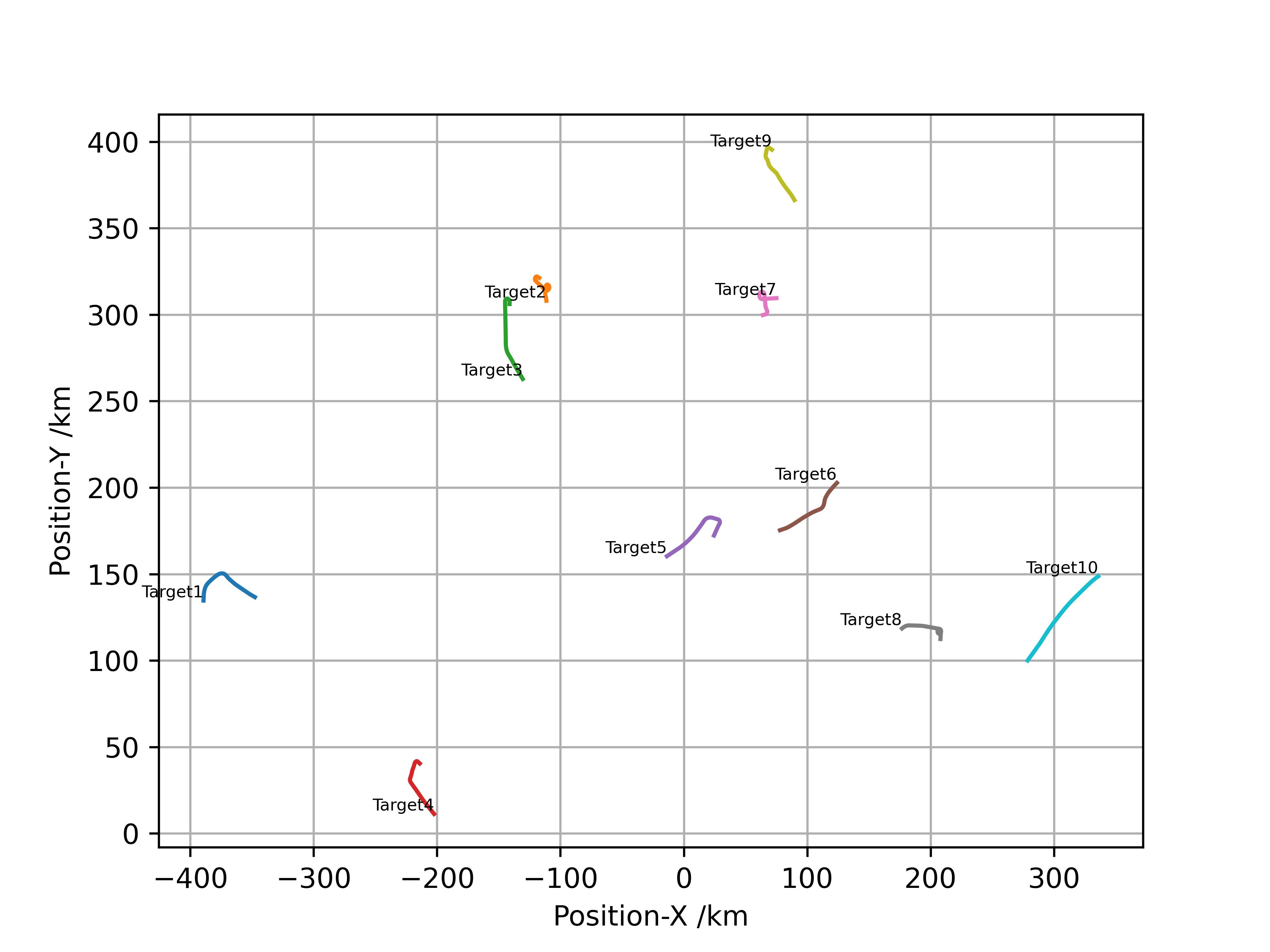}
    \captionof{figure}{Target Trajectories.}
    \label{fig:23}
\end{minipage}
\end{figure*}


\begin{figure*}[!htbp]
\centering
\subfloat[Position RMSE of 100 MC tracking for target 1.]{
    \includegraphics[width=0.5\linewidth]{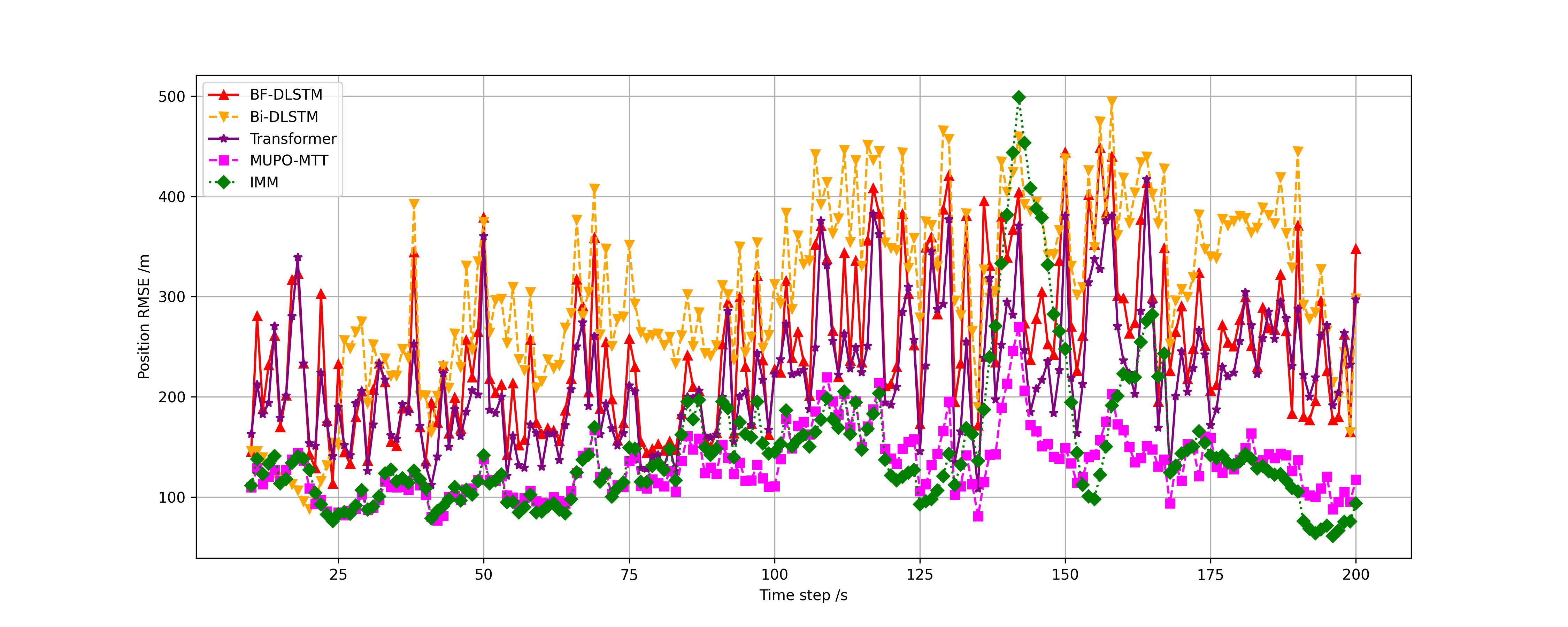}}
\subfloat[Position RMSE of 100 MC tracking for target 2.]{
    \includegraphics[width=0.5\linewidth]{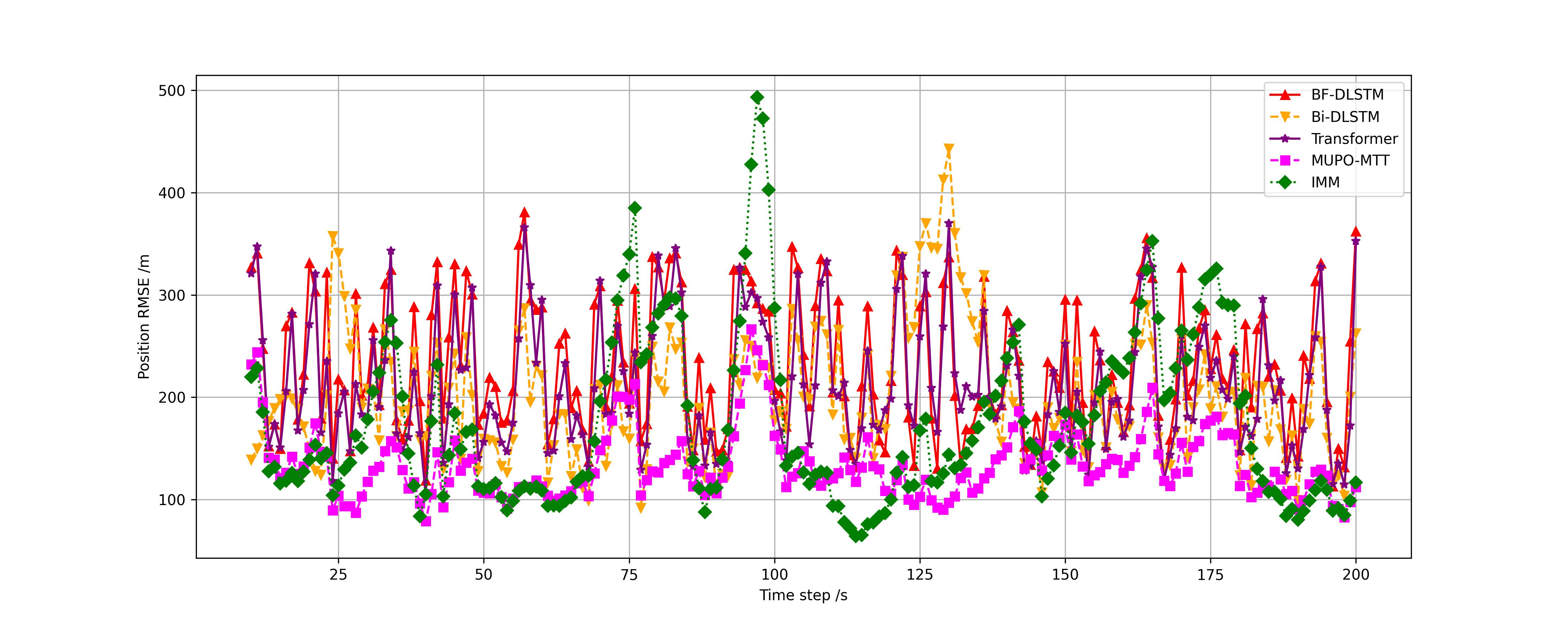}}\\
\caption{Tracking performance of different tracking methods.}
\label{fig:24}
\end{figure*}

\begin{figure*}[!htbp]
\centering
\subfloat[Position RMSE of 100 MC tracking for target 3.]{
    \includegraphics[width=0.5\linewidth]{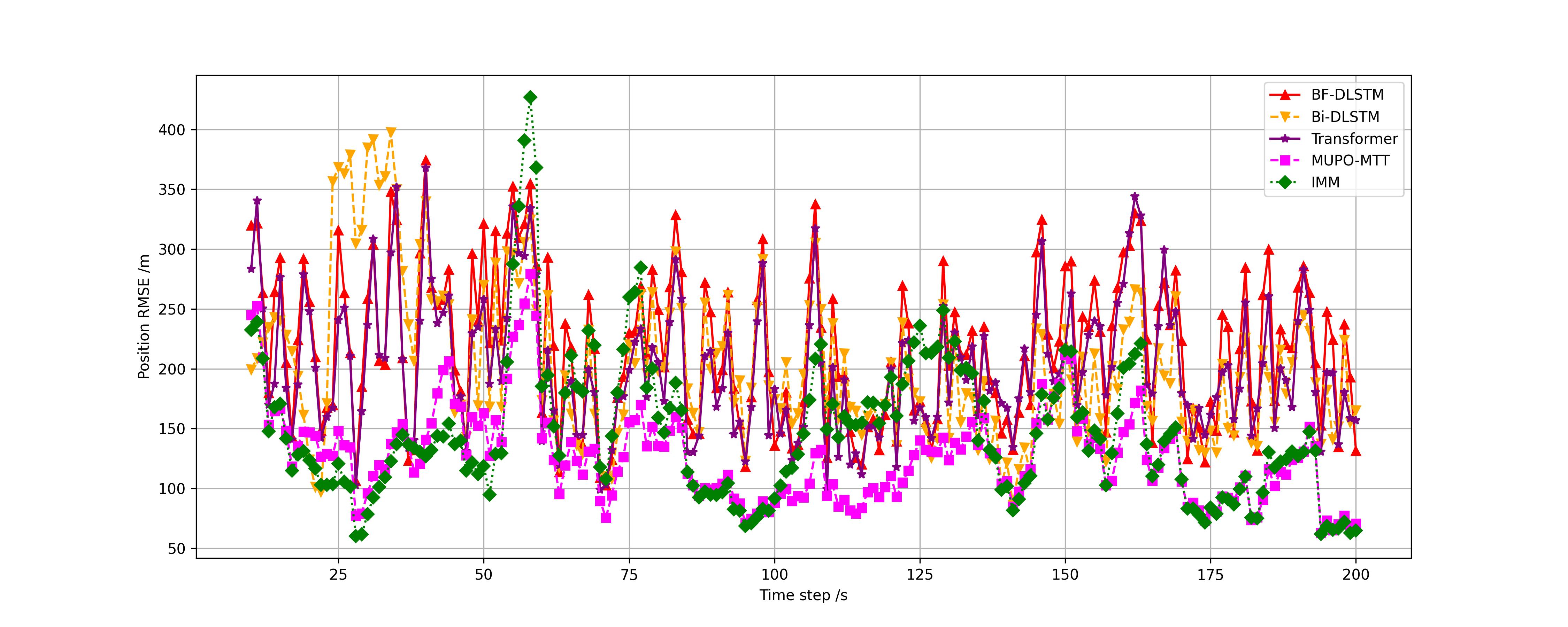}}
\subfloat[Position RMSE of 100 MC tracking for target 4.]{
    \includegraphics[width=0.5\linewidth]{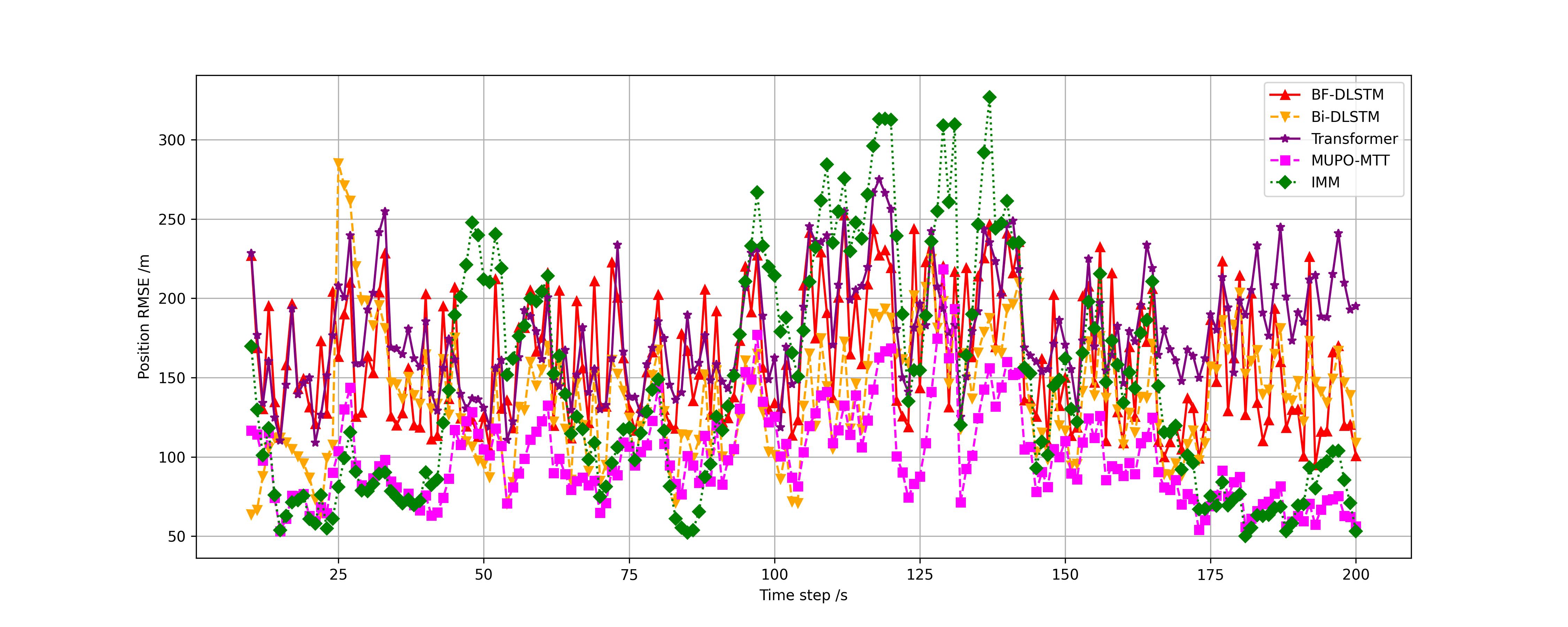}}\\
\caption{Tracking performance of different tracking methods.}
\end{figure*}

\begin{figure*}[!htbp]
\centering
\subfloat[Position RMSE of 100 MC tracking for target 5.]{
    \includegraphics[width=0.5\linewidth]{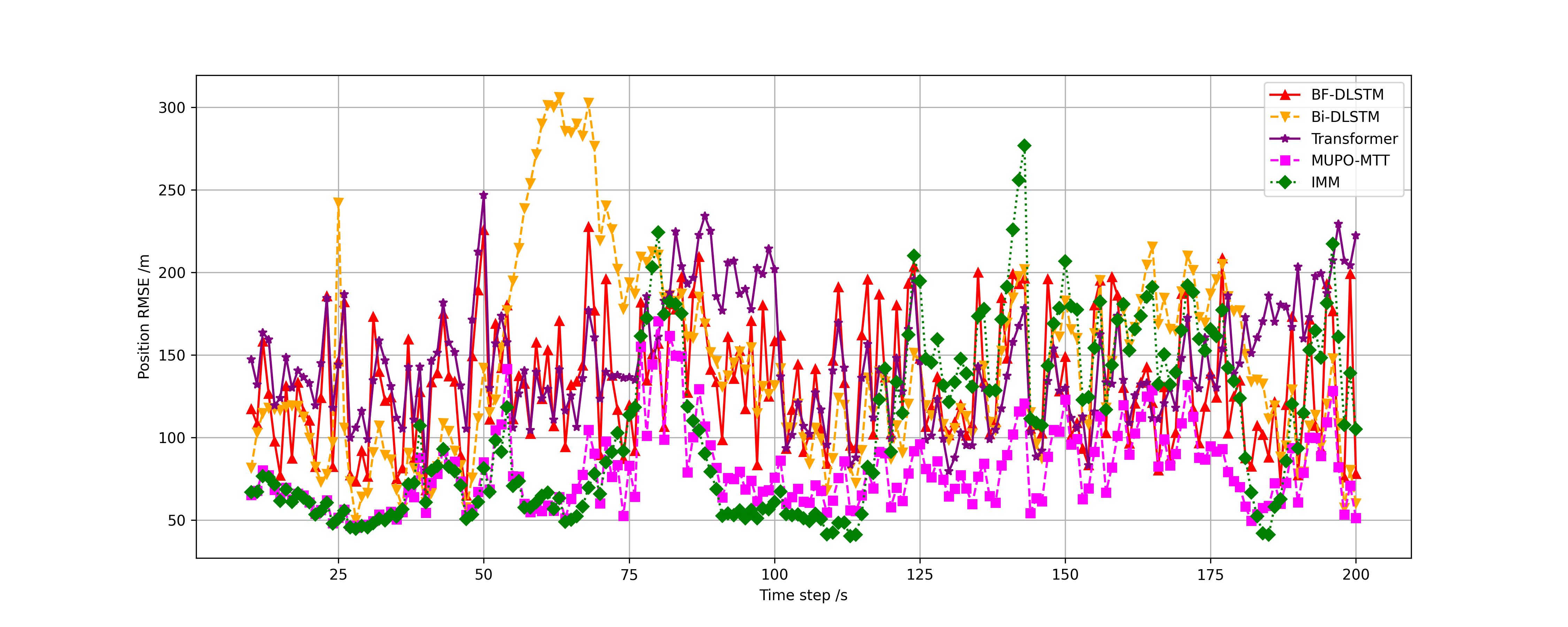}}
\subfloat[Position RMSE of 100 MC tracking for target 6.]{
    \includegraphics[width=0.5\linewidth]{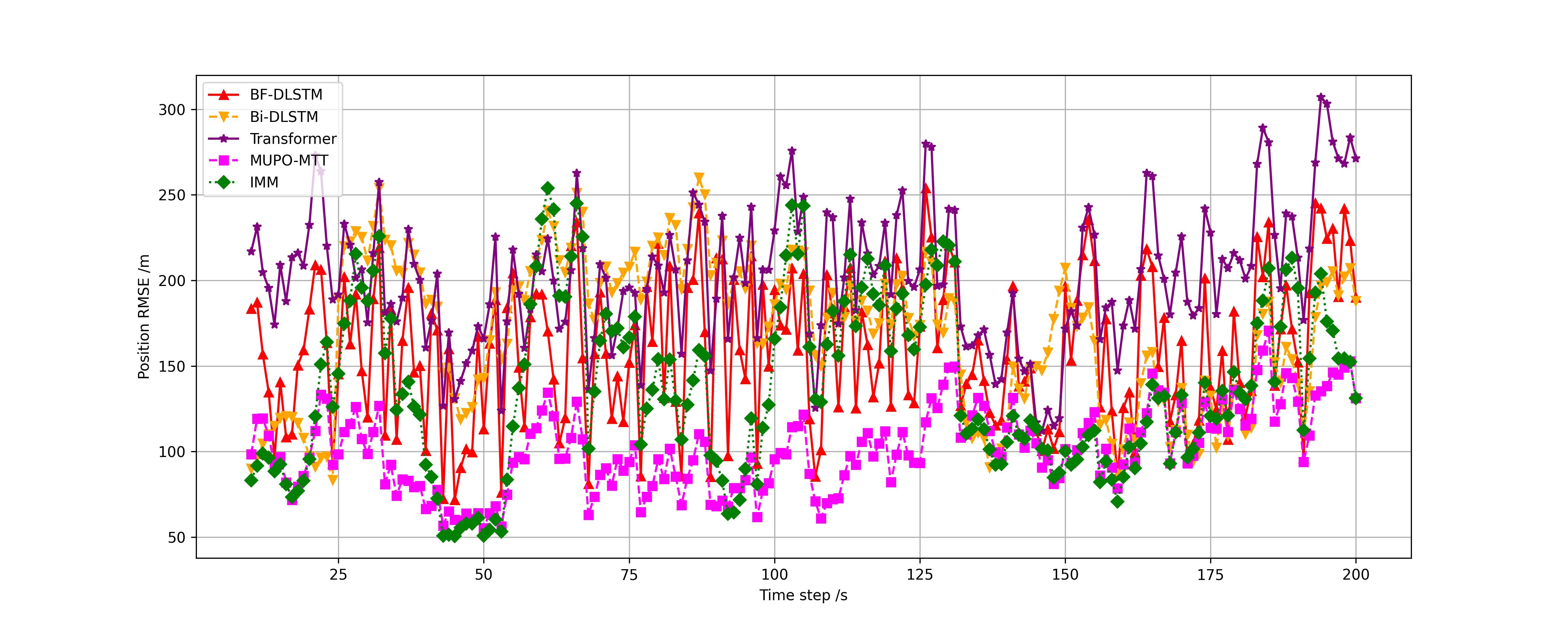}}\\
\caption{Tracking performance of different tracking methods.}
\end{figure*}

\begin{figure*}[!htbp]
\centering
\subfloat[Position RMSE of 100 MC tracking for target 7.]{
    \includegraphics[width=0.5\linewidth]{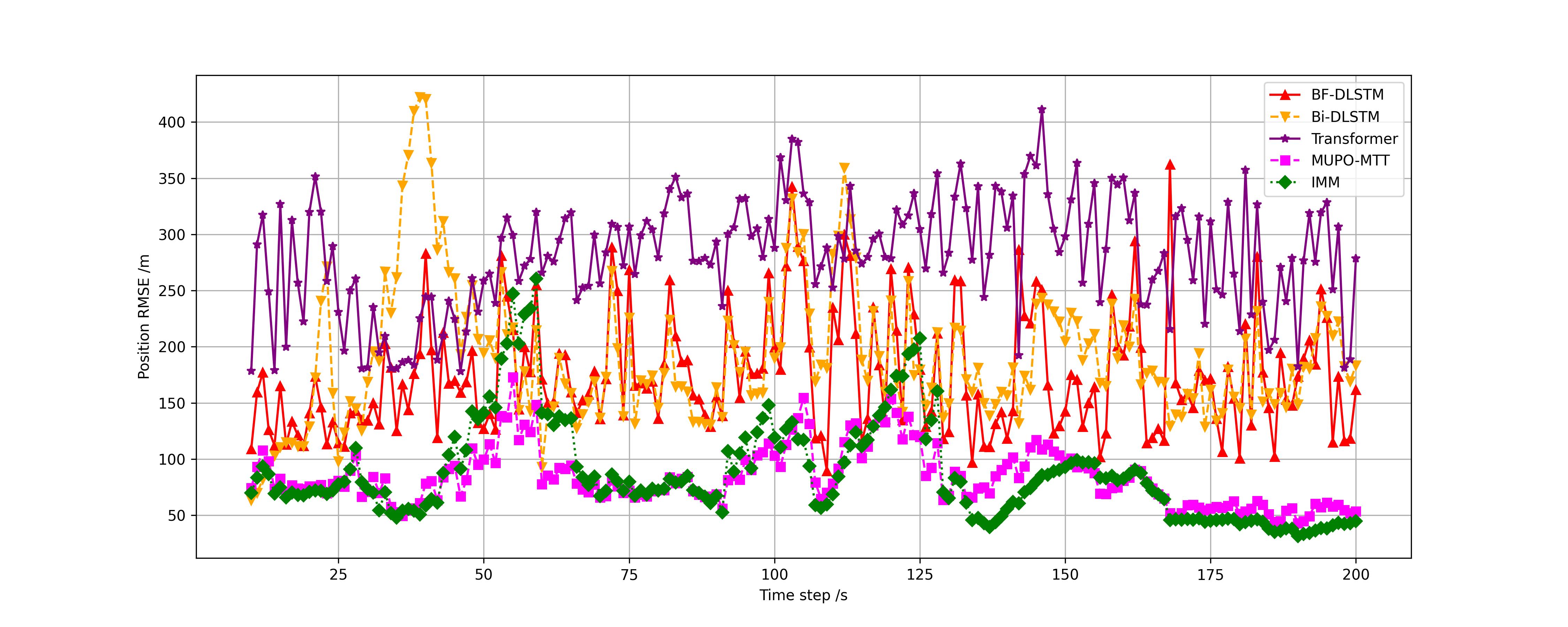}}
\subfloat[Position RMSE of 100 MC tracking for target 8.]{
    \includegraphics[width=0.5\linewidth]{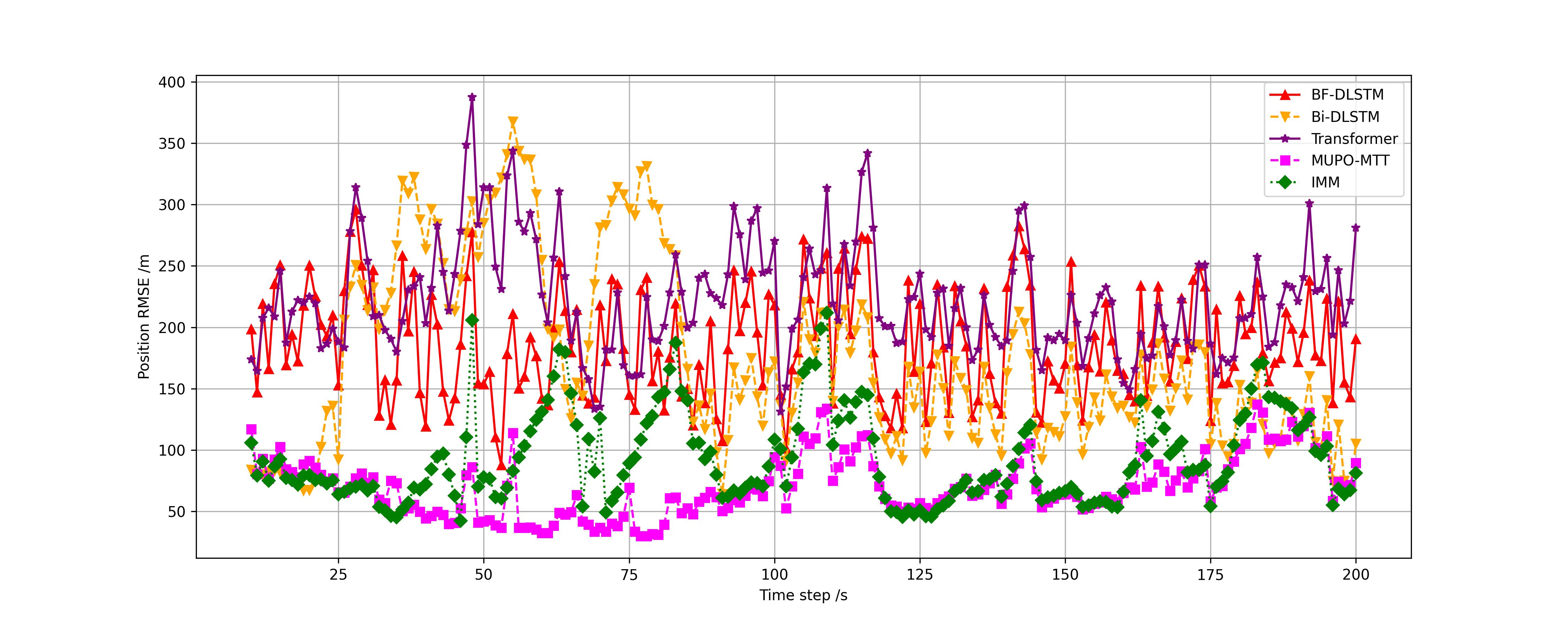}}\\
\caption{Tracking performance of different tracking methods.}
\end{figure*}
\begin{figure*}[!htbp]
\centering
\subfloat[Position RMSE of 100 MC tracking for target 9.]{
    \includegraphics[width=0.5\linewidth]{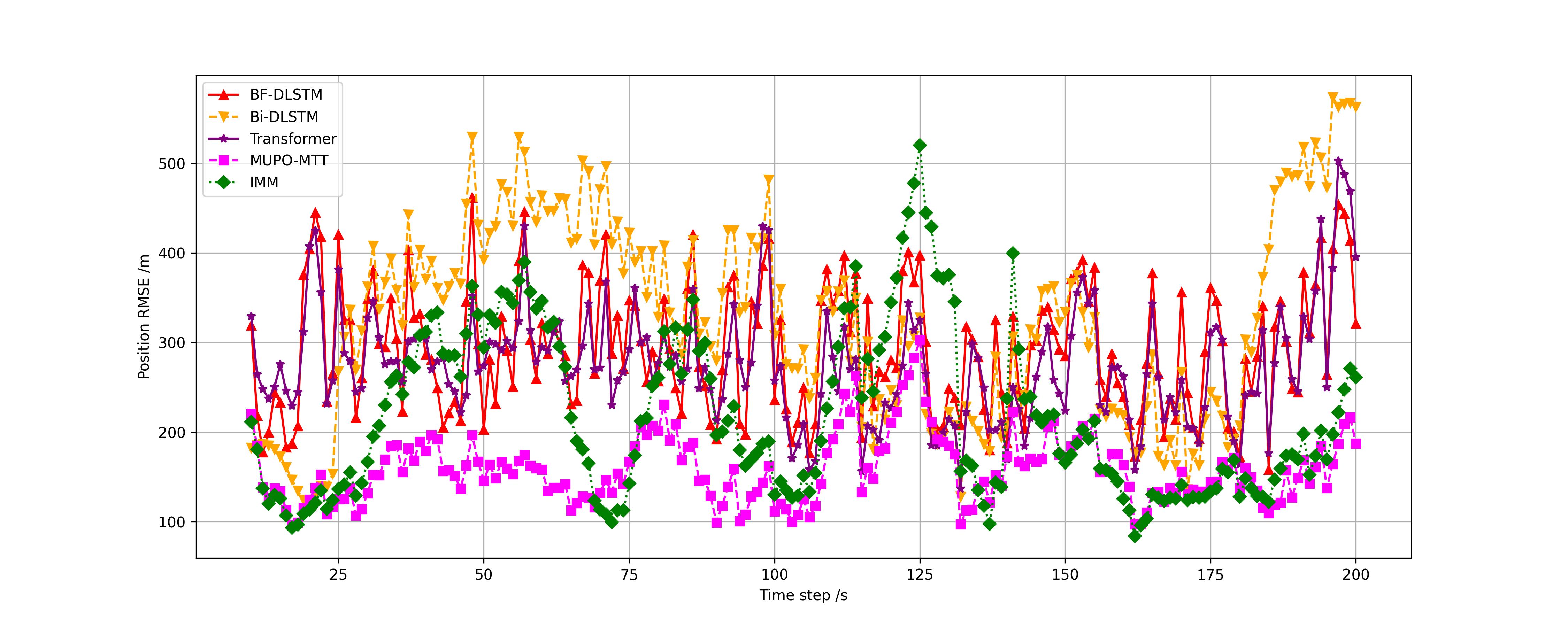}} 
\subfloat[Position RMSE of 100 MC tracking for target 10.]{
    \includegraphics[width=0.5\linewidth]{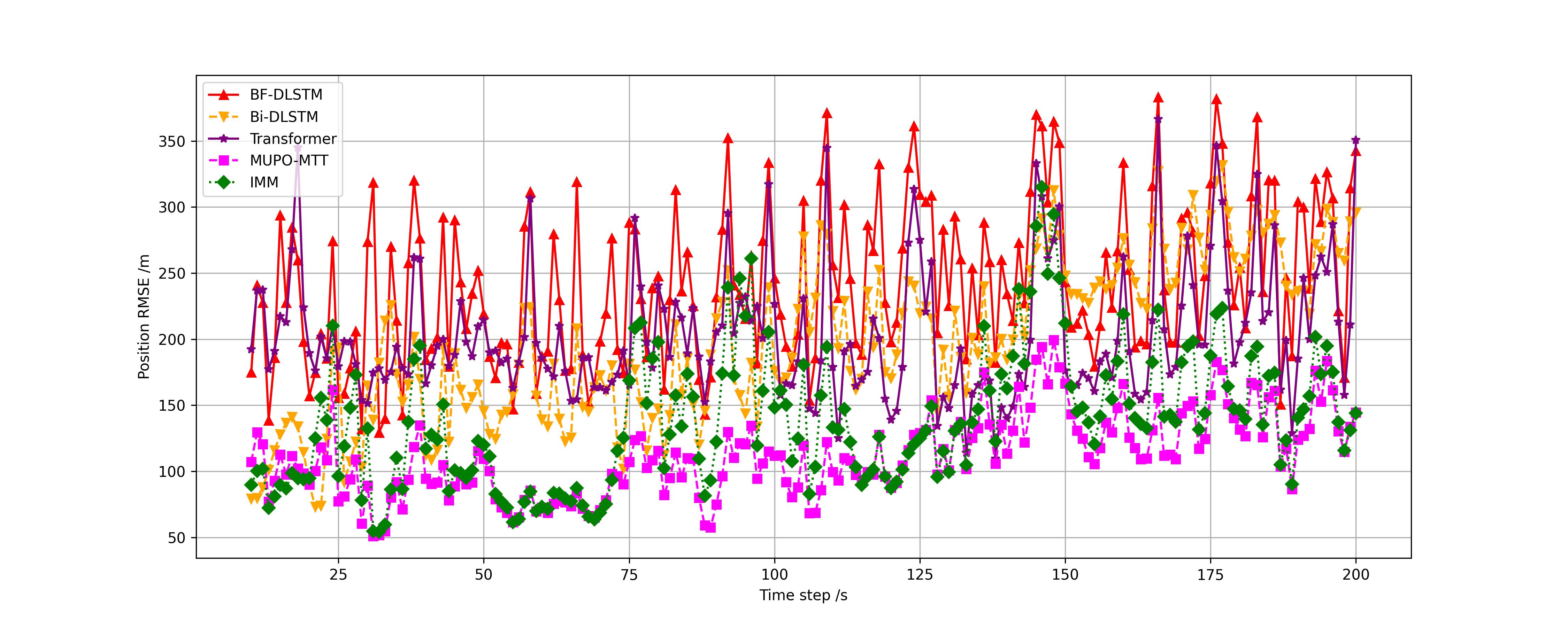}}
\caption{Tracking performance of different tracking methods.}
\label{fig:28}
\end{figure*}

\newpage
\newpage
{\small
\bibliographystyle{IEEEtran}
\bibliography{mybib} 
}

\end{document}